\documentclass[twocolumn]{openjournal}

\usepackage{enumitem}
\usepackage[T1]{fontenc}
\usepackage{ae,aecompl}
\usepackage{amsbsy,amsfonts,amsthm,color}
\usepackage{graphicx}		
\usepackage{amsmath}		
\usepackage{amssymb}		
\usepackage[utf8]{inputenc}
\usepackage{qtree}
\usepackage[normalem]{ulem}
\usepackage{gensymb}
\usepackage[table]{xcolor}
\usepackage{newtxtext,newtxmath}
\usepackage{verbatim}		
\usepackage{calrsfs}
\usepackage[cal=pxtx]{mathalfa}
\usepackage{natbib}
\usepackage{amstext}
\usepackage{lipsum}
\usepackage[colorlinks,linkcolor=blue,citecolor=blue,urlcolor=blue ]{hyperref}
\usepackage{float}
\usepackage[caption=false]{subfig}
\usepackage{upgreek}

\renewcommand*{\arraystretch}{1.1}
\newcommand\blfootnote[1]{%
  \begingroup
  \renewcommand\thefootnote{}\footnote{#1}%
  \addtocounter{footnote}{-1}%
  \endgroup
}

\newcommand{\alphasqrbullet}{\hbox{$\alpha^2_\bullet$}}

\newcommand{\Cisco}{\mbox{$\mathcal{C}_\mathrm{isco}$}}
\newcommand{\cs}{\mbox{$c_\mathrm{s}$}}

\newcommand{\ccircledastsqr}{\mbox{$c_\circledast^2$}}

\newcommand{\dE}{\mbox{$\dot{E}$}}
\newcommand{\Disco}{\mbox{$\mathcal{D}_\mathrm{isco}$}}

\newcommand{\dJcircledcircvec}{\mbox{$\boldsymbol{\dot{J}}_\circledcirc$}}

\newcommand{\dMbullet}{\mbox{$\dot{M}_\bullet$}}
\newcommand{\dmbullet}{\mbox{$\dot{m}_\bullet$}}
\newcommand{\dMcircledastarrowcircledcirc}{\mbox{$\dot{M}_{\circledast \rightarrow \circledcirc}$}}

\newcommand{\dMcircledcircarrowbullet}{\mbox{$\dot{M}_{\circledcirc \rightarrow \bullet}$}}

\newcommand{\dMEdd}{\mbox{$\dot{M}_\mathrm{Edd}$}}
\newcommand{\dSbulletaccvec}{\mbox{$\boldsymbol{\dot{S}}_\mathrm{\bullet ,acc}$}}
\newcommand{\dSbulletBPvec}{\mbox{$\boldsymbol{\dot{S}}_\mathrm{\bullet ,BP}$}}
\newcommand{\dSbulletvec}{\mbox{$\boldsymbol{\dot{S}}_\bullet$}}

\newcommand{\Eisco}{\mbox{$E_\mathrm{isco}$}}
\newcommand{\epsilonf}{\mbox{$\varepsilon_\mathrm{f}$}}
\newcommand{\epsilonr}{\mbox{$\varepsilon_\mathrm{r}$}}

\newcommand{\fEdd}{\mbox{$f_\mathrm{Edd}$}}

\newcommand{\gisco}{\mbox{$g_\mathrm{isco}$}}

\newcommand{\hisco}{\mbox{$h_\mathrm{isco}$}}

\newcommand{\jcirc}{\mbox{$j_\mathrm{circ}$}}

\newcommand{\jcircledcirchat}{\mbox{$\boldsymbol{\hat{j}}_\circledcirc$}}

\newcommand{\Jisco}{\mbox{$J_\mathrm{isco}$}}

\newcommand{\kB}{\mbox{$k_\mathrm{B}$}}
\newcommand{\kappaFF}{\mbox{$\kappa_\mathrm{ff}$}}
\newcommand{\kappaES}{\mbox{$\kappa_\mathrm{es}$}}
\newcommand{\kappatot}{\mbox{$\kappa_\mathrm{tot}$}}

\newcommand{\Lobs}{\mbox{$L_\mathrm{obs}$}}

\newcommand{\Mgas}{\mbox{$M_\mathrm{gas}$}}

\newcommand{\mpr}{\mbox{$m_\mathrm{pr}$}}

\newcommand{\Omegaisco}{\mbox{$\Omega_\mathrm{isco}$}}

\newcommand{\pgas}{\mbox{$p_\mathrm{gas}$}}
\newcommand{\Pisco}{\mbox{$\mathcal{P}_\mathrm{isco}$}}
\newcommand{\prad}{\mbox{$p_\mathrm{rad}$}}
\newcommand{\ptot}{\mbox{$p_\mathrm{tot}$}}

\newcommand{\rBondi}{\mbox{$r_\mathrm{Bondi}$}}
\newcommand{\rcapt}{\mbox{$r_\mathrm{capt}$}}

\newcommand{\Rgrav}{\mbox{$R_\mathrm{grav}$}}
\newcommand{\RHL}{\mbox{$R_\mathrm{HL}$}}
\newcommand{\Rinit}{\mbox{$R_\mathrm{init}$}}
\newcommand{\Risco}{\mbox{$R_\mathrm{isco}$}}
\newcommand{\Router}{\mbox{$R_\mathrm{outer}$}}
\newcommand{\Rph}{\mbox{$R_\mathrm{ph}$}}
\newcommand{\Rcalisco}{\mbox{$\mathcal{R}_\mathrm{isco}$}}

\newcommand{\Sbulletvec}{\mbox{$\textbf{\textit{S}}_\bullet$}}
\newcommand{\sbullethat}{\mbox{$\boldsymbol{\hat{s}}_\bullet$}}

\newcommand{\sigmaT}{\mbox{$\sigma_\mathrm{T}$}}
\newcommand{\sigmaSB}{\mbox{$\sigma_\mathrm{SB}$}}

\newcommand{\tBondi}{\mbox{$t_\mathrm{Bondi}$}}
\newcommand{\tBP}{\mbox{$t_\mathrm{BP}$}}

\newcommand{\tdep}{\mbox{$t_\mathrm{dep}$}}
\newcommand{\tprec}{\mbox{$t_\mathrm{prec}$}}

\newcommand{\tvisc}{\mbox{$t_\mathrm{visc}$}}

\newcommand{\ucirc}{\mbox{$u_\mathrm{circ}$}}

\newcommand{\xisco}{\mbox{$x_\mathrm{isco}$}}

\newcommand{\astropy}{\mbox{\textsc{Astropy}}}

\newcommand{\ketju}{\mbox{\textsc{Ketju}}}
\newcommand{\matplotlib}{\mbox{\textsc{Matplotlib}}}
\newcommand{\numpy}{\mbox{\textsc{NumPy}}}

\newcommand{\scipy}{\mbox{\textsc{SciPy}}}

\newcommand{\Msun}{\mbox{$\mathrm{M_\odot}$}}

\newenvironment{shortitem}
{\begin{list}{$\bullet$}{\topsep=0pt\itemsep=0pt\parsep=0pt\parskip=0pt\leftmargin=12pt}}
{\end{list}}
\newcommand{\bsi}{\begin{shortitem}}
\newcommand{\esi}{\end{shortitem}}
\newenvironment{shortsubitem}
{\begin{list}{$\circ$}{\topsep=0pt\itemsep=0pt\parsep=0pt\parskip=0pt\leftmargin=12pt}}
{\end{list}}
\newcommand{\bssi}{\begin{shortsubitem}}
\newcommand{\essi}{\end{shortsubitem}}

\newcommand{\App}[1]{Appendix~\ref{app:#1}}
\newcommand{\Eq}[1]{equation~(\ref{eq:#1})}
\newcommand{\Eqno}[1]{(\ref{eq:#1})}
\newcommand{\Fig}[1]{Fig.~\ref{fig:#1}}
\newcommand{\Sec}[1]{Section~\ref{sec:#1}}
\newcommand{\Tab}[1]{Table~\ref{tab:#1}}
\DeclareRobustCommand{\VAN}[3]{#2}
\let\VANthebibliography\thebibliography
\def\thebibliography{\DeclareRobustCommand{\VAN}[3]{##3}\VANthebibliography}

\begin{document}

\title{RABBITS -- III. Modelling relativistic accretion discs around spinning black holes \\ in galaxy formation simulations\vspace{-3em}}

\author{Dimitrios Irodotou$^{1\dagger}$} 
\author{Shihong Liao$^{2\dagger}$} 
\author{Theodoros Nakas$^{3}$} 
\author{Geoffrey Comp\`ere$^{4}$} 
\author{Roberto Oliveri$^{5}$} 

\author{Jessica M. Hislop$^{1}$} 
\author{Alexander Rawlings$^{6}$} 
\author{Sonja Soininen$^{6}$} 
\author{Aswin P. Vijayan$^{7}$} 

\affiliation{$^1$The Institute of Cancer Research, 123 Old Brompton Road, London SW7 3RP, UK}
\affiliation{$^2$ Key Laboratory for Computational Astrophysics, National Astronomical Observatories, Chinese Academy of Sciences, Beijing 100101, China}
\affiliation{$^3$Cosmology, Gravity, and Astroparticle Physics Group, Center for Theoretical Physics of the Universe, Institute for Basic Science (IBS), Daejeon 34126, Korea}
\affiliation{$^4$Universit\'e Libre de Bruxelles, BLU-ULB Brussels Laboratory of the Universe, C.P. 231, B-1050 Bruxelles, Belgium}
    \affiliation{$^5$LUTH, Laboratoire Univers et Theories, Observatoire de Paris, 5 place Jules Janssen, 92190 Meudon, France}
\affiliation{$^6$Department of Physics, University of Helsinki, Gustaf Hällströmin katu 2, FI-00014, Helsinki, Finland}
\affiliation{$^7$Astronomy Centre, University of Sussex, Falmer, Brighton BN1 9QH, UK}

\blfootnote{$^{\dagger}$ Corresponding authors: di.irodotou@gmail.com \& shliao@nao.cas.cn}

\begin{abstract}
In this third study of the `\textbf{R}esolving superm\textbf{A}ssive \textbf{B}lack hole \textbf{B}inaries \textbf{I}n galac\textbf{T}ic hydrodynamical \textbf{S}imulations' (\textbf{\textsc{Rabbits}}) series we develop and implement a geometrically thin relativistic accretion disc model, which self--consistently evolves the mass and spin vector of black holes via analytically modelling the structure of steady--state accretion discs. The model employs a suite of relativistic, local solutions where pressure is dominated by either gas or radiation, while opacity is primarily governed by either electron scattering or free-free absorption. These local solutions are piece--wisely combined to form the global structure of the accretion disc based on each solution's range of validity. By explicitly modelling the structure of accretion discs, the model mitigates the stochasticity inherent in Bondi-type prescriptions, resulting in an approach where every episode of black hole mass accretion is derived from first principles. For the first time, our model enables galaxy formation simulations to place constraints on accretion disc sizes and structures. In addition, flux and temperature radial profiles can be directly extracted from the simulation, enabling the generation of spectral energy distributions (SED). Consequently, by incorporating the thermal structure and spacetime geometry around spinning black holes, our model more accurately captures the energetic output of quasars, overcoming critical limitations of classical approaches. Along with this manuscript, we make public a C version of the model appropriate to be used as a module in simulations, a Python version of the model that can be used independently to post--process any simulation and build mock accretion discs, and an updated version of the \textsc{Relagn} model that has the capability of producing SEDs by building an accretion disc for a given set of parameters and extracting its surface density, temperature, and opacity profiles.
\end{abstract}

\maketitle

\section{Introduction} \label{sec:Introduction}

Supermassive black holes (SMBHs) are thought to reside at the centres of nearly all massive galaxies, as supported by compelling observational evidence from stellar--dynamical measurements in the Galactic Centre \citep[e.g.][]{GKM98,RAB98,GEG2010} as well as from direct imaging of black hole shadow of M87 and of Sagittarius A* \citep{EHT19,EHT22}. Even though supermassive black holes and their accretion discs occupy a relatively microscopic part of the galactic centre \citep[a few milliparsec,][]{MKM10}, the energetic and gravitational phenomena arising from these extreme environments have been found to affect the formation and evolution of galaxies \citep[e.g.][]{MTR98,FM00,TGB02,MH03,MHD03,F12,KH13,HR04,MM13}.

It has been more than half a century since the pioneering work of \cite{S64} and \cite{L69} who suggested that: as gas accumulates around massive objects, it forms flat accretion discs which are able to produce the observed powerful emissions by converting a fraction of their gravitational potential energy into radiation \citep{S82}. Soon after that, the first accretion disc models were formulated \citep[e.g.][]{PR72,NT73,SS73,SLE76,I77} in an attempt to capture the physical processes governing these discs (e.g. angular momentum transport, energy dissipation, vertical structure and outflows). Understanding these accretion processes is crucial not only for explaining high-energy astrophysical phenomena, but also for testing predictions of gravity theories \citep[e.g.][]{ABN96,PRP13,ARU23,MBI24}. Hence, black holes and their surroundings provide a unique testing ground for advancing our understanding of spacetime, plasma physics, thermodynamics, and particle physics \citep{L16}.

Over the past two decades, hydrodynamical simulations of galaxy formation have been incorporating models for black hole accretion and feedback processes \citep{SD15,VMT20,HSL22}. However, due to the limited temporal and spatial resolution \citep[but see also][]{HGS24,HSS24}, simulations rely on sub--grid models to capture the unresolved black hole processes \citep[e.g.][]{CSW06,BS09,DDS12,HDS14,SVG15,TKG17,WSH17,DAN19,HOB22}. The pioneering work of \cite{WL03,VHM03,DSH05,SDH05} paved the way for these models. In the \cite{SDH05} parametrisation, gas particles close to the black hole are stochastically absorbed by the black hole at a rate defined by the Bondi-Hoyle-Lyttleton formula \citep{HL39,BH44,B52}. In addition, accretion--related radiation was assumed to be generated at a fixed efficiency of 0.1 \citep[i.e. the mean value of the mass to energy conversion in a geometrically thin disk around a Schwarzschild black hole][]{SS73}.

Attempts to capture not just the black hole's mass evolution due to accretion but also its spin evolution have been introduced by many authors \citep[e.g.][]{SSH09,DDS10,MDP13,DVS14,FSP18,BS19,DBB21,HLS22,SVB24}. These models are usually based on the \cite{SS73} prescription for modelling the accretion disc, which -- unlike the \cite{NT73} model -- neglects relativistic effects \citep[but see also][]{FSP18,BS19,HLS23,KSS24}. As a result, the \cite{SS73} model describes Newtonian discs and thus inaccurately captures the energy dissipation and angular momentum transport in the inner regions of the accretion disc around spinning black holes \citep{WM03}. Black hole spins not only affect galactic evolution by regulating star formation \citep[due to the spin--dependent radiative efficiency of accretion][]{T74}, but also influence gravitational wave signatures from black hole mergers and have a strong influence on black hole recoil velocity following the merger \citep[e.g.][]{CLZ07,GHS07,LCZ10}. Therefore, incorporating a physically motivated and relativistically consistent model of accretion disc evolution and black hole spin dynamics is essential for capturing the interplay between black holes and their host galaxies, and for interpreting upcoming multi--messenger observations.

In this work -- for the first time -- we combine (i) a relativistic accretion disc model \citep{NT73,CO17}, (ii) the \ketju\ integrator that resolves small--scale dynamics around black holes \citep{RPJ17,RPM20}, and (iii) a smoothed--particle hydrodynamics (SPH) galaxy formation simulation \citep{STW05,STW06,AWN13,NON17}. Combining these three components allows us to simultaneously (i) model the accretion disc physics and evolve black hole properties, (ii) follow the black hole--stellar particles interactions with high precision, and (iii) self--consistently simulate the evolution of a galaxy.

This paper is organised as follows. In \Sec{Model}, we present the equations governing the structure and evolution of a black hole--accretion disc (BH--AD) system.\footnote{See also \App{Implementation} and \ref{app:Testing} where for the sake of minimizing the length of \Sec{Model}, we keep most technical details regarding, respectively, the numerical implementation and testing of the model.} We split this section into four subsections: 
\begin{enumerate}[wide=0pt,labelindent=10pt,labelwidth=10pt]
\item \Sec{Model:Properties} describes black hole (\Sec{Model:Properties:BH}) and accretion disc (Sections \ref{sec:Model:Properties:AD boundaries}, \ref{sec:Model:Properties:AD}, and \ref{sec:Model:Properties:AD structure}) properties, and illustrates the construction of an accretion disc through an example (\Sec{Model:Properties:Example}). 
\item \Sec{Model:Evolve} describes the structural evolution of the accretion disc (\Sec{Model:Evolve:AD}) and of the black hole (\Sec{Model:Evolve:BH}). 
\item \Sec{Model:Feedback} presents the updated AGN feedback model. 
\item \Sec{Model:Summary} provides a summary of the BH--AD model developed in this work.
\end{enumerate}

In \Sec{Results}, we perform a parameter space exploration to quantify the effects of different viscosity parameters; black hole spins, masses, and accretion rate; and different orientations (co-- and counter--rotation) on the structure (\Sec{Results:Validity}), surface density (\Sec{Results:Surface density}), and stability (\Sec{Results:Stability}) of the accretion disc. In \Sec{Applications}, we implement the model in an SPH galaxy formation simulation and present the evolution of an isolated galaxy (\Sec{Applications:Isolated}) along with our unique half--light radii and spectral energy distribution (SED) predictions (\Sec{Applications:Predictions}). Finally, in \Sec{Discussion}, we provide a summary of the model and discuss potential applications and future enhancements for it.

\section{Relativistic accretion discs around spinning black holes} \label{sec:Model}

Before we present our model, we summarise in \Tab{properties} the key properties and constants used to describe black holes and accretion discs throughout this work, along with their corresponding units and symbols. To more easily distinguish between different components we use as subscripts: a bullet ($\bullet$) for black hole properties, a circled circle ($\circledcirc$) for accretion disc properties, and a circled asterisk ($\circledast$) for gas particle properties.
\begin{table*}
	\centering
	\caption{Key properties and constants used to describe black holes and accretion discs in the main text. \App{Corrections} and \App{Implementation} contain additional properties which for simplicity are not included in this table.}
	\label{tab:properties}
	\begin{tabular}{lccc}
		\hline
	    Black hole properties & Symbols & Units & Equations \\
		\hline
		Spin & \Sbulletvec & cm$^2$ g s$^{-1}$ & \Eqno{Sbulletvec} \\
		Mass accretion rate & \dMbullet\ or \dMcircledcircarrowbullet & g s$^{-1}$ & \Eqno{dMcircledcircarrowbullet} \\
		Eddington fraction & \fEdd & -- & \Eqno{fEdd} \\
		Angular momentum rate & \dSbulletvec & cm$^2$ g s$^{-2}$ & \Eqno{dSbulletvec} \\
		Feedback energy & \dE & erg s$^{-1}$ & \Eqno{dE} \\
		\hline
	    Accretion disc properties & Symbols & Units & Equations \\
		\hline
		ISCO radius & \Risco & cm & \Eqno{Risco} \\
		Total angular momentum & $J_\circledcirc$ & cm$^2$ g s$^{-1}$ & \Eqno{Jcircledcirc} \\
		Gas--Electron Scattering surface density & $\Sigma \big|^\mathrm{gas}_\mathrm{es}$ & g cm$^{-2}$ & \Eqno{SigmaGES} \\
		Gas--Electron Scattering opening angle & $h \big|^\mathrm{gas}_\mathrm{es}$ & -- & \Eqno{hGES} \\		
		Gas--Electron Scattering pressure ratio & $\frac{\prad}{\pgas} \bigg|^\mathrm{gas}_\mathrm{es}$ & -- & \Eqno{pradpgasGES} \\
		Gas--Electron Scattering opacity ratio & $\frac{\kappaFF}{\kappaES} \bigg|^\mathrm{gas}_\mathrm{es}$ & -- & \Eqno{kappaFFkappaESGES} \\
		Radiation--Electron Scattering surface density & $\Sigma \big|^\mathrm{rad}_\mathrm{es}$ & g cm$^{-2}$ & \Eqno{SigmaRES} \\
		Radiation--Electron Scattering opening angle & $h \big|^\mathrm{rad}_\mathrm{es}$ & -- & \Eqno{hRES} \\		
		Radiation--Electron Scattering pressure ratio & $\frac{\pgas}{\prad} \bigg|^\mathrm{rad}_\mathrm{es}$ & -- & \Eqno{pgaspradRES} \\
		Radiation--Electron Scattering opacity ratio & $\frac{\kappaFF}{\kappaES} \bigg|^\mathrm{rad}_\mathrm{es}$ & -- & \Eqno{kappaFFkappaESRES} \\
		Gas--Free Free surface density & $\Sigma \big|^\mathrm{gas}_\mathrm{ff}$ & g cm$^{-2}$ & \Eqno{SigmaGFF} \\
		Gas--Free Free opening angle & $h \big|^\mathrm{gas}_\mathrm{ff}$ & -- & \Eqno{hGFF} \\
		Gas--Free Free pressure ratio & $\frac{\prad}{\pgas} \bigg|^\mathrm{gas}_\mathrm{ff}$ & -- & \Eqno{pradpgasGFF} \\
		Gas--Free Free opacity ratio & $\frac{\kappaES}{\kappaFF} \bigg|^\mathrm{gas}_\mathrm{ff}$ & -- & \Eqno{kappaESkappaFFGFF} \\
		BH--AD alignment condition & cos$(\theta)$ & -- & \Eqno{costheta} \\
		ISCO specific angular momentum & \Jisco & cm$^2$ s$^{-1}$ & \Eqno{Jisco} \\
		Radiative efficiency & \epsilonr & -- & \Eqno{Lobs} \\
		ISCO angular velocity & \Omegaisco & s$^{-1}$ & \Eqno{Omegacirc} \\
		\hline
	    Constants & Symbols & Value & Units \\
		\hline
		Gravity & $G$ & 6.67 $\times\ 10^{-8}$ & cm$^3$ g$^{-1}$ s$^{-2}$ \\
		Light speed & $c$ & 3.00 $\times\ 10^{10}$ & cm s$^{-1}$ \\
		Solar mass & \Msun & 1.99 $\times\ 10^{33}$ & g \\
		Proton mass & \mpr & 1.67 $\times\ 10^{-24}$ & g \\
		Thomson cross section & \sigmaT & 6.65 $\times\ 10^{-25}$ & cm$^2$ \\
		\hline
	\end{tabular}
\end{table*}

\subsection{Properties of the black hole--accretion disc system} \label{sec:Model:Properties}

\subsubsection{Black hole properties} \label{sec:Model:Properties:BH}

The `no--hair' theorem \citep[e.g.][]{H72} states that black holes can be entirely described by their mass \citep{E16,C70}, spin angular momentum \citep[][hereafter, spin]{K63}, and electric charge \citep{NCC65,NJ65}. For the astrophysical black holes considered in this study, the latter property can be taken to be zero (e.g. \citealt{AF13,CSL21}, but see also \citealt{ZT19}). Hence, a black hole's mass ($M_\bullet$) and spin vector (\Sbulletvec) are the two fundamental properties required to fully characterize a black hole. 

In this work, the spin vector is given by
\begin{flalign} \label{eq:Sbulletvec}
\Sbulletvec = S_\bullet\ \sbullethat\ \equiv \alpha_\bullet\ M_\bullet\ c\ \Rgrav\ \sbullethat  = \alpha_\bullet \frac{G\ M_\bullet^2}{c} \sbullethat \;,
\end{flalign}
where $\alpha_\bullet$ is the dimensionless spin parameter which describes the ratio between the angular momentum of the black hole and its maximum possible value \citep[i.e. ensures that the weak cosmic censorship conjecture of ][is not violated]{P69}, $c$ is the speed of light, $\Rgrav = G\ M_\bullet\ c^{-2}$ is the gravitational radius of the black hole, $G$ is the gravitational constant, and \sbullethat\ is the direction of \Sbulletvec. 

\subsubsection{Accretion disc boundaries} \label{sec:Model:Properties:AD boundaries}

As we move away from the ring singularity of a spinning black hole \citep{C68,K23}, there are three particle/photon orbits in the Kerr geometry that are relevant for the astrophysical black holes studied in this work \citep{BPT72}.
\begin{enumerate}[wide=0pt,labelindent=10pt,labelwidth=10pt]
\item The photon radius (\Rph) represents the smallest possible circular orbit of a particle with zero rest mass, which implies that accretion disc equations break down at the photon radius \citep{PSM12}. In this work, we assume that the accretion disc starts just outside the photon radius \citep[the so--called plunging or intra--ISCO region e.g.][]{AK00,P21,MB23} which is given by
\begin{flalign} \label{eq:Rph}
\Rph = 2 \Rgrav\ \left(1 + \mathrm{cos} \left[\frac{2 \mathrm{arccos}(\mp \alpha_\bullet)}{3} \right] \right) \;,
\end{flalign}
where the upper sign refers to co--rotating and the lower to counter--rotating orbits with respect to \Sbulletvec\ [see \Eq{costheta} for a description of how we quantify that].
\item The innermost stable circular orbit (ISCO) represents the minimum distance from the black hole allowed for a massive particle on a stable circular orbit (i.e. particles revolving around the black hole on orbits inside the ISCO will plunge into the black hole). In this work, the ISCO radius is given by
\begin{flalign} \label{eq:Risco}
\Risco = \Rgrav\ Z \;,
\end{flalign}
where
\begin{flalign} \label{eq:Z}
&Z = 3 + Z_2 \mp \sqrt{(3 - Z_1)(3 + Z_1 + 2Z_2)} \;, \\
&Z_1 = 1 + \left(1 - \alpha_\bullet^2 \right)^{1/3} \left[ (1 + \alpha_\bullet)^{1/3} + (1 - \alpha_\bullet)^{1/3} \right] \;, \\
&Z_2 = \sqrt{3\alpha_\bullet^2 + Z_1^2} \;.
\end{flalign}

\item The outermost radius \Router\ represents the outer edge of the accretion disc. As discussed by \cite{BW71}, relativistic gas discs are more prone to fragmentation than their Newtonian counterparts and their stability can be quantified by local criteria \citep[e.g.][]{GL65,T64}. It has been previously reported \citep[][]{P78,G03,HC13,TD14,R15} that a massive accretion disc will be subject to significant self--gravity -- especially at the outer parts -- which will trigger gravitational instabilities. Since in this work we model the structure of the accretion disc, we can quantify \Router\ (i.e. the stability radius of a thin accretion disc against the growth of small over-densities) by evaluating the `relativistic' Toomre parameter (see also discussion after \Eq{Jcircledcirc2}) 
\begin{flalign} \label{eq:Q}
Q = \frac{\kappa\ \cs}{\uppi\ G\ \Sigma} \;,
\end{flalign}
where $\kappa$ is the epicyclic frequency, \cs\ is the sound speed, and $\Sigma$ is the surface density profile. In \cite{BL67} coordinates, the relativistic epicyclic frequency can be written as \citep{G04}
\begin{flalign} \label{eq:kappa}
\kappa = \Omega \left[1 - 6\left( \frac{r}{\Rgrav} \right)^{-1} \pm 8\alpha_\bullet \left( \frac{r}{\Rgrav} \right)^{-3/2} - 3\alphasqrbullet \left( \frac{r}{\Rgrav} \right)^{-2} \right]^{1/2} \;,
\end{flalign}
where
\begin{flalign} \label{eq:Omega}
\Omega = \pm \frac{c^3}{G\ M_\bullet} \left[ \left( \frac{r}{\Rgrav} \right)^{3/2} \pm \alpha_\bullet \right]^{-1} \;,
\end{flalign}
is the relativistic Keplerian angular velocity \citep{BPT72} and $r$ is the Boyer--Lindquist radial coordinate. Finally, the (isothermal) sound speed can be written as a pressure--to--density ratio as
\begin{flalign} \label{eq:cs}
\cs = \sqrt{\frac{p}{\rho}} \;.
\end{flalign}
In the following subsection, we detail how the aforementioned radial profiles of the surface density, pressure, and density are calculated in our model.
\end{enumerate}

\subsubsection{Accretion disc properties} \label{sec:Model:Properties:AD}

In this work, we use a relativistic approach \cite[e.g.][]{BPT72,NT73,PT74} to model an accretion disc whose radial properties we describe below. As in most relativistic $\alpha$--disc models, physical properties of the accretion disc are functions of four parameters: three black hole properties, namely the mass, spin, and accretion rate; and the dimensionless viscosity coefficient $\alpha \leq 1$ introduced by \cite{SS73}.

A fundamental accretion disc property that regulates the properties of the BH--AD system [see \Eq{costheta}, (\ref{eq:fEdd}), and (\ref{eq:dSbulletvec})] is the total angular momentum magnitude of the accretion disc, which can be written as
\begin{flalign} \label{eq:Jcircledcirc}
J_\circledcirc = \int_{\circledcirc} 2\uppi\ r\ \Sigma\ \widetilde{J_\circledcirc}\ \mathrm{d}r \;,
\end{flalign}
where $\widetilde{J_\circledcirc}$ is the relativistic specific angular momentum given by
\begin{flalign} \label{eq:tildeJcircledcirc}
\widetilde{J_\circledcirc} = \sqrt{G\ M_\bullet\ r}\ \mathcal{C}^{-1/2}\ \mathcal{F} \;,
\end{flalign}
where $\mathcal{C}$ and $\mathcal{F}$ are the relativistic corrections introduced by \cite{BPT72} and depend on $\alpha_\bullet$, on the alignment or not between the angular momenta of the black hole and the accretion disc, and on the distance from the black hole (i.e. approach $\pm1$ far away from the black hole, see \Fig{correction_functions_all}). We provide all correction functions and show their dependencies on the aforementioned three parameters in \App{Corrections}.

In order to evaluate \Eq{Jcircledcirc}, we need to know the global surface density profile of the accretion disc, which can be constructed by piece--wisely combining locally valid surface density profiles (see \Fig{regional_solutions}, \ref{fig:surface_density}, and \ref{fig:accretion_disc}). Identifying and combining local solutions is only possible when one term/mechanism dominates the total pressure (i.e. the sum of the radiation pressure $\prad$ and the gas pressure $\pgas$)
\begin{flalign} \label{eq:ptot}
\ptot &= \prad + \pgas \\
&= \frac{4\sigmaSB}{3c}T^4 + \frac{\kB\ \rho}{\mpr}T \;,
\end{flalign}
where \sigmaSB\ is the Stefan--Boltzmann constant, $T$ is the mid--plane gas temperature, \kB\ is the Boltzmann constant, and \mpr\ is the proton mass; and the total optical opacity (i.e. the sum of opacity terms arising from free--free absorption $\kappaFF$ and electron scattering $\kappaES$) which are given by \citep{NT73}
\begin{flalign} \label{eq:kappatot}
\kappatot &= \kappaFF + \kappaES \\
&= \left( 0.64 \times 10^{23} \mathrm{cm}^2\ \mathrm{g}^{-1} \right) \left( \frac{\rho}{\mathrm{g}\ \mathrm{cm}^{-3}} \right) \left( \frac{T}{\mathrm{K}} \right)^{-7/2} + 0.40\ \mathrm{cm}^2\ \mathrm{g}^{-1} \;,
\end{flalign}
where a pure hydrogen disc without quantum mechanical corrections (i.e. Gaunt factor of 1) has been assumed \citep[e.g.][]{ST83}.

Since free--free absorption (i.e. inverse `Bremsstrahlung' absorption) is strongly inversely proportional to temperature but also linearly proportional to density, it can become the main source of opacity near the black hole and/or in the outer/colder/under--dense regions of the accretion disc. For the same reason, gas pressure can dominate over radiation pressure around the black hole and/or in the outer parts, which led \cite{SS73} to introduce three independent local solutions (hereafter, regimes) which are different combinations of the above four mechanisms (two for pressure and two for opacity): 
\begin{enumerate}[wide=0pt,labelindent=10pt,labelwidth=10pt]
\item \textbf{Gas--ES} where \ptot\ = \pgas\ and $\kappa = \kappaES$;
\item \textbf{Rad--ES} where \ptot\ = \prad\ and $\kappa = \kappaES$;
\item \textbf{Gas--FF} where \ptot\ = \pgas\ and $\kappa = \kappaFF$.
\end{enumerate}
For these three cases, \cite{CO17} provide local solutions for a plethora of physical properties including the surface density, pressure, density, and opening angle ($h \equiv$ half--thickness / radial coordinate) profiles which can be written as \\ 
\noindent \\
\textbf{Gas--ES}
\begin{flalign} \label{eq:SigmaGES}
\Sigma \big|^\mathrm{gas}_\mathrm{es} =& \left( 5.00 \times 10^4\ \mathrm{g\ cm^{-2}} \right) \left (\alpha^{-4/5}\ m_\bullet^{-2/5}\ \dmbullet^{3/5}\ x^{-9/5} \right) \nonumber \\ 
&\times \mathcal{C}^{1/5}\ \mathcal{D}^{-4/5}\ \mathcal{P}^{3/5} \;,
\end{flalign}
\begin{flalign} \label{eq:pGES}
p \big|^\mathrm{gas}_\mathrm{es} =& \left( 1.80 \times 10^{17}\ \mathrm{dyn\ cm^{-2}} \right) \left( \alpha^{-9/10}\ m_\bullet^{-17/10}\ \dmbullet^{4/5}\ x^{-59/10} \right) \nonumber \\ 
&\times \mathcal{C}^{1/10}\ \mathcal{D}^{-9/10}\ \mathcal{P}^{4/5}\ \mathcal{R}^{1/2} \;,
\end{flalign}
\begin{flalign} \label{eq:rhoGES}
\rho \big|^\mathrm{gas}_\mathrm{es} =& \left( 8.10\ \mathrm{g\ cm^{-3}} \right) \left( \alpha^{-7/10}\ m_\bullet^{-11/10}\ \dmbullet^{2/5}\ x^{-37/10} \right) \nonumber \\ 
&\times \mathcal{C}^{3/10}\ \mathcal{D}^{-7/10}\ \mathcal{P}^{2/5}\ \mathcal{R}^{1/2} \;,
\end{flalign}
\begin{flalign} \label{eq:hGES}
h \big|^\mathrm{gas}_\mathrm{es} =& \left( 7.00 \times 10^{-3} \right) \left( \alpha^{-1/10}\ m_\bullet^{-3/10}\ \dmbullet^{1/5}\ x^{-1/10} \right) \nonumber \\ 
&\times \mathcal{C}^{-1/10}\ \mathcal{D}^{-1/10}\ \mathcal{P}^{1/5}\ \mathcal{R}^{-1/2} \;, 
\end{flalign}
\begin{flalign} \label{eq:pradpgasGES}
\frac{\prad}{\pgas} \bigg|^\mathrm{gas}_\mathrm{es} =&\; 69.00 \left( \alpha^{1/10}\ m_\bullet^{-7/10}\ \dmbullet^{4/5}\ x^{-29/10} \right) \nonumber \\ 
&\times \mathcal{C}^{-9/10}\ \mathcal{D}^{1/10}\ \mathcal{P}^{4/5}\ \mathcal{R}^{-1/2} \;, 
\end{flalign}
\begin{flalign} \label{eq:kappaFFkappaESGES}
\frac{\kappaFF}{\kappaES} \bigg|^\mathrm{gas}_\mathrm{es} =& \left( 4.40 \times 10^{-6} \right) \left( m_\bullet\ \dmbullet^{-1}\ x^{4} \right) \mathcal{C}\ \mathcal{P}^{-1}\ \mathcal{R}^{1/2} \;, 
\end{flalign}

\noindent \\
\textbf{Rad--ES}
\begin{flalign} \label{eq:SigmaRES}
\Sigma \big|^\mathrm{rad}_\mathrm{es} =& \left( 10.00\ \mathrm{g\ cm^{-2}} \right) \left( \alpha^{-1}\ m_\bullet\ \dmbullet^{-1}\ x^{4} \right) \nonumber \\ 
&\times \mathcal{C}^{2}\ \mathcal{D}^{-1}\ \mathcal{P}^{-1}\ \mathcal{R} \;,
\end{flalign}
\begin{flalign} \label{eq:pRES}
p \big|^\mathrm{rad}_\mathrm{es} =& \left( 2.60 \times 10^{15}\ \mathrm{dyn\ cm^{-2}} \right) \left( \alpha^{-1}\ m_\bullet^{-1}\ x^{-3} \right) \nonumber \\ 
&\times \mathcal{C}\ \mathcal{D}^{-1}\ \mathcal{R} \;,
\end{flalign}
\begin{flalign} \label{eq:rhoGRS}
\rho \big|^\mathrm{rad}_\mathrm{es} =& \left( 2.50 \times 10^{-5}\ \mathrm{g\ cm^{-3}} \right) \left( \alpha^{-1}\ m_\bullet\ \dmbullet^{-2}\ x^{5} \right) \nonumber \\ 
&\times \mathcal{C}^{3}\ \mathcal{D}^{-1}\ \mathcal{P}^{-2}\ \mathcal{R}^{2} \;,
\end{flalign}
\begin{flalign} \label{eq:hRES}
h \big|^\mathrm{rad}_\mathrm{es} =& \left( 5.00 \times 10^{-1} \right) \left( m_\bullet^{-1}\ \dmbullet\ x^{-3} \right) \mathcal{C}^{-1}\ \mathcal{P}\ \mathcal{R}^{-1} \;, 
\end{flalign}
\begin{flalign} \label{eq:pgaspradRES}
\frac{\pgas}{\prad} \bigg|^\mathrm{rad}_\mathrm{es} =& \left( 2.60 \times 10^{-5} \right) \left( \alpha^{-1/4}\ m_\bullet^{7/4}\ \dmbullet^{-2}\ x^{29/4} \right) \nonumber \\ 
&\times \mathcal{C}^{9/4}\ \mathcal{D}^{-1/4}\ \mathcal{P}^{-2}\ \mathcal{R}^{5/4} \;, 
\end{flalign}
\begin{flalign} \label{eq:kappaFFkappaESRES}
\frac{\kappaFF}{\kappaES} \bigg|^\mathrm{rad}_\mathrm{es} =& \left( 2.20 \times 10^{-8} \right) \left( \alpha^{-1/8}\ m_\bullet^{15/8}\ \dmbullet^{-2}\ x^{61/8} \right)  \nonumber \\ 
&\times \mathcal{C}^{17/8}\ \mathcal{D}^{-1/8}\ \mathcal{P}^{-2}\ \mathcal{R}^{9/8} \;, 
\end{flalign}

\noindent \\
\textbf{Gas--FF}
\begin{flalign} \label{eq:SigmaGFF}
\Sigma \big|^\mathrm{gas}_\mathrm{ff} =& \left( 1.70 \times 10^5\ \mathrm{g\ cm^{-2}} \right) \left( \alpha^{-4/5}\ m_\bullet^{-1/2}\ \dmbullet^{7/10}\ x^{-11/5} \right) \nonumber \\ 
&\times \mathcal{C}^{1/10}\ \mathcal{D}^{-4/5}\ \mathcal{P}^{7/10}\ \mathcal{R}^{-1/20} \;,
\end{flalign}
\begin{flalign} \label{eq:pGFF}
p \big|^\mathrm{gas}_\mathrm{ff} =& \left( 3.30 \times 10^{17}\ \mathrm{dyn\ cm^{-2}} \right) \left( \alpha^{-9/10}\ m_\bullet^{-7/4}\ \dmbullet^{17/20}\ x^{-61/10} \right) \nonumber \\ 
&\times \mathcal{C}^{1/20}\ \mathcal{D}^{-9/10}\ \mathcal{P}^{17/20}\ \mathcal{R}^{19/40} \;,
\end{flalign}
\begin{flalign} \label{eq:rhoGFF}
\rho \big|^\mathrm{gas}_\mathrm{ff} =& \left( 51.00\ \mathrm{g\ cm^{-3}} \right) \left( \alpha^{-7/10}\ m_\bullet^{-5/4}\ \dmbullet^{11/20}\ x^{-43/10} \right) \nonumber \\ 
&\times \mathcal{C}^{3/20}\ \mathcal{D}^{-7/10}\ \mathcal{P}^{11/20}\ \mathcal{R}^{17/40} \;,
\end{flalign}
\begin{flalign} \label{eq:hGFF}
h \big|^\mathrm{gas}_\mathrm{ff} =& \left( 3.80 \times 10^{-3} \right) \left( \alpha^{-1/10}\ m_\bullet^{-1/4}\ \dmbullet^{3/20}\ x^{1/10} \right) \nonumber \\ 
&\times \mathcal{C}^{-1/20}\ \mathcal{D}^{-1/10}\ \mathcal{P}^{3/20}\ \mathcal{R}^{-19/40} \;, 
\end{flalign}
\begin{flalign} \label{eq:pradpgasGFF}
\frac{\prad}{\pgas} \bigg|^\mathrm{gas}_\mathrm{ff} =&\; 0.27 \left( \alpha^{1/10}\ m_\bullet^{-1/4}\ \dmbullet^{7/20}\ x^{-11/10} \right) \nonumber \\ 
&\times \mathcal{C}^{-9/20}\ \mathcal{D}^{1/10}\ \mathcal{P}^{7/20}\ \mathcal{R}^{-11/40} \;,
\end{flalign}
\begin{flalign} \label{eq:kappaESkappaFFGFF}
\frac{\kappaES}{\kappaFF} \bigg|^\mathrm{gas}_\mathrm{ff} =& \left( 4.80 \times 10^{2} \right) \left( m_\bullet^{-1/2}\ \dmbullet^{1/2}\ x^{-2} \right) \mathcal{C}^{-1/2}\ \mathcal{P}^{1/2}\ \mathcal{R}^{-1/4} \;, 
\end{flalign}
where
\begin{flalign} \label{eq:mbullet}
m_\bullet \equiv \frac{M_\bullet}{3\Msun} \;,
\end{flalign}
\begin{flalign} \label{eq:dmbullet}
&\dmbullet \equiv \frac{\dot{M}_{\circledcirc \rightarrow \bullet}}{10^{17} \mathrm{g}\; \mathrm{s}^{-1}} \;,
\end{flalign}
are respectively the normalised black hole mass and accretion rate (subscript indicates mass being transferred from the accretion disc to the black hole), $x \equiv (r / \Rgrav)^{1/2}$ is the dimensionless radial coordinate, and the $\mathcal{C}, \mathcal{D},\mathcal{P},$ and $\mathcal{R}$ correction functions can be found in \App{Corrections}.

In addition to the above three cases which are valid from the ISCO outwards, \cite{MB23} have recently developed -- for the first time -- analytic relativistic solutions that smoothly combine the extra--ISCO to the intra--ISCO region, where the surface density profile can be written as a function of the surface density at the ISCO\footnote{See Appendix A in \cite{MB23} on the smooth combination of extra-- and intra--ISCO solutions.} as \\
\noindent \\
\textbf{intra--ISCO}
\begin{flalign} \label{eq:SigmaintraISCO}
\Sigma \big|^\mathrm{intra}_\mathrm{isco} = \Sigma_\mathrm{isco} \frac{\Risco / r}{\left( \frac{u_\mathrm{isco}}{c} \sqrt{\frac{3\Risco}{2\Rgrav}} \right)^{-1} \left( \frac{\Risco}{r} - 1 \right)^{3/2} + 1} \;,
\end{flalign}
where $u_\mathrm{isco}$ is the radial velocity at the ISCO.

Hence, given that \cite{CO17} derived equations for the global structure of a relativistic accretion disc by following the sonic--ISCO boundary condition of \cite{PSM12} \citep[i.e. by relaxing the unphysical\footnote{\cite{NT73} by imposing a zero--torque boundary condition at the ISCO introduced an inconsistency in the model since the velocity of the fluid at the ISCO diverges whereas observables like magnetic fields exist inside the ISCO \citep[e.g.][]{K99a}.} zero--torque assumption of][]{NT73}, accretion disc material should exist between the ISCO and the photon radius (although arguably not significant amounts, see for example \Fig{surface_density} and \Sec{Model:Evolve:AD}). In this work, by using the \cite{CO17} in conjunction with the \cite{MB23} solutions, we model accretion discs by combining extra--ISCO with intra--ISCO accretion theories, which allows for a more physically complete modelling of relativistic accretion discs.

\subsubsection{Accretion disc structure} \label{sec:Model:Properties:AD structure}

In order to build and evolve an accretion disc, apart from quantifying where the accretion disc starts and stops (i.e. calculating the radial boundaries  of \Sec{Model:Properties:AD boundaries}) one needs to know which of the local solutions (\Sec{Model:Properties:AD}) are valid and where. These regions of validity for the \textbf{Gas--ES}, \textbf{Rad--ES}, and \textbf{Gas--FF} regimes can be calculated by quantifying for all $r \geq \Risco$ the inequalities:
\begin{enumerate}[wide=0pt,labelindent=10pt,labelwidth=10pt]
\item $\frac{\prad}{\pgas} \bigg|^\mathrm{gas}_\mathrm{es} < 1\ \&\ \frac{\kappaFF}{\kappaES} \bigg|^\mathrm{gas}_\mathrm{es} < 1$ for the \textbf{Gas--ES} regime;
\item $\frac{\pgas}{\prad} \bigg|^\mathrm{rad}_\mathrm{es} < 1\ \&\ \frac{\kappaFF}{\kappaES} \bigg|^\mathrm{rad}_\mathrm{es} < 1$ for the \textbf{Rad--ES} regime;
\item $\frac{\prad}{\pgas} \bigg|^\mathrm{gas}_\mathrm{ff} < 1\ \&\ \frac{\kappaES}{\kappaFF} \bigg|^\mathrm{gas}_\mathrm{ff} < 1$ for the \textbf{Gas--FF} regime;
\end{enumerate}
where the pressure and opacity ratios are given by \Eq{pradpgasGES} and (\ref{eq:kappaFFkappaESGES}), \Eq{pgaspradRES} and (\ref{eq:kappaFFkappaESRES}), and \Eq{pradpgasGFF} and (\ref{eq:kappaESkappaFFGFF}), respectively. Thus, the solutions to the aforementioned sets of inequalities provides the boundaries of the corresponding regimes, which allow one to calculate the global structure of the accretion disc by combining locally valid equations. Finally, the \textbf{intra--ISCO} equations are valid for $r \in (\Rph, \Risco)$.

\subsubsection{A practical example} \label{sec:Model:Properties:Example}

\begin{figure}
\includegraphics[width=0.48\textwidth]{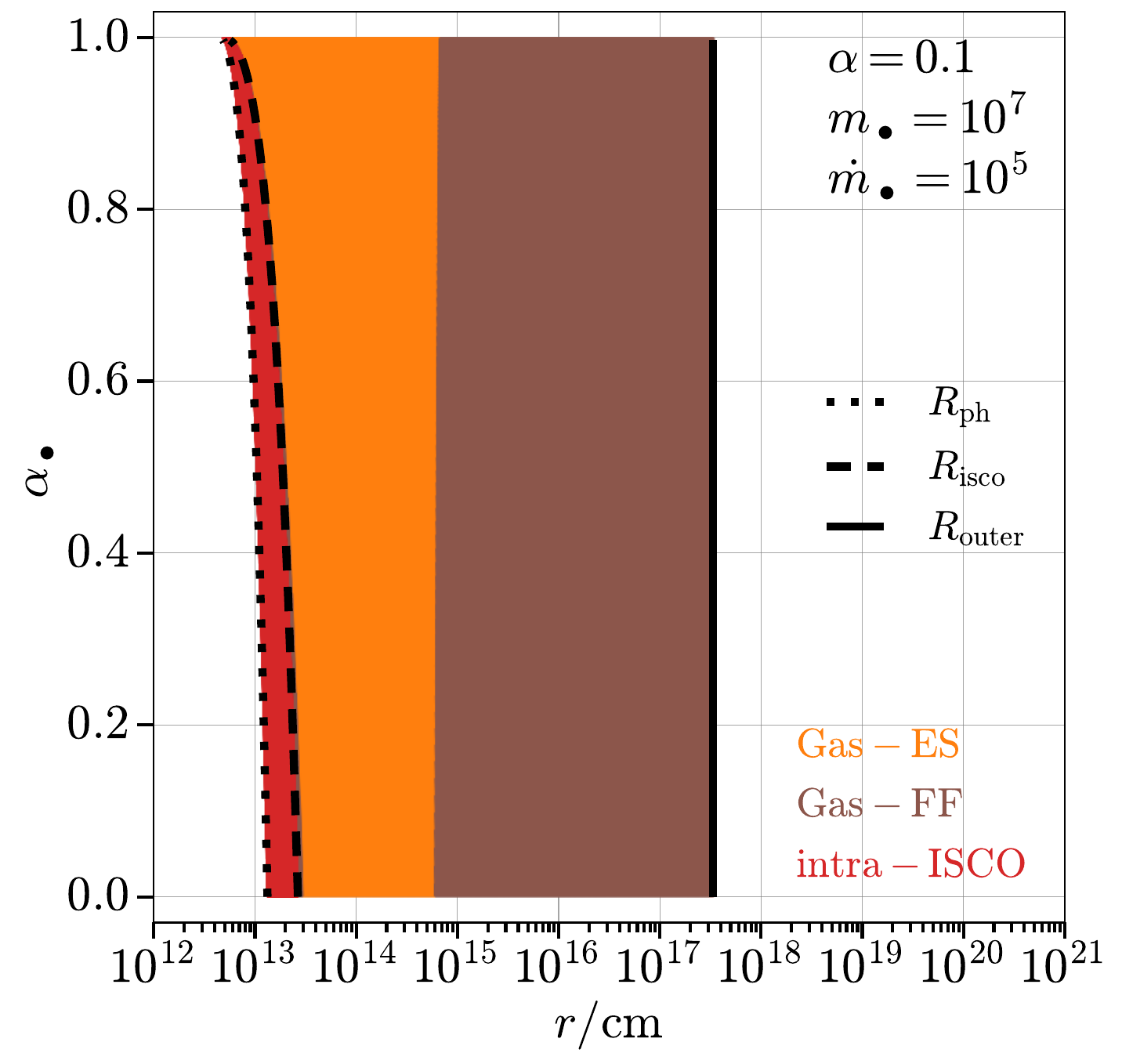}
\caption{Regions of validity on the spin--radius plane for $\alpha = 0.1$, $m_\bullet = 10^7$, and $\dmbullet = 10^5$, created by solving the sets of inequalities (see text for more details) for $\alpha_\bullet \in [0, 0.998]$ and $r \in (\Rph, \Router]$. The dotted, dashed, and solid lines represent, respectively, the photon radius of \Eq{Rph}, the ISCO radius of \Eq{Risco}, and the outer radius which is calculated from the Toomre parameter of \Eq{Q} assuming a fixed $\alpha_\bullet=0.5$ (see \Fig{surface_density} and the relevant discussion about \Router).}
\label{fig:regional_solutions}
\end{figure}

As an example of the above processes, in \Fig{regional_solutions} we show the regions of validity in the spin--radius plane. As described above and can be seen from the equations in \Sec{Model:Properties:AD}, the four physical variables that determine the structure of the accretion disc are the black hole mass, spin, and accretion rate; and the viscosity parameter. Since adding to these four parameters the radial coordinate $r$ will require plotting in 5D space, we set for this example: $\alpha = 0.1$, $m_\bullet=10^7$ (i.e. $M_\bullet = 3 \times 10^7\ \Msun$), $\dmbullet = 10^5$ (i.e. $\dot{M}_{\circledcirc \rightarrow \bullet} \sim 1.5 \times 10^{-4}\ \Msun /$yr), and also assume a co--rotating accretion disc.

By solving the aforementioned sets of inequalities: \Eq{pradpgasGES} and (\ref{eq:kappaFFkappaESGES}); \Eq{pgaspradRES} and (\ref{eq:kappaFFkappaESRES}); and \Eq{pradpgasGFF} and (\ref{eq:kappaESkappaFFGFF}) for $\alpha_\bullet \in [0, 0.998]$ and $r \in (\Rph, \Router]$, we can identify the transitioning radii between the different regimes of the accretion disc (i.e. \textbf{Gas--ES}, \textbf{Rad--ES}, and \textbf{Gas--FF}). 

In this example, gas pressure dominates everywhere over radiation pressure. For accretion discs that obey the \cite{NT73,SS73,PT74} assumptions -- like the ones in this work -- the temperature of the accretion disc is anti--correlated with the black hole mass but positively correlated with the accretion rate of the black hole \citep[e.g.][]{CO17}. Hence, around slowly accreting supermassive black holes (like the one in this example) the accretion discs cannot reach temperatures high enough for the radiation pressure to dominate over the gas pressure (see \Sec{Results}). Furthermore, as far as the opacity mechanisms are concerned, the main source of opacity in the ISCO and outer regions is free--free absorption, whilst in the region in--between, electron scattering is the dominant opacity mechanism. 

\begin{figure}
\includegraphics[width=0.48\textwidth]{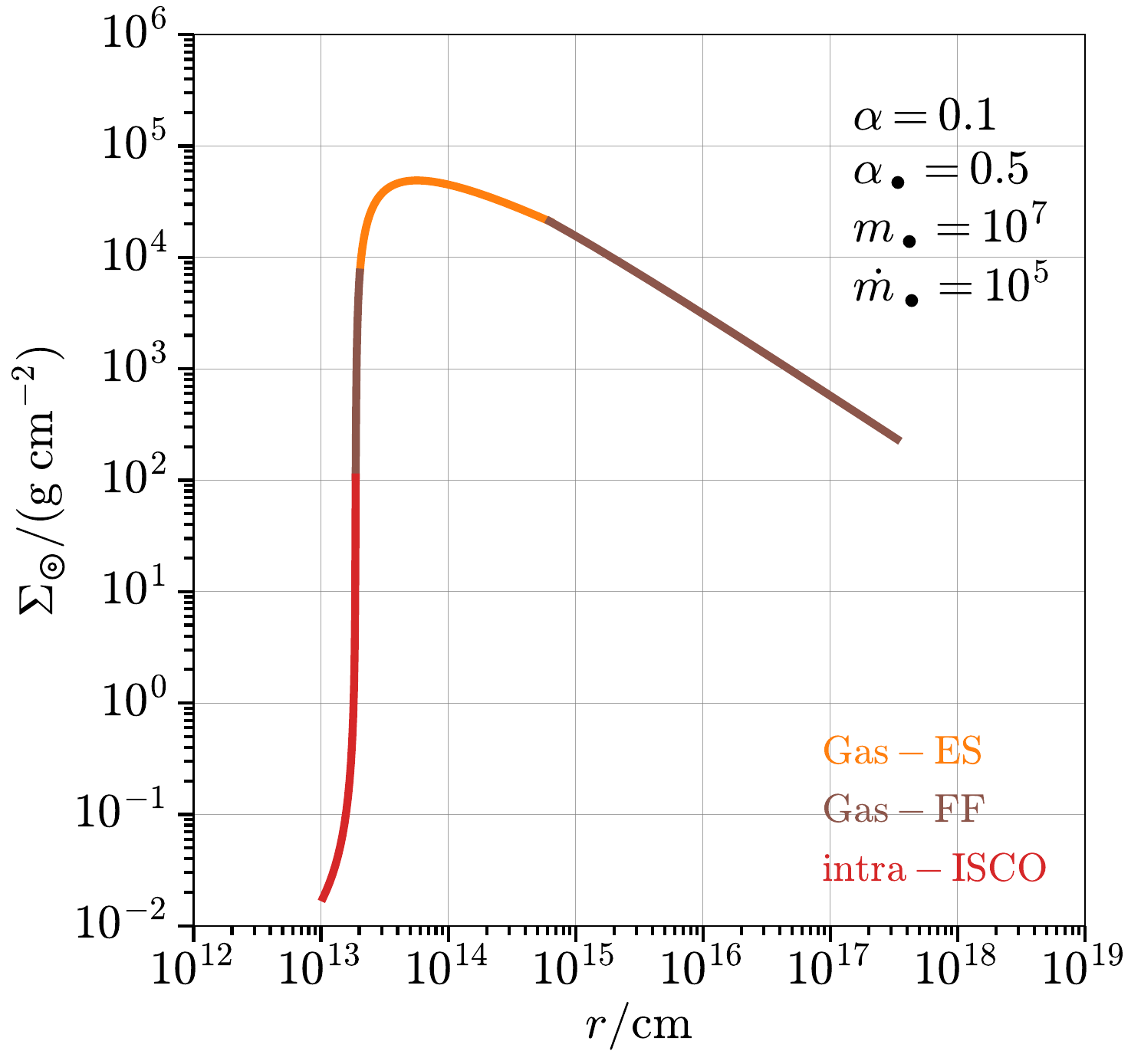}
\caption{The global surface density profile of a co--rotating accretion disc as a combination of locally valid profiles for $\alpha = 0.1$, $\alpha_\bullet = 0.5$, $m_\bullet = 10^7$, and $\dmbullet = 10^5$.}
\label{fig:surface_density}
\end{figure}

Knowing from \Fig{regional_solutions} which solutions are expressed and where, allows one to build for a given spin parameter $\alpha_\bullet$ the global structure of the accretion disc by combining the locally valid profiles of \Sec{Model:Properties:AD} based on their region(s) of validity. For example, the surface density profile of a co--rotating accretion disc with $\alpha = 0.1$, $m_\bullet = 10^7$, and $\dmbullet = 10^5$, assuming also $\alpha_\bullet = 0.5$ (i.e. fixing the values for all four parameters)\footnote{In \Fig{regional_solutions}, it was necessary to provide a spin parameter value to calculate \Router\ via \Eq{Q}. Even though a value of 0.5 was chosen, the exact value of $\alpha_\bullet$ does not practically affect \Router\ (see \Sec{Results}).} can be seen in \Fig{surface_density}. This global profile can be then integrated to provide the total angular momentum of the accretion disc (whose importance was discussed in \Sec{Model:Properties:AD}) following \Eq{Jcircledcirc}
\begin{flalign} \label{eq:Jcircledcirc2}
J_\circledcirc =& \int_{\circledcirc} 2\uppi\ r\ \Sigma\ \widetilde{J_\circledcirc}\ \mathrm{d}r \nonumber \\
=& \int_{R_\mathrm{ph}}^{R_\mathrm{isco}} 2\uppi\ r\ \Sigma \big|^\mathrm{intra}_\mathrm{isco}\ \widetilde{J_\circledcirc}\ \mathrm{d}r + \int_{R_\mathrm{isco}}^{R_\mathrm{Gas-ES}} 2\uppi\ r\ \Sigma \big|^\mathrm{gas}_\mathrm{ff}\ \widetilde{J_\circledcirc}\ \mathrm{d}r \nonumber \\
+& \int_{R_\mathrm{Gas-ES}}^{R_\mathrm{Gas-FF}} 2\uppi\ r\ \Sigma \big|^\mathrm{gas}_\mathrm{es}\ \widetilde{J_\circledcirc}\ \mathrm{d}r + \int_{R_\mathrm{Gas-FF}}^{R_\mathrm{outer}} 2\uppi\ r\ \Sigma \big|^\mathrm{gas}_\mathrm{ff}\ \widetilde{J_\circledcirc}\ \mathrm{d}r \;,
\end{flalign}
where the integration limits are practically: (\Rph, \Risco), [\Risco, $R_\mathrm{Gas-ES}$), [$R_\mathrm{Gas-ES}, R_\mathrm{Gas-FF}$), and [$R_\mathrm{Gas-FF}$, \Router]; and $R_\mathrm{Gas-ES}$, $R_\mathrm{Rad-ES}$, and $R_\mathrm{Gas-FF}$ represent the starting radii of the corresponding regime. Note also that in our model determining \Router\ is an iterative process where an accretion disc is built with an initial size and then repeatedly extended until a self--gravitational structure has been achieved (see \App{Implementation} for more details). For the initial radial extent we use the observed relation of \cite{MKM10} who derived -- by fitting a power--law to a sample of eleven quasars -- a correlation between accretion disc half--light radii and black hole masses
\begin{flalign} \label{eq:Rinit}
\log_{10} \left( \frac{\Rinit}{\mathrm{cm}} \right) = 15.78 + 0.80 \log_{10} \left( \frac{M_\bullet}{10^9 M_\odot} \right) \;.
\end{flalign}

\begin{figure*}
\includegraphics[width=0.49\textwidth]{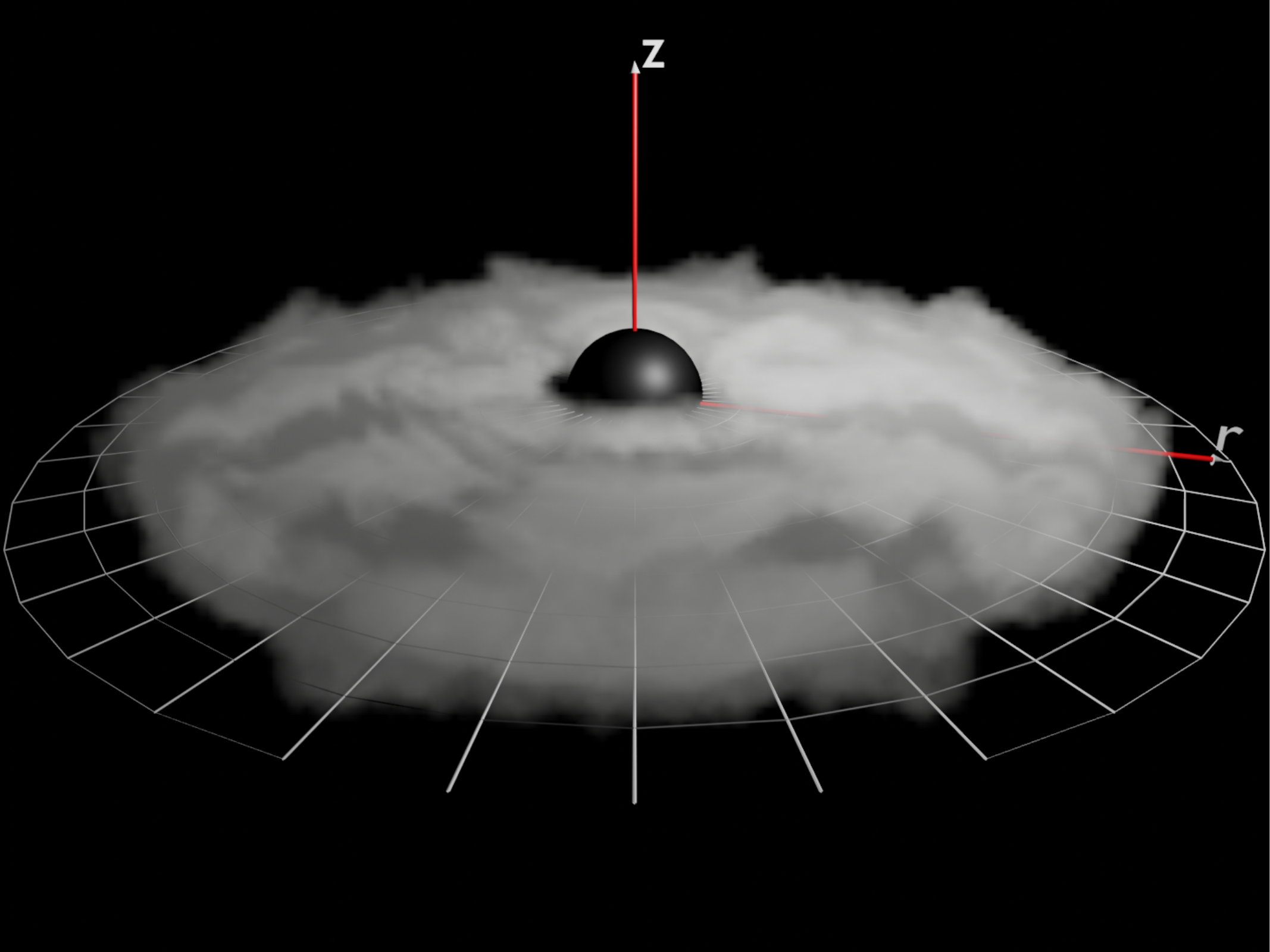} \includegraphics[width=0.49\textwidth]{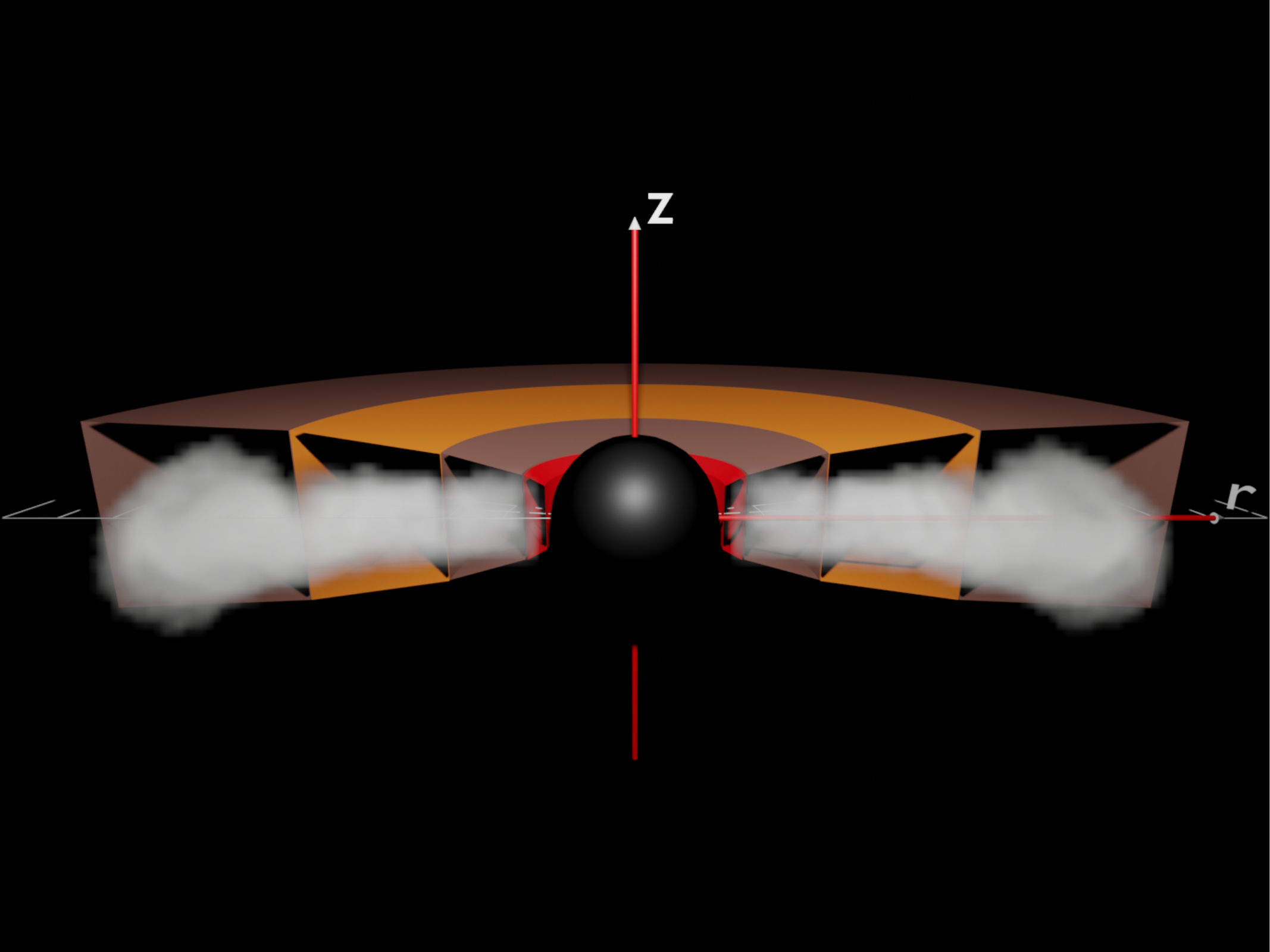}
\caption{A schematic representation (i.e. not to scale) of the global structure of an accretion disc (left--hand panel) that can be constructed as a combination of locally valid solutions (right--hand panel), where the colours of the regions represent those of \Fig{surface_density}.}
\label{fig:accretion_disc}
\end{figure*}

In \Fig{accretion_disc}, we demonstrate with a schematic representation (i.e. not to scale) how the global structure of an accretion disc (left--hand panel) can be constructed as a combination of locally valid solutions (right--hand panel), where the red, brown, and orange colours of the regions represent the \textbf{intra--ISCO}, \textbf{Gas--FF}, and \textbf{Gas--ES} regimes, respectively.

\subsection{Evolution of the black hole–accretion disc system} \label{sec:Model:Evolve}

In the previous subsection, we described the properties of black holes and accretion discs in our model, and how one can calculate the global structure of an accretion disc by combining locally valid solutions. In this subsection, we detail how we follow the evolution of accretion discs (\Sec{Model:Evolve:AD}) and black holes (\Sec{Model:Evolve:BH}).

\subsubsection{Evolving the accretion disc} \label{sec:Model:Evolve:AD}

\begin{figure*}
\includegraphics[width=0.49\textwidth]{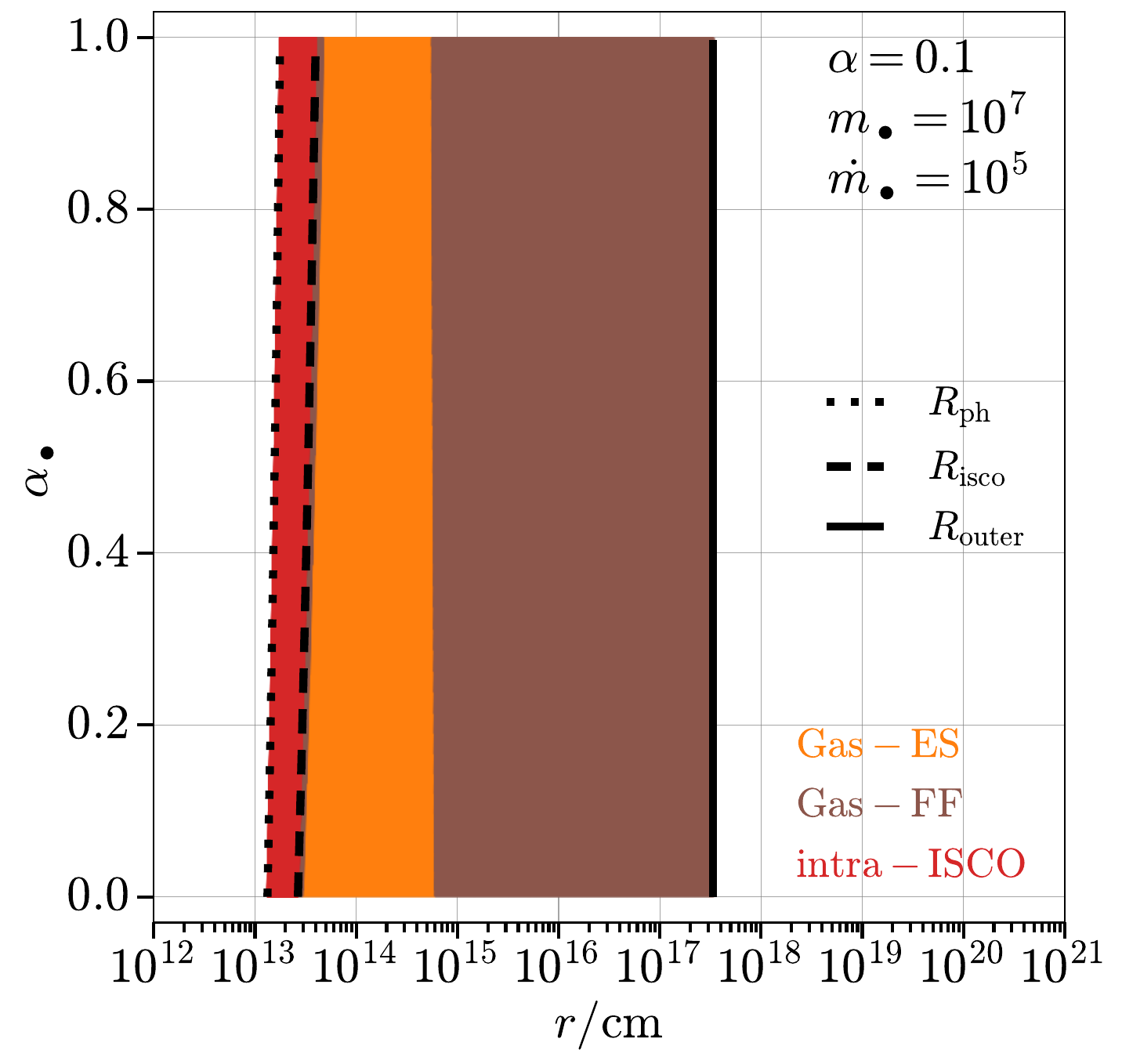} \includegraphics[width=0.49\textwidth]{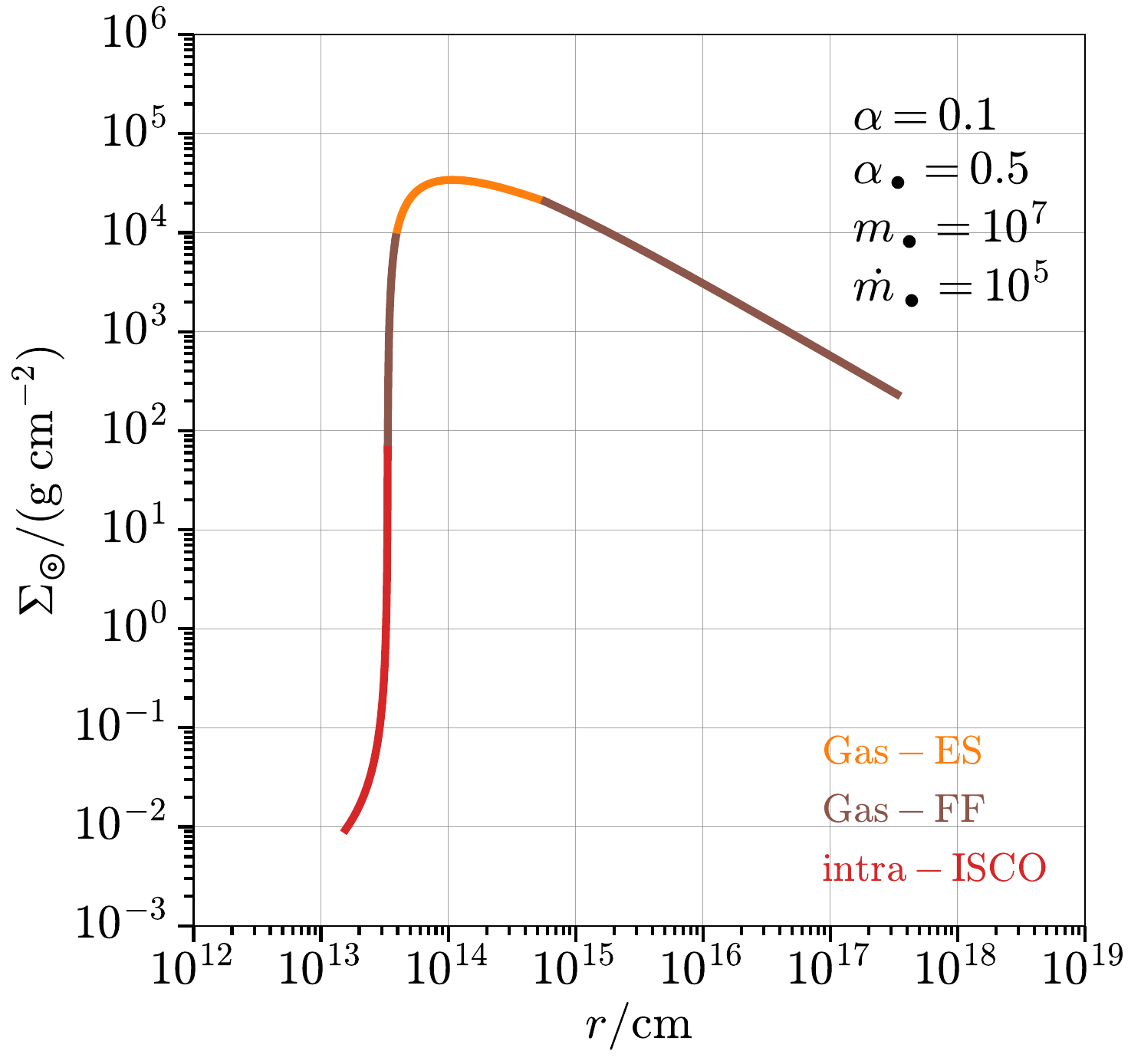}
\caption{An illustration of the effect counter--rotation has on the regions of validity (left--hand panel, equivalent to \Fig{regional_solutions}) and global surface density profile (right--hand panel, equivalent to \Fig{surface_density}) of the accretion disc discussed in \Sec{Model:Properties:Example}.}
\label{fig:regional_solutions_counter}
\end{figure*}

As discussed above, the alignment or not between the accretion disc and the black hole directly affects \Eq{Rph}, \Eq{Risco} [via \Eq{Z}], \Eq{kappa} and (\ref{eq:Omega}), and the relativistic correction equations (\ref{eq:C}), (\ref{eq:F}), (\ref{eq:G}), and (\ref{eq:P}). Thus, it fundamentally alters the structure of the accretion disc as a whole. To demonstrate that, in \Fig{regional_solutions_counter} we reproduce \Fig{regional_solutions} (left--hand panel) and \Fig{surface_density} (right--hand panel) by keeping all black hole and accretion disc properties the same (i.e.  $\alpha = 0.1$, $\alpha_\bullet = 0.5$, $m_\bullet = 10^7$, and $\dmbullet = 10^5$) but assuming a counter--rotating accretion disc. Both the starting radius of the accretion disc and the regions of validity of the \textbf{intra--ISCO}, \textbf{Gas--FF}, and \textbf{Gas--ES} regimes are affected, leading to a different global surface density profile.

In order to assess if the angular momentum of the accretion disc is aligned or counter--aligned with respect to the spin of the black hole, we use the criterion developed by \cite{KLO05} which states that alignment exists if and only if
\begin{flalign} \label{eq:costheta}
\cos(\theta) \equiv \sbullethat \cdot \jcircledcirchat > - J_\circledcirc /2S_\bullet \;.
\end{flalign}
Hence, in this work, knowing the structure of the accretion disc and being able to calculate the integral in \Eq{Jcircledcirc} as was done for example in \Eq{Jcircledcirc2}, allows us to directly evaluate the above inequality (see also text where \Fig{bh_spins} is discussed).

Furthermore, the structure of the accretion disc depends via \Eq{Risco} both on the mass and spin of the black hole which as they evolve (see \Sec{Model:Evolve:BH}) will further influence the material in the accretion disc. Therefore, accretion discs in our model have a dynamical structure which allows to accurately follow the co--evolution between black holes and accretion discs.

Apart from the changes in the structure of the accretion disc induced by the black hole, there are internal mechanisms that affect the distribution of its material. Since the accretion disc rotates differentially, fluid elements in adjacent annuli that were initially at the same azimuth will become (azimuthally) separated. This separation will introduce shear forces which -- given that the angular momentum increases outwards -- will transfer angular momentum from the inner fluid elements to the outer ones. Therefore, the differential rotation will also introduce radial separations between fluid elements. For magnetized, rotating flows this will trigger magnetorotational instabilities \citep[MRI,][]{BH91} which can facilitate angular momentum transport in accretion disks and inwards mass migration. Even though dissipative processes like the aforementioned MRI--generated viscous stresses and the turbulence of the gas flow are not captured by our model, their effect can be parametrised by the \cite{SS73} $\alpha$ parameter which in practice encapsulates our ignorance regarding the physical mechanism of viscosity \citep{FKR02}.

Once the aforementioned migrating gas elements reach the ISCO they plunge towards the black hole with a radial velocity that is defined by the local sound speed and increase the black hole's mass by an amount equal to the specific binding energy at the ISCO over $c^2$ \cite[i.e. we follow the sound--ISCO assumption of][]{PSM12,CO17}. 

We calculate the rate at which mass migrates (due to an effective viscosity with $\alpha$--parameter fixed at 0.1) from the accretion disc to the black hole as
\begin{flalign} \label{eq:dMcircledcircarrowbullet}
\dMcircledcircarrowbullet \equiv \dMbullet = \dMEdd \times \mathrm{max}(0.01,\mathrm{min}(\fEdd,1)) \;,
\end{flalign}
where the Eddington accretion rate is given by
\begin{flalign} \label{eq:dMEdd}
\dMEdd = \frac{4\uppi\ G\ \mpr}{\sigmaT\ c} \frac{M_\bullet}{\epsilonr} \;,
\end{flalign}
where \sigmaT\ is the Thomson cross section, and \epsilonr\ is the radiative efficiency parameter [see \Eq{Lobs}] which describes how efficiently material in the accretion disc releases its rest mass energy as thermal energy (i.e. radiation) as it inspirals towards the black hole via a succession of circular orbits \citep{P81}. In addition, the Eddington fraction $\fEdd \equiv \dMbullet/\dMEdd$ can be written as
\begin{flalign} \label{eq:fEdd}
\fEdd = 0.76 \left( \frac{\epsilonr}{0.1} \right) \left( \frac{M_\circledcirc}{10^4 \Msun} \right)^5 \left( \frac{M_\bullet}{10^6 \Msun} \right)^{-47/7} \left( \frac{\alpha_\bullet\ |J_\circledcirc |}{3S_\bullet} \right)^{-25/7} \;.
\end{flalign}
This equation was derived by \cite{FSP18} (see their Appendix A)\footnote{Note that in the final term we take the absolute value of $J_\circledcirc$. As it can be seen from \Eq{F}, when material is counter--rotating with respect to the black hole; \Eq{tildeJcircledcirc} -- which describes the $\phi$ component of the four--velocity -- becomes negative. Unless corrected, this results in negative \fEdd.} by only taking into account the so--called `c' regime of the \cite{SS73} model which describes the region very close to the black hole and the extensive outermost part where the pressure is dominated by the gas and the main mechanism of opacity is free--free absorption (i.e. the \textbf{Gas--FF} regime). We emphasise that within the steady--state \cite{NT73} framework adopted in this work, the mass accretion rate is conserved with radius and is therefore regulated at large radii. In the supermassive black hole mass range considered here, the outer regions of thin accretion discs are gas--pressure dominated with free-–free absorption as the dominant opacity source (see e.g. \Fig{regional_solutions}), consistent with the regime assumed by \cite{FSP18} when deriving \Eq{fEdd}. We therefore adopt the $\fEdd$ prescription from \cite{FSP18} solely as a parametrization for the mass inflow rate at unresolved scales, and not as a model for the disc structure, while the local thermodynamic state, pressure regime, opacity source, and radiative efficiency of the disc are computed self--consistently using the piece-wise solution described above.

Additionally, following previous theoretical studies \citep[e.g][]{RBS16,FSP18,CPS20,WHC22,MDB23} we impose in \Eq{dMcircledcircarrowbullet} a lower limit -- one per cent of the \dMEdd\ -- in order to ensure that the assumptions of the standard accretion disc model \citep{NT73,SS73} are not violated and our accretion discs can be approximated by a radiatively--efficient, geometrically thin disc. For significantly sub--Eddington accretion rates advection--dominated, thicker discs emerge since low accretion rates result in decreased densities hence less efficient cooling processes \citep{ACL88,NY94,NY95,CAL97,BB99}. On the other hand, the upper limit in \Eq{dMcircledcircarrowbullet} ensures that the accretion rate is kept below the Eddington limit, hence the accretion disc luminosity does not violate the thin disc approximation by increasing the opening angle to values higher than 1 due to radiation pressure \citep[e.g.][]{PMN10,MTB12,LVS16,SGL21}. However, in a future work we intend to incorporate both types of accretion discs in our model \citep[e.g.][]{BS19,HLS23,KSS24}

Finally, the accretion disc's total angular momentum vector can be affected by three processes. The first process involves the accretion of angular momentum from the ISM (see \App{Implementation}). The second process is the feeding of the black hole: as mass is leaving the accretion disc and plunges into the black hole at a rate given by \Eq{dMcircledcircarrowbullet}, the structure of the accretion disc will be affected and so will its angular momentum magnitude [see \Eq{dSbulletaccvec}]. The final process that affects the total angular momentum of the accretion disc is due to the distortion a spinning black hole introduces on the space--time around it [see next subsection and \Eq{dSbulletBPvec}].

\subsubsection{Evolving the black hole} \label{sec:Model:Evolve:BH}


Accretion of gas onto the black hole will cause -- apart from mass changes -- its angular momentum to evolve. Assuming that gas at the ISCO maintains its corresponding energy and angular momentum whilst it radially falls into the black hole, we can calculate the angular momentum accretion rate as
\begin{flalign} \label{eq:dSbulletaccvec}
\dSbulletaccvec = \dMbullet\ \Jisco\ \sbullethat \;,
\end{flalign}
where
\begin{flalign} \label{eq:Jisco}
\Jisco = \pm \frac{G\ M_\bullet}{c} \frac{Z^2 \mp 2\alpha_\bullet \sqrt{Z} + \alpha_\bullet^2}{Z\sqrt{Z - 3 \pm 2\alpha_\bullet\ Z^{-1/2}}} \;,
\end{flalign}
is the specific angular momentum at the ISCO \citep{BPT72,FSP18} where the upper sign refers to co--rotating and the lower to counter--rotating orbits with respect to \Sbulletvec, similar to \Eq{Z}.

Since \dSbulletaccvec\ has the same direction as \Sbulletvec, the above process only alters the magnitude of \Sbulletvec. However, the direction can also be affected by BH--AD interactions. The non--spherically symmetric gravitational field of a spinning black hole induces a precession \citep{LT18,W72} of the orbital plane of a circular orbit about the spin axis of the black hole with a frequency
\begin{flalign} \label{eq:omega}
\omega = \frac{2G\ S_\bullet}{c^2\ r^3} \;,
\end{flalign}
and causes test particles that are closer to the black hole to precess faster as $\omega \propto r^{-3}$. In order to conserve the BH--AD systems' angular momentum, an opposite torque is acting on the black hole which modifies the direction of its spin axis and results in the alignment or counter--alignment of the two components \citep{BP75}. Following \cite{P92}, the rate at which the spin direction changes due to the Bardeen--Petterson effect can be written as
\begin{flalign} \label{eq:dSbulletBPvec}
\dSbulletBPvec &= -\ \boldsymbol{\omega} \times\ \boldsymbol{J}_\circledcirc \nonumber \\
&= \frac{4\uppi\ G}{c^2} \int_\circledcirc \frac{\Sigma_\circledcirc(r)\ \widetilde{J_\circledcirc}(r)\ S_\bullet\ \sin(\theta)}{r^2} \mathrm{d}r\ \jcircledcirchat \times\ \sbullethat \;,
\end{flalign}
where $\boldsymbol{\omega} = \omega\ \sbullethat$ and $\boldsymbol{J}_\circledcirc = J_\circledcirc\ \jcircledcirchat$, and since \dSbulletBPvec\ is a cross product of \Sbulletvec, it only changes the direction of \Sbulletvec\ and not its magnitude. In summary, the total evolution of the black hole spin vector is described by
\begin{flalign} \label{eq:dSbulletvec}
\dSbulletvec &= - \dJcircledcircvec \nonumber  \\ 
&= \dMbullet\ \Jisco\ \sbullethat + \frac{4\uppi\ G}{c^2} \int_\circledcirc \frac{\Sigma_\circledcirc(r)\ \widetilde{J_\circledcirc}(r)\ S_\bullet\ \sin(\theta)}{r^2} \mathrm{d}r\ \jcircledcirchat \times\  \sbullethat \;.
\end{flalign}
As discussed in \Sec{Model:Properties:AD} and demonstrated in the example of \Sec{Model:Properties:Example}, knowing the exact profile of the accretion disc's surface density allows us to directly calculate the integral in \Eq{dSbulletvec}, instead of using analytic expressions \citep[e.g.][]{KLO05,MPT07,PDV09}. This explicit calculation of the BH--AD torques is an additional advantage of our model.

\subsection{Modelling the black hole--accretion disc feedback} \label{sec:Model:Feedback}

In this work, we only include thermal/quasar mode feedback \citep{SR98} which is assumed to be generated by a radiatively efficient accretion disc that is surrounding an SMBH which is experiencing high mass accretion rates (i.e. \fEdd\ > 0.01). For the geometrically thin discs considered in this work, the spin--dependent efficiency at which gravitational rest mass energy can be extracted from the accretion disc and released as thermal radiation to the surroundings can be written as
\begin{flalign} \label{eq:epsilonralphabullet}
\varepsilon_{\alpha_\bullet} = 1 - \Eisco = 1 - \sqrt{1 - \frac{2\Rgrav}{3\Risco}} \;,
\end{flalign}
where \Eisco\ is the dimensionless, specific energy at the ISCO \citep{R16} derived for bound, circular orbits on the equatorial plane of the black hole \citep{ST83}. 

In addition to $\varepsilon_{\alpha_\bullet}$, \cite{CO17} followed \cite{PSM12} and relaxed the `zero--torque at the ISCO' assumption of \cite{NT73}, which allows one to include an extra component to the energy radiated per unit time (i.e. luminosity) from the accretion disc generated by that additional torque at the ISCO. Hence, the total luminosity generated by the accretion disc that an observer at infinity will measure is
\begin{flalign} \label{eq:Lobs}
\Lobs &= \left( \varepsilon_{\alpha_\bullet} + \frac{G\ \gisco\ \Omegaisco}{c\ \dMbullet} \right) \dMbullet\ c^2 \nonumber \\  
& \equiv \epsilonr\ \dMbullet\ c^2 \;,
\end{flalign}
where \gisco\ is the additional torque at the ISCO contributing to the radiative efficiency of the accretion disc \citep{L02a} and can be written as
\begin{flalign} \label{eq:gcirc}
\gisco &=  M_\bullet\ \dMbullet\ \Pisco \left(c^2 \Eisco - \Omegaisco\ \Jisco \right)^{-1} \;,
\end{flalign}
where \Pisco\ is a relativistic parameter introduced by \cite{CO17} (see our \App{Corrections}) and \Omegaisco\ is the relativistic angular velocity \citep{K90,DN20} evaluated at the ISCO
\begin{flalign} \label{eq:Omegacirc} 
\Omegaisco = \pm \frac{\sqrt{G\ M_\bullet}}{\Risco^{3/2} \pm \alpha_\bullet\ \Rgrav^{3/2}} \;.
\end{flalign}

A significant implication from the above feedback recipe is that faster spinning black holes can radiate more efficiently -- especially when they are aligned with the accretion disc -- since the faster the black hole is spinning the closer \Risco\ moves to the photon radius\footnote{The reason for this behaviour is that the space–-time geometry around faster--spinning black holes can support stable co--rotating circular orbits at progressively smaller radii, closer to the event horizon \citep{BPT72}. In contrast, for counter--rotating accretion discs the innermost stable circular orbit moves to larger radii as the black hole spin increases. This behaviour is illustrated in \Fig{regional_solutions} and the left-hand panels of \Fig{regional_solutions_counter}, which show the dependence of 
\Rph and \Risco on the spin parameter $\alpha_\bullet$.}. Hence, in order for mass elements to reach closer to the horizon, they have to radiate away more of their rest--mass energy \citep{B70}. Therefore, both the magnitude and the orientation of the black hole spin can influence (via \epsilonr) the growth of SMBHs since for fixed black hole mass; faster spinning, aligned black holes have lower Eddington accretion rates than counter--aligned, which restrict their maximum allowed mass accretion rates [see \Eq{dMEdd}].

Furthermore, the fact that \gisco\ depends linearly on the accretion rate \dMbullet\ and the second term of \epsilonr\ is inversely proportional to \dMbullet, implies that an accretion disc around a spinning black hole can radiate -- due to torques -- even in cases when there is no accretion of mass \citep{L02a}, although admittedly this radiation is a few orders of magnitude lower than the radiation produced due to $\varepsilon_{\alpha_\bullet}$ (see \Sec{Applications:Isolated:Results}). Hence, our definition of the radiative efficiency differs slightly from previous studies, \citep[e.g.][]{FSP18}; however, this difference affects only a subdominant correction factor and does not alter our qualitative conclusions.

In addition to \epsilonr, a second efficiency parameter is needed to describe how effectively the released thermal energy can be absorbed by the surrounding gas. In this work, the fraction of the thermal energy that can couple to the gas is assumed to be \epsilonf = 0.05 \citep{SDH05}. Thus, the rate at which thermal feedback energy is deposited into the surrounding gas is given by 
\begin{flalign} \label{eq:dE}
\dE = \epsilonf\ \epsilonr\ \dMbullet\ c^2 \;.
\end{flalign}

\subsection{Summary of the black hole--accretion disc model} \label{sec:Model:Summary}

Here, we provide a short summary of the method detailed in \Sec{Model} which describes our approach of modelling relativistic accretion discs around spinning black holes.

For a black hole of a given mass and spin parameter [\Eq{Sbulletvec}], an accretion disc (potentially) exists that starts at the photon radius [\Eq{Rph}] and ends at its relativistic self--gravitating radius [evaluated via \Eq{Q}]. The accretion disc's mass is distributed in the space between these two boundaries following a surface density profile that is a combination of locally valid profiles. 

In the \textbf{intra--ISCO} region between the photon radius and the ISCO radius [\Eq{Risco}], the surface density is given by \Eq{SigmaintraISCO}. For larger radial distances, the surface density profile is a synthesis of three different regimes that represent solutions for three combinations of two different pressure [\Eq{ptot}] and two different opacity [\Eq{kappatot}] mechanisms. The region(s) of validity for each of the aforementioned three regimes is calculated by solving a set of two inequalities for each regime [\Eq{pradpgasGES} and (\ref{eq:kappaFFkappaESGES}) for \textbf{Gas--ES}, \Eq{pgaspradRES} and (\ref{eq:kappaFFkappaESRES}) for \textbf{Rad--ES}, and \Eq{pradpgasGFF} and (\ref{eq:kappaESkappaFFGFF}) for \textbf{Gas--FF}]. Once each valid regime and its boundaries have been identified [see e.g. \Fig{regional_solutions}], the corresponding surface density profile [\Eq{SigmaGES}, \Eq{SigmaRES}, and \Eq{SigmaGFF}, respectively] can be used to build the global surface density profile [see e.g. \Fig{surface_density} and \Fig{accretion_disc}]. 

Once the accretion disc has been constructed and its orientation with respect to the black hole spin axis has been evaluated [\Eq{costheta}], we are able to calculate the total angular momentum of the accretion disc [\Eq{Jcircledcirc}] which allows us to calculate the fraction of the accretion disc's mass that migrates inwards and feeds the black hole [\Eq{dMcircledcircarrowbullet}]. During this accretion process, the black hole's and accretion disc's spin magnitude and direction are updated [\Eq{dSbulletvec}] to conserve the total angular momentum and mass of the system, and radiation is being emitted [\Eq{Lobs}] to conserve the total energy.

\section{Parameter space demonstration} \label{sec:Results}

\begin{figure}
\includegraphics[width=0.48\textwidth]{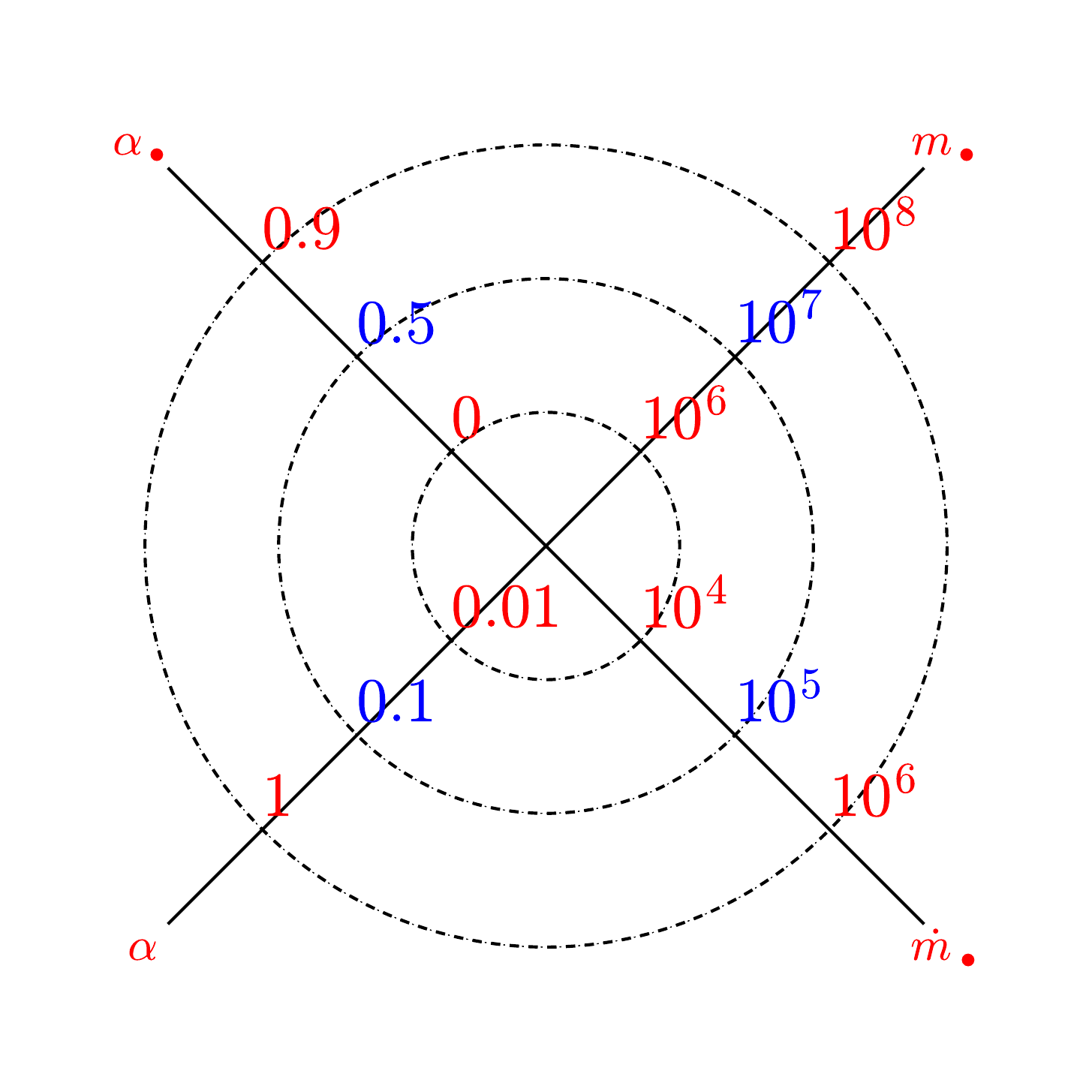}
\caption{An illustration of the parameter space explored. Moving clockwise from the bottom left--hand corner, the axes represent the accretion disc viscosity parameter $\alpha$, the black hole spin parameter $\alpha_\bullet$, the dimensionless black hole mass $m_\bullet$, and the dimensionless black hole accretion rate $\dot{m}_\bullet$. Blue values represent the ones used in the example in \Sec{Model:Properties:Example}.}
\label{fig:parameter_space}
\end{figure}

In this section, we build upon the example presented in \Sec{Model:Properties:Example} of a co--rotating accretion disc with $\alpha = 0.1$, $\alpha_\bullet = 0.5$, $m_\bullet = 10^7$, and $\dmbullet = 10^5$; and provide a detailed but concise parameter space exploration to demonstrate how our model behaves for different values of: (i) accretion disc viscosity parameter; (ii) black hole spin parameter [\Eq{Sbulletvec}]; (iii) dimensionless black hole mass [\Eq{mbullet}]; (iv) dimensionless black hole accretion rate [\Eq{dmbullet}]; and (v) accretion disc orientation. We explore two different orientations (co-- and counter--rotation) and vary independently each one of the aforementioned four numerical parameters namely: 
\begin{enumerate}[wide=0pt,labelindent=10pt,labelwidth=10pt]
\item $\alpha = \left\{ 0.01, 0.1, 1 \right\}$
\item $\alpha_\bullet = \left\{ 0.0, 0.5, 0.9 \right\}$
\item $m_\bullet = \left\{ 10^6, 10^7, 10^8 \right\}$
\item $\dmbullet = \left\{ 10^4, 10^5, 10^6 \right\}$
\end{enumerate}
as can be also seen in the radar chart of \Fig{parameter_space}.

\subsection{Regions of validity} \label{sec:Results:Validity}

\begin{figure*}
\includegraphics[width=0.32\textwidth]{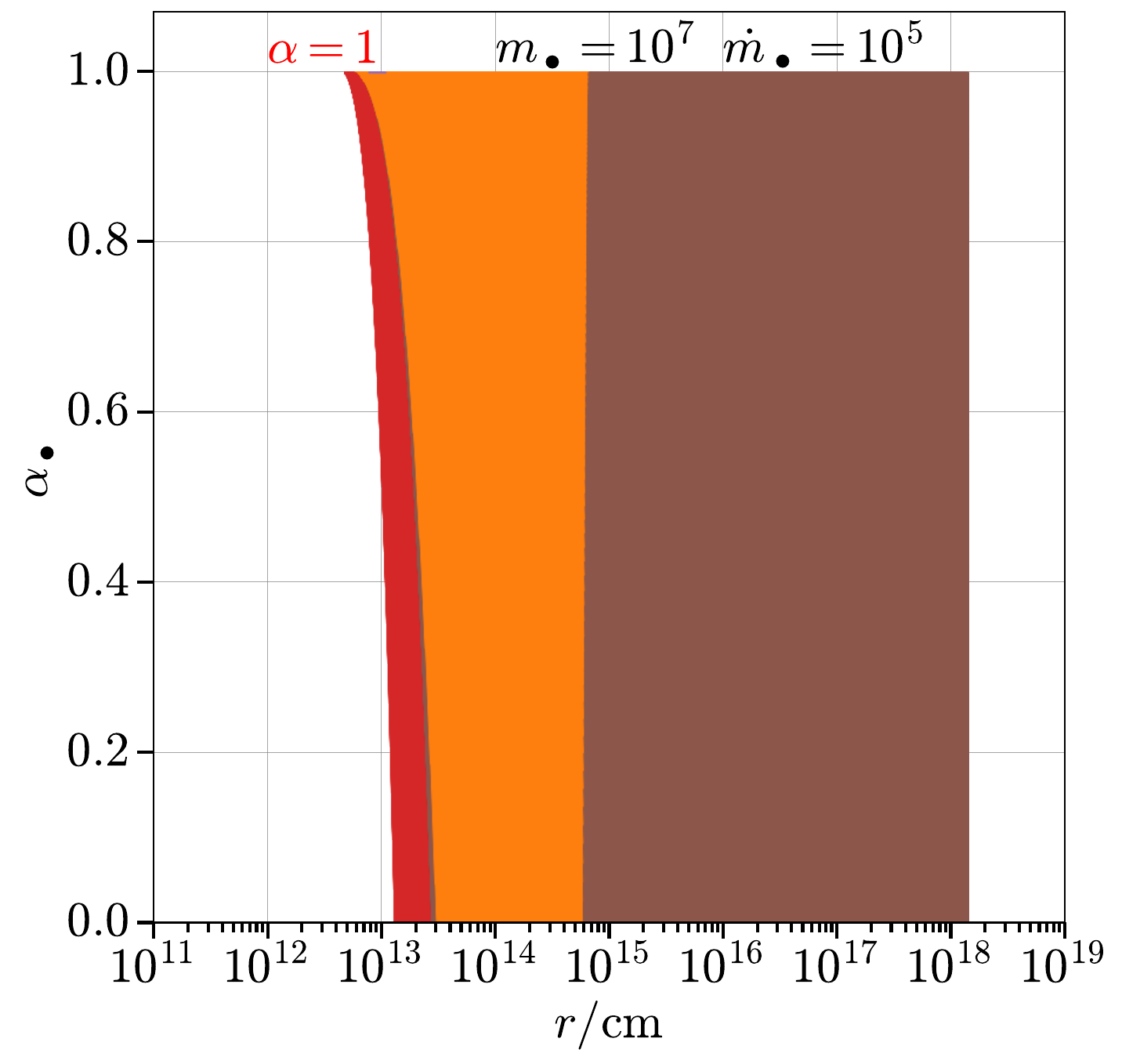} \includegraphics[width=0.32\textwidth]{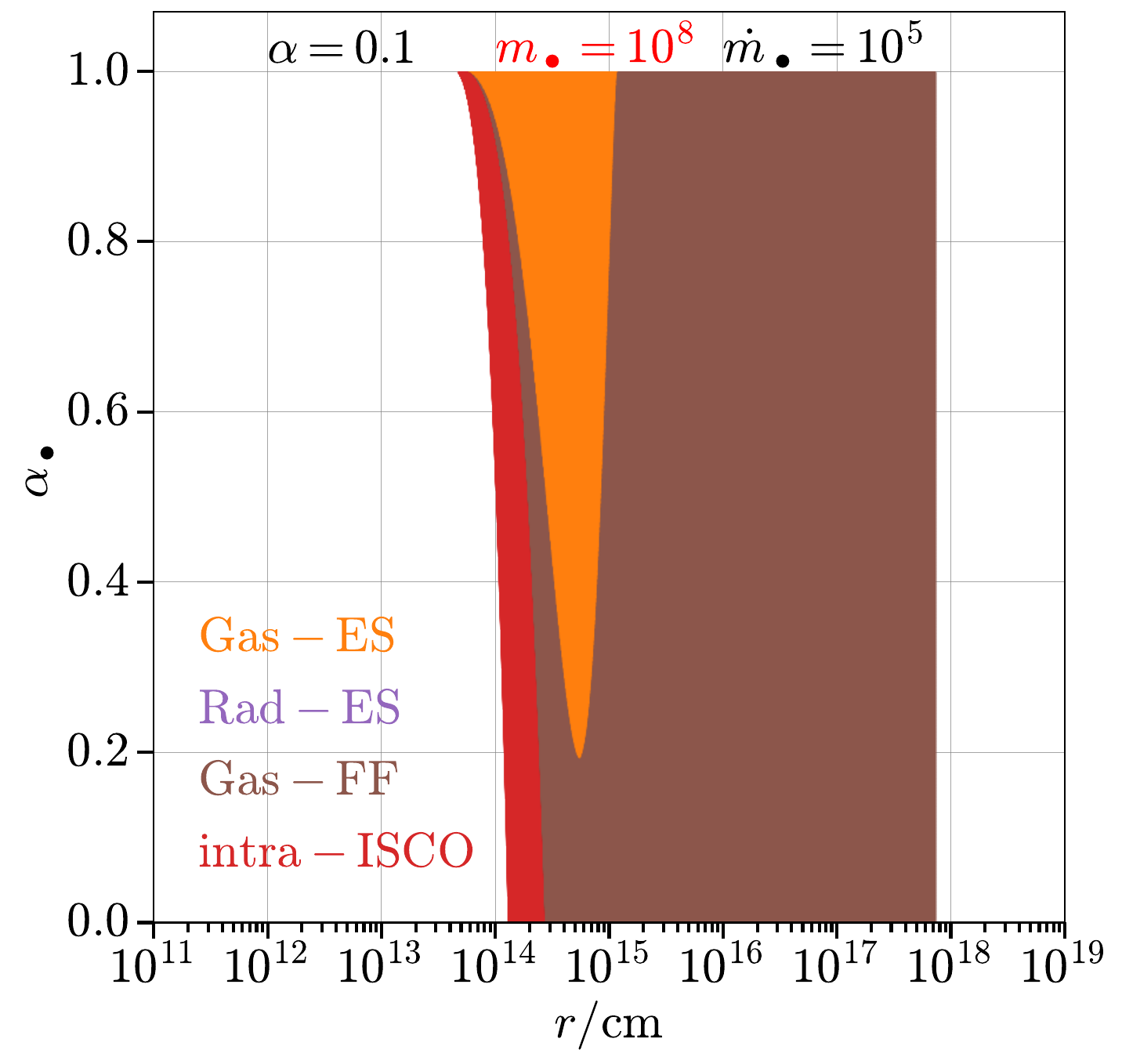} 
\includegraphics[width=0.32\textwidth]{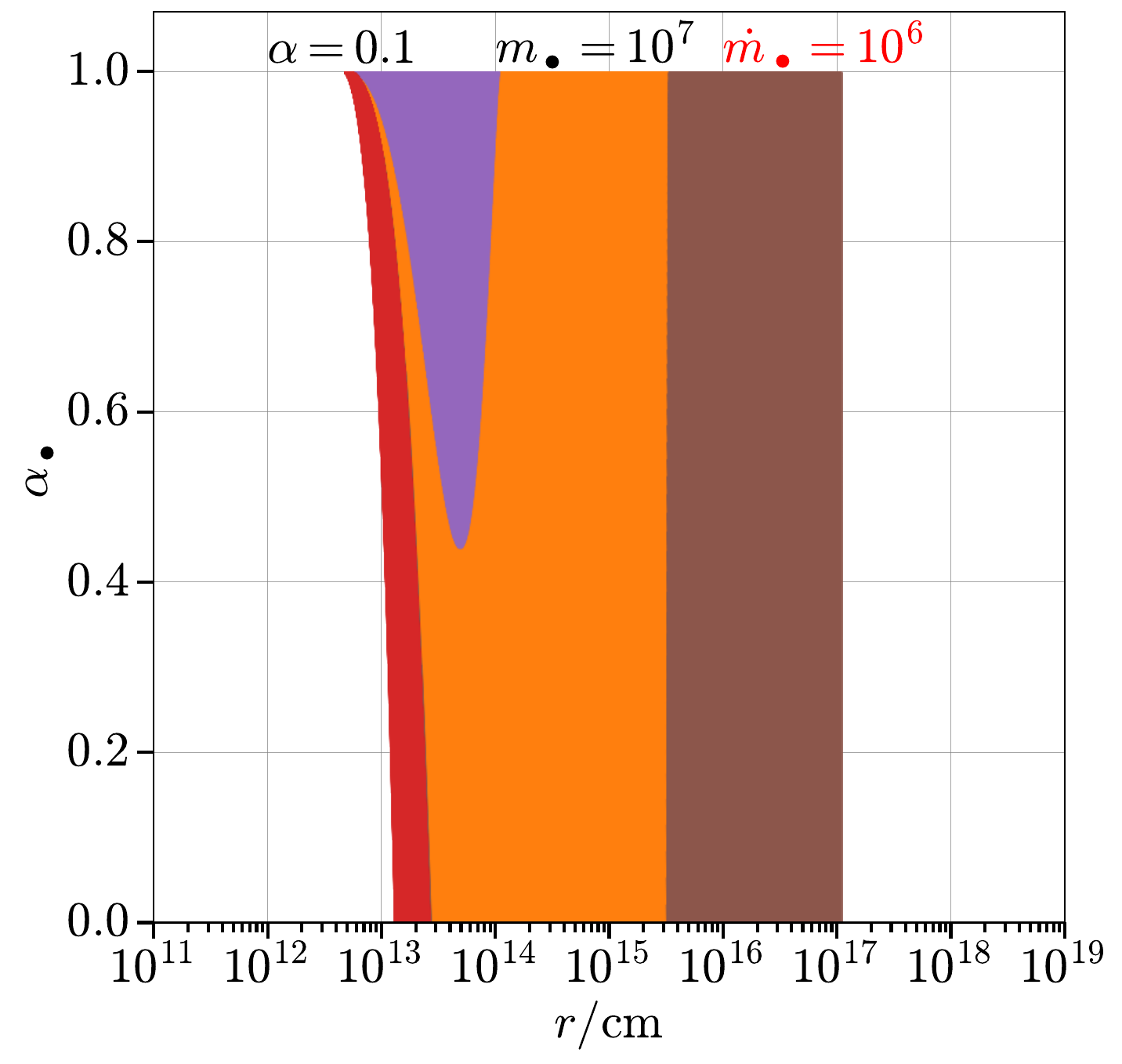} \\
\includegraphics[width=0.32\textwidth]{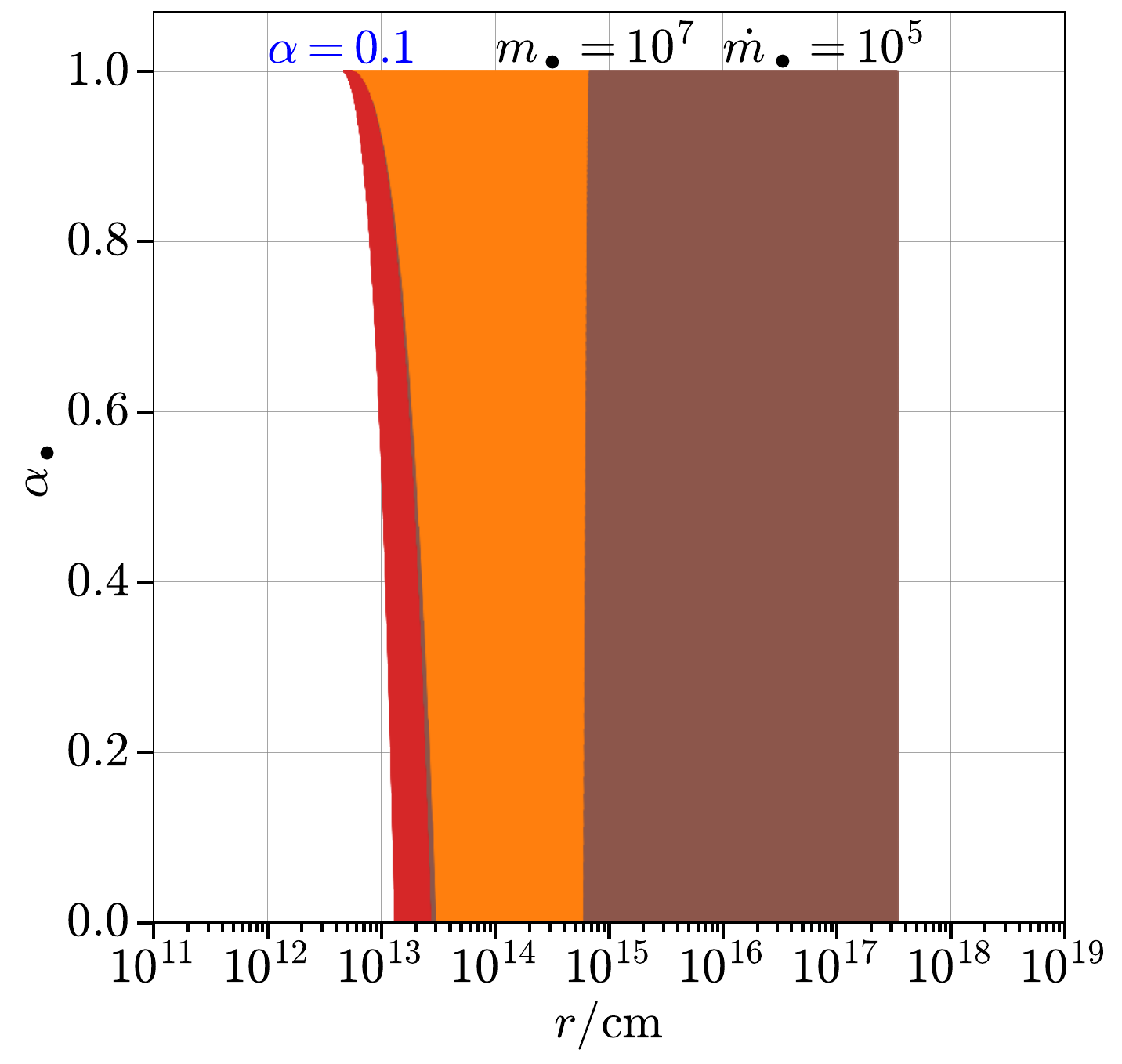} \includegraphics[width=0.32\textwidth]{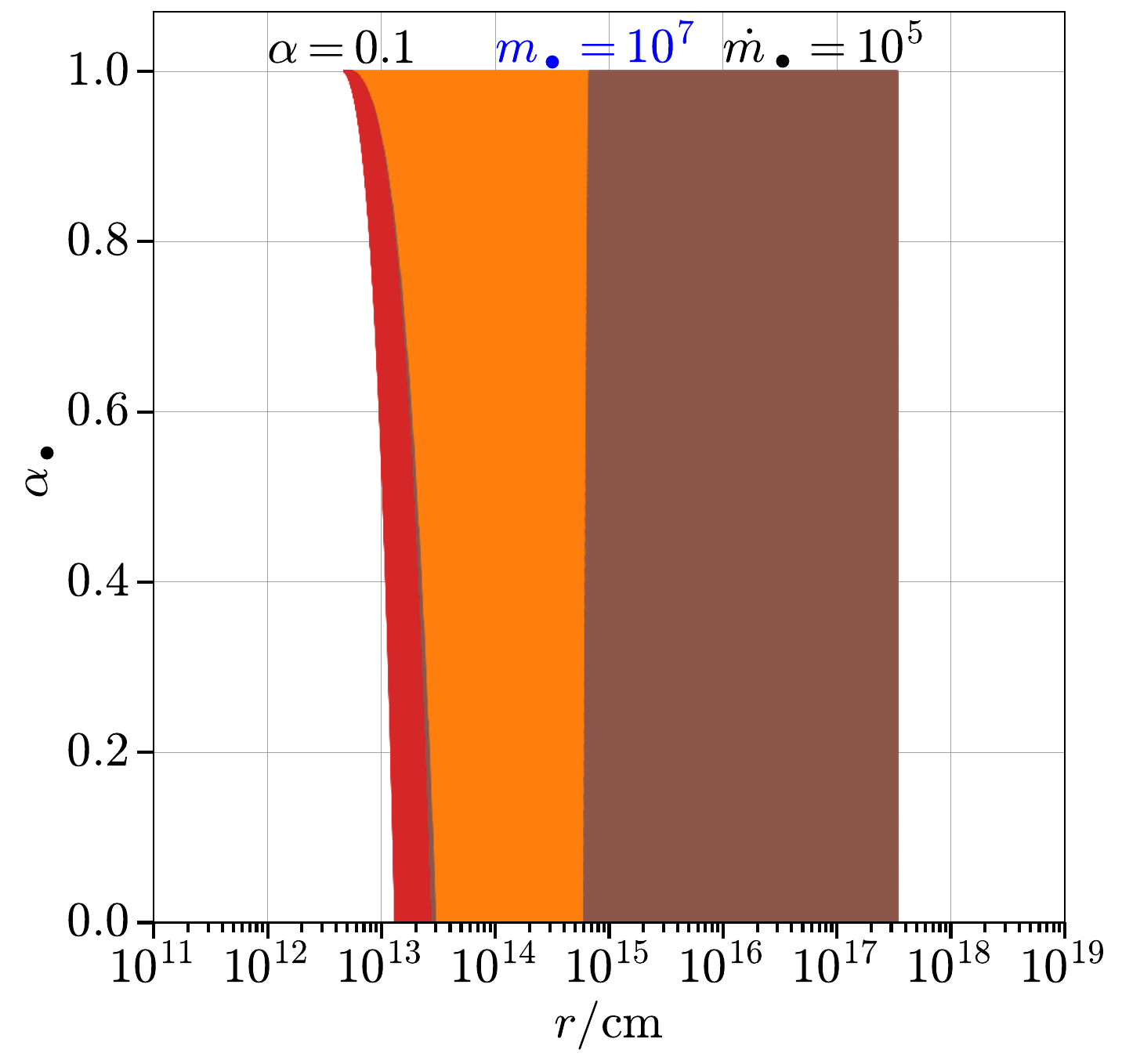}
\includegraphics[width=0.32\textwidth]{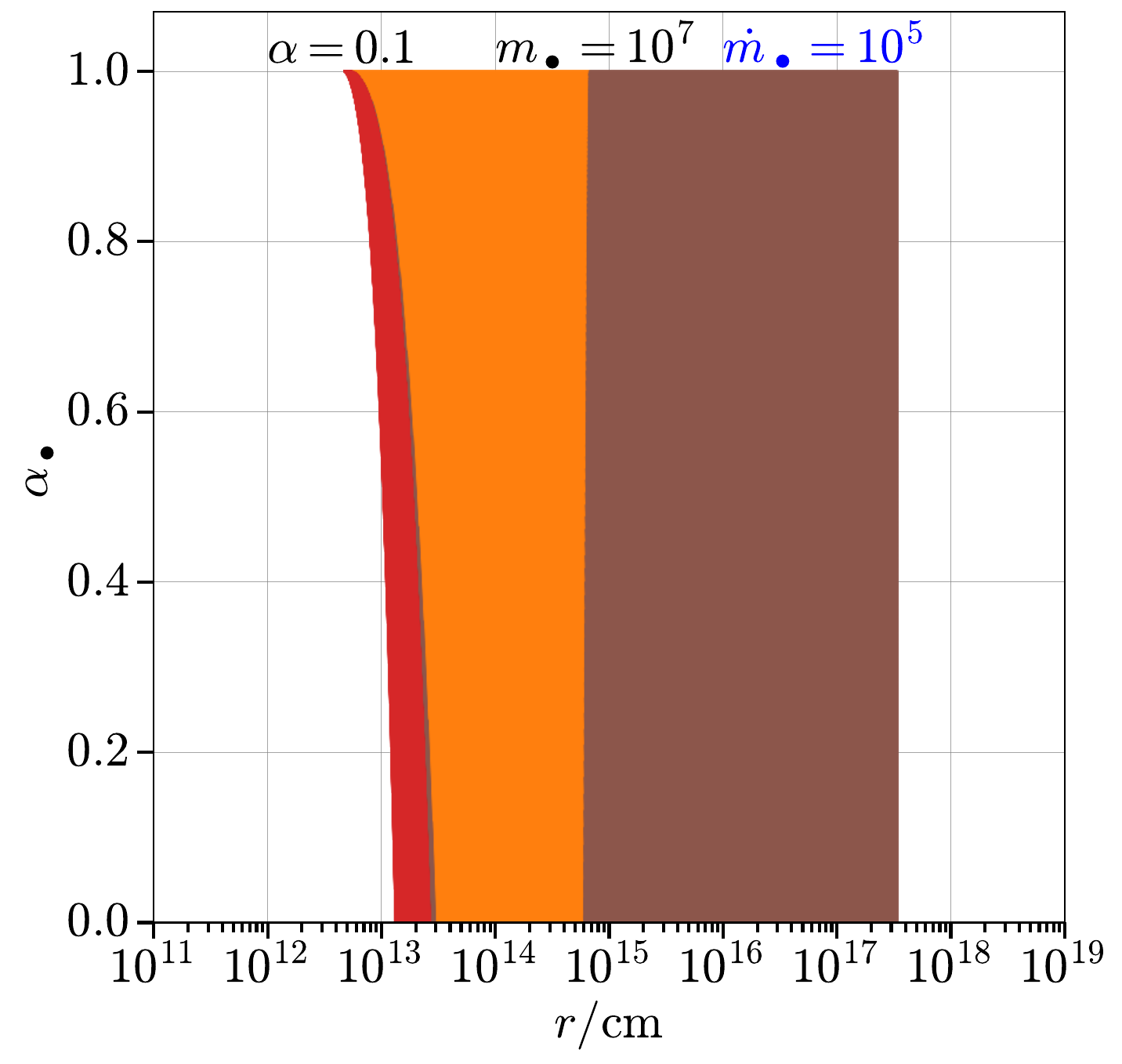} \\
\includegraphics[width=0.32\textwidth]{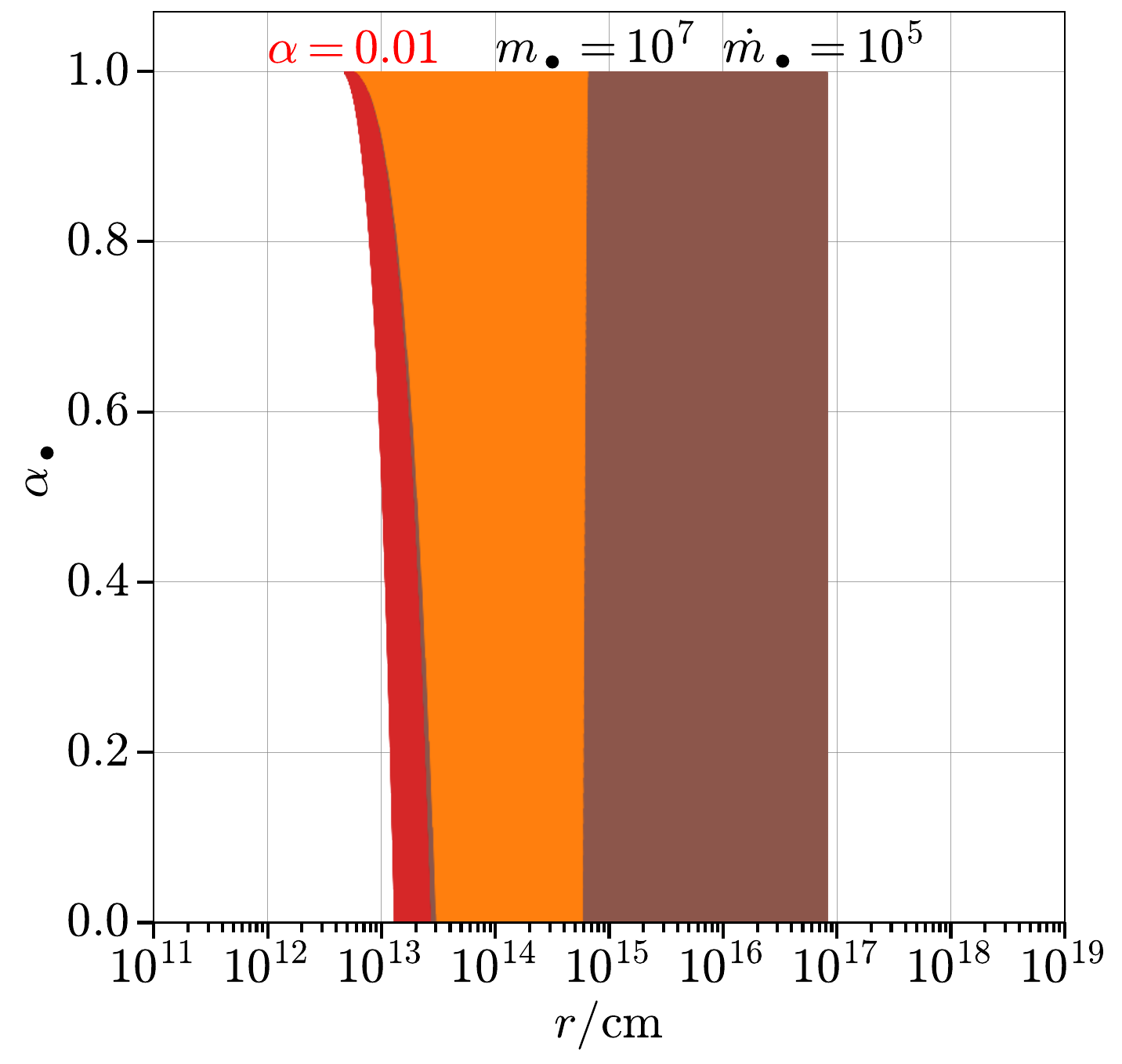} \includegraphics[width=0.32\textwidth]{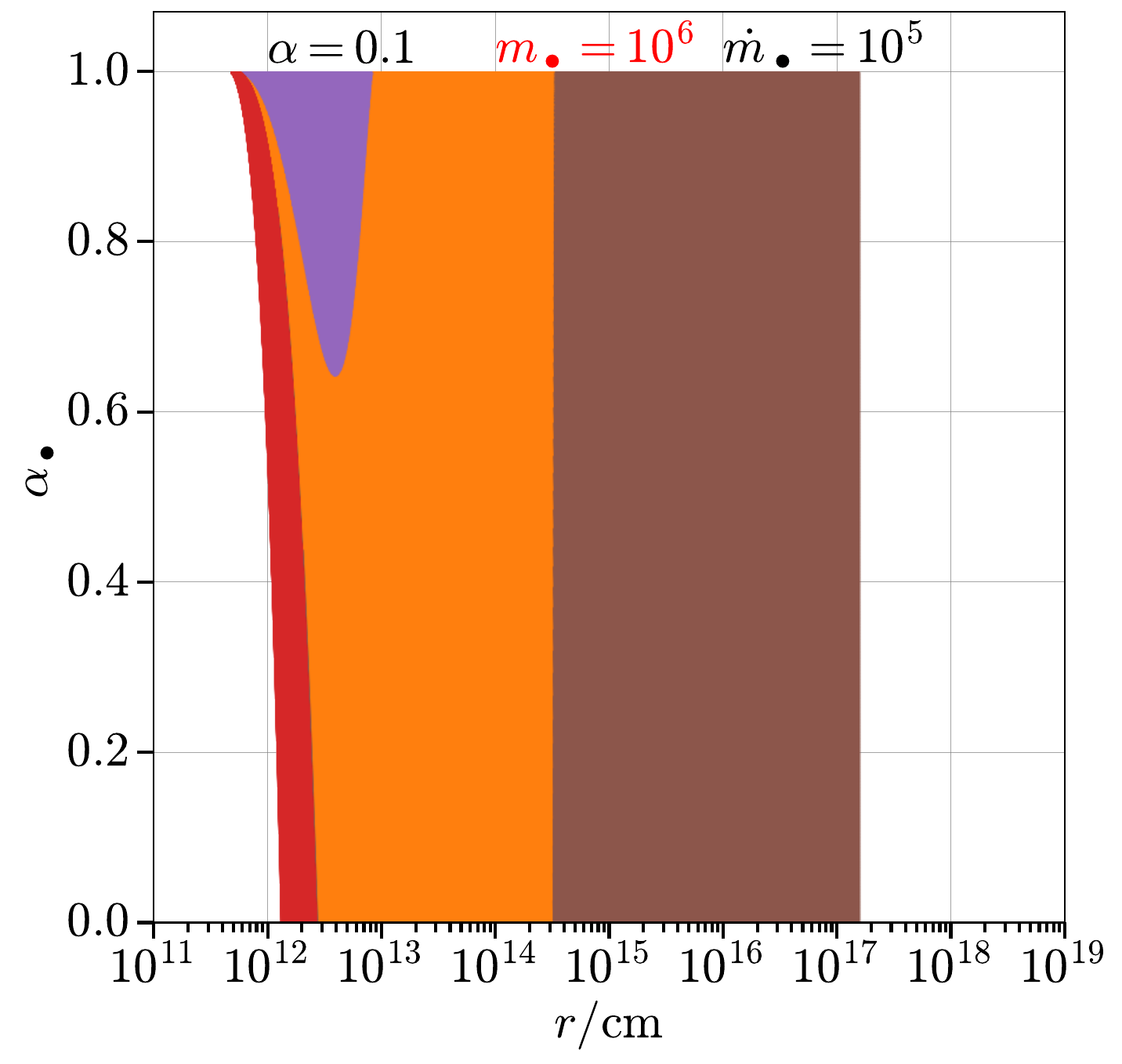}
\includegraphics[width=0.32\textwidth]{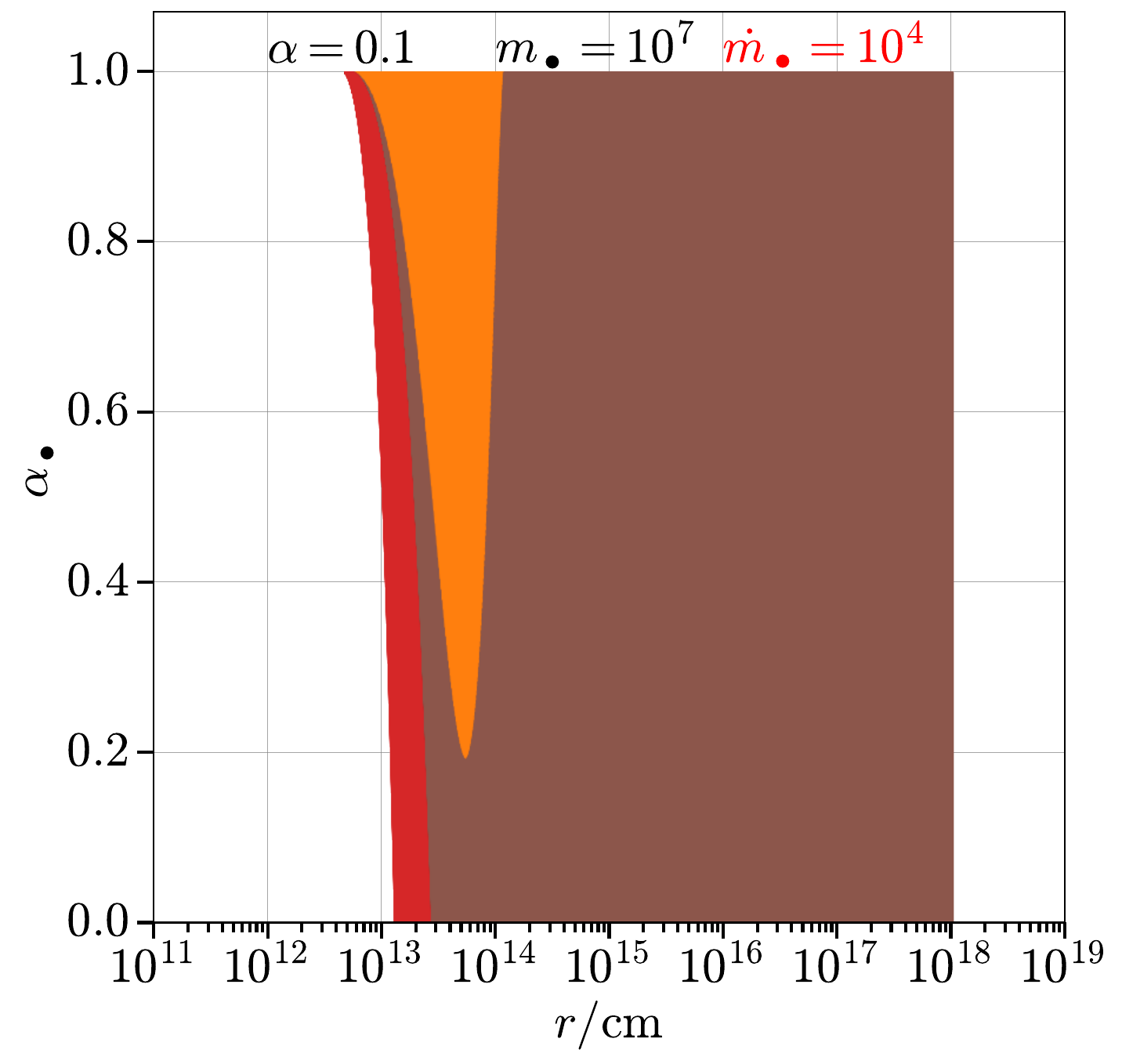}
\caption{Regions of validity on the spin--radius plane for different combinations of parameters. The left--hand, middle, and right--hand columns correspond to different values of viscosities, dimensionless black hole masses, and dimensionless black hole accretion rates, respectively, as indicated by the text in red and blue inside each panel.}
\label{fig:regional_solutions_all}
\end{figure*}

In \Fig{regional_solutions_all}, we show how the regions of validity (e.g. \Fig{regional_solutions})\footnote{As in \Fig{regional_solutions}, a spin parameter value is only necessary to calculate the extent of the accretion disc, however \Router\ is not sensitive to the exact value of $\alpha_\bullet$ (see top right--hand panel in \Fig{surface_densities_all} and \Sec{Results:Stability}).} behave for different combinations of parameters. The left--hand column shows an exploration of the viscosity parameter where the bottom, middle, and top panels correspond to $\alpha = \left\{ 0.01, 0.1, 1 \right\}$, respectively, whilst the rest of the parameters are kept fixed (i.e. $\alpha_\bullet = 0.5$, $m_\bullet = 10^7$, and $\dmbullet = 10^5$). Visually, the three panels on the left--hand side column look almost identical, apart from the fact that the outer edge of the accretion disc (i.e. its self--gravitational radius) becomes larger for higher $\alpha$ values, which implies that accretion discs with higher viscosity are able to support more mass at their outskirts before they become gravitationally unstable (see also \Sec{Results:Surface density} and \Fig{surface_densities_all}). The only difference is an extremely thin \textbf{Rad--ES} regime (purple colour) that appears around $(r/\mathrm{cm},\alpha_\bullet) \sim (10^{13}, 0.998)$ in the top panel where $\alpha = 1$, an arguably extreme value for the viscosity parameter \citep{SS73}. The weak dependence of the local solutions -- which dictate the global structure of the accretion disc -- on the viscosity parameter can been seen from the set of equations that are solved in order to identify which regime is valid and where [\Eq{pradpgasGES} and (\ref{eq:kappaFFkappaESGES}) for \textbf{Gas--ES}, \Eq{pgaspradRES} and (\ref{eq:kappaFFkappaESRES}) for \textbf{Rad--ES}, and \Eq{pradpgasGFF} and (\ref{eq:kappaESkappaFFGFF}) for \textbf{Gas--FF}], which have either very small or even non--existent dependency on $\alpha$. However, as we show below, even though the regions of validity of each regime do not strongly depend on the viscosity parameter, the properties of the accretion disc e.g. its surface density, show a strong dependency.

The middle column shows an exploration of the dimensionless black hole mass [$m_\bullet = M_\bullet/(3\Msun$)] where the bottom, middle, and top panels correspond to $m_\bullet = \left\{ 10^6, 10^7, 10^8 \right\}$, respectively, whilst the rest of the parameters are kept fixed (i.e. $\alpha = 0.1$, $\alpha_\bullet = 0.5$, and $\dmbullet = 10^5$). A strong dependence on different values of $m_\bullet$ is present, which both shifts the accretion disc further away from the black hole (i.e. more massive black holes have larger \Rph) and also drives the outer edge farther out. However, the inner edge is proportional to the black hole mass [see \Eq{Rph}], hence, \Rph\ increases linearly with $m_\bullet$, whilst \Router\ follows a shallower than linear correlation. A similar trend was reported in quasars by \cite{MKM10}, where the observed sizes of accretion discs scale as $\propto M_\bullet^{0.8}$ [see \Eq{Rinit}]. In \Sec{Applications:Isolated}, we demonstrate how with our model we can derive the half--light radius of an accretion disc and compare with observations; this is a unique feature of our model which arises from our approach to model the accretion disc's structure. Lastly, reducing $m_\bullet$ to $10^6$ whilst keeping $\dmbullet = 10^5$ (i.e. bottom panel of middle column) results in accretion rates high enough to cause the accretion disc to heat up -- especially for $\alpha_\bullet \gtrsim 0.65$ -- thus a \textbf{Rad--ES} regime appears.

Finally, the right--hand column shows an exploration of the dimensionless black hole accretion rate [$\dmbullet = \dot{M}_\bullet / (10^{17} \mathrm{g}\; \mathrm{s}^{-1})$] where the bottom, middle, and top panels correspond to $\dmbullet = \left\{ 10^4, 10^5, 10^6 \right\}$, respectively, whilst the rest of the parameters are kept fixed (i.e. $\alpha = 0.1$, $m_\bullet = 10^7$, and $\alpha_\bullet = 0.5$). As discussed in the previous paragraph, high accretion rates significantly change the structure of the accretion disc as they result in radiation pressure dominating over gas pressure for specific values of $\alpha_\bullet$, due to the more significant build up of mass on the accretion disc (see bottom left--hand panel in \Fig{surface_densities_all}). For the same reason, the accretion disc's extent decreases for higher accretion rates since the local sound speed of the \textbf{Gas--FF} regime has weaker dependence on the accretion rate compared to the surface density, resulting in the accretion disc becoming gravitationally unstable further in [see \Eq{Q}].

For the sake of minimising the length of \Sec{Results}, we refrain from reproducing \Fig{regional_solutions_all} for counter--rotating accretion discs. However, the effect of different orientations on the regions of validity can be indirectly inferred from \Fig{surface_densities_all} or \Fig{stabilities_all} based on which regimes are expressed and where.

\subsection{Surface density profiles} \label{sec:Results:Surface density}

\begin{figure*}
\includegraphics[width=0.32\textwidth]{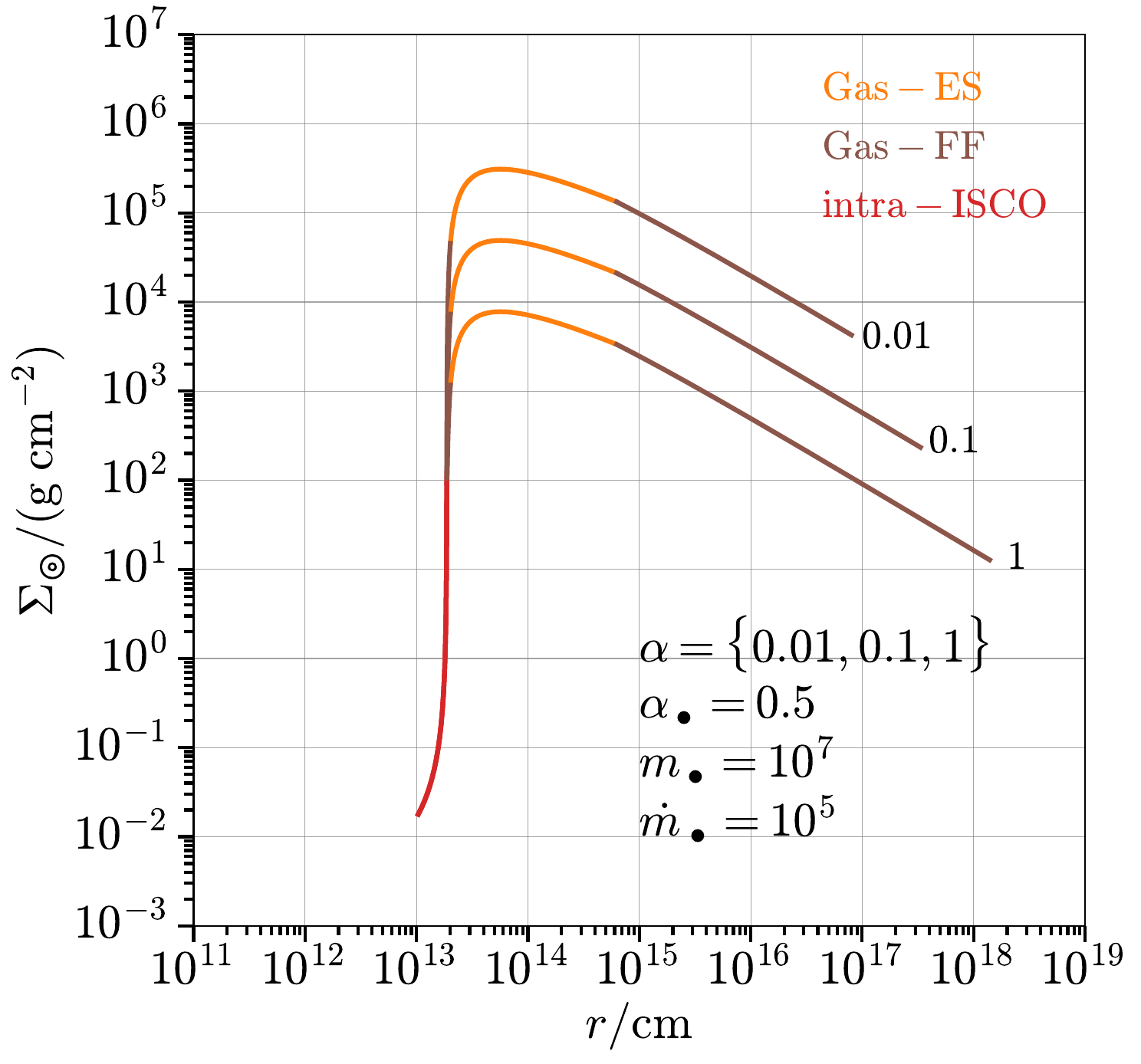} \includegraphics[width=0.32\textwidth]{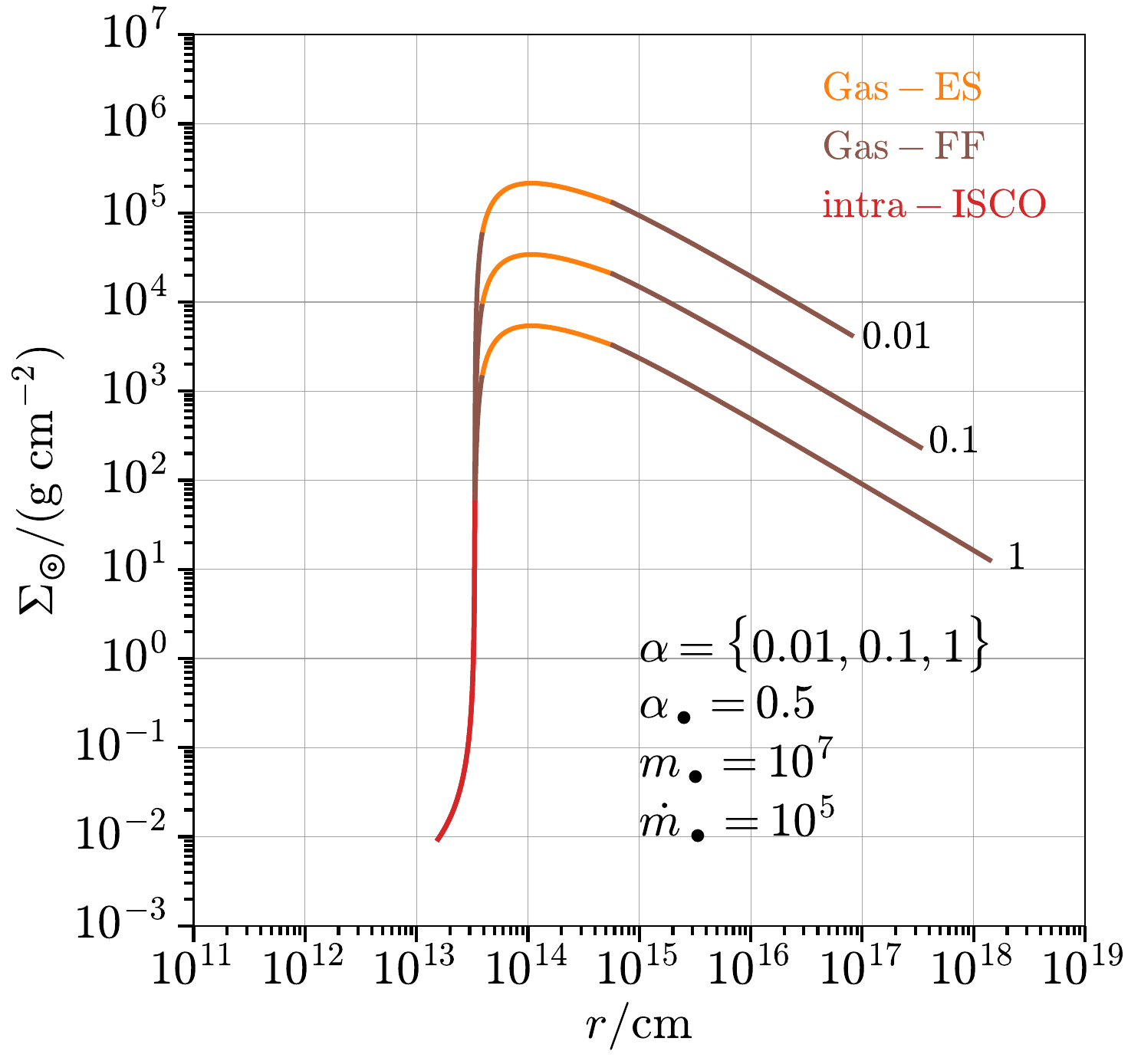} \\
\includegraphics[width=0.32\textwidth]{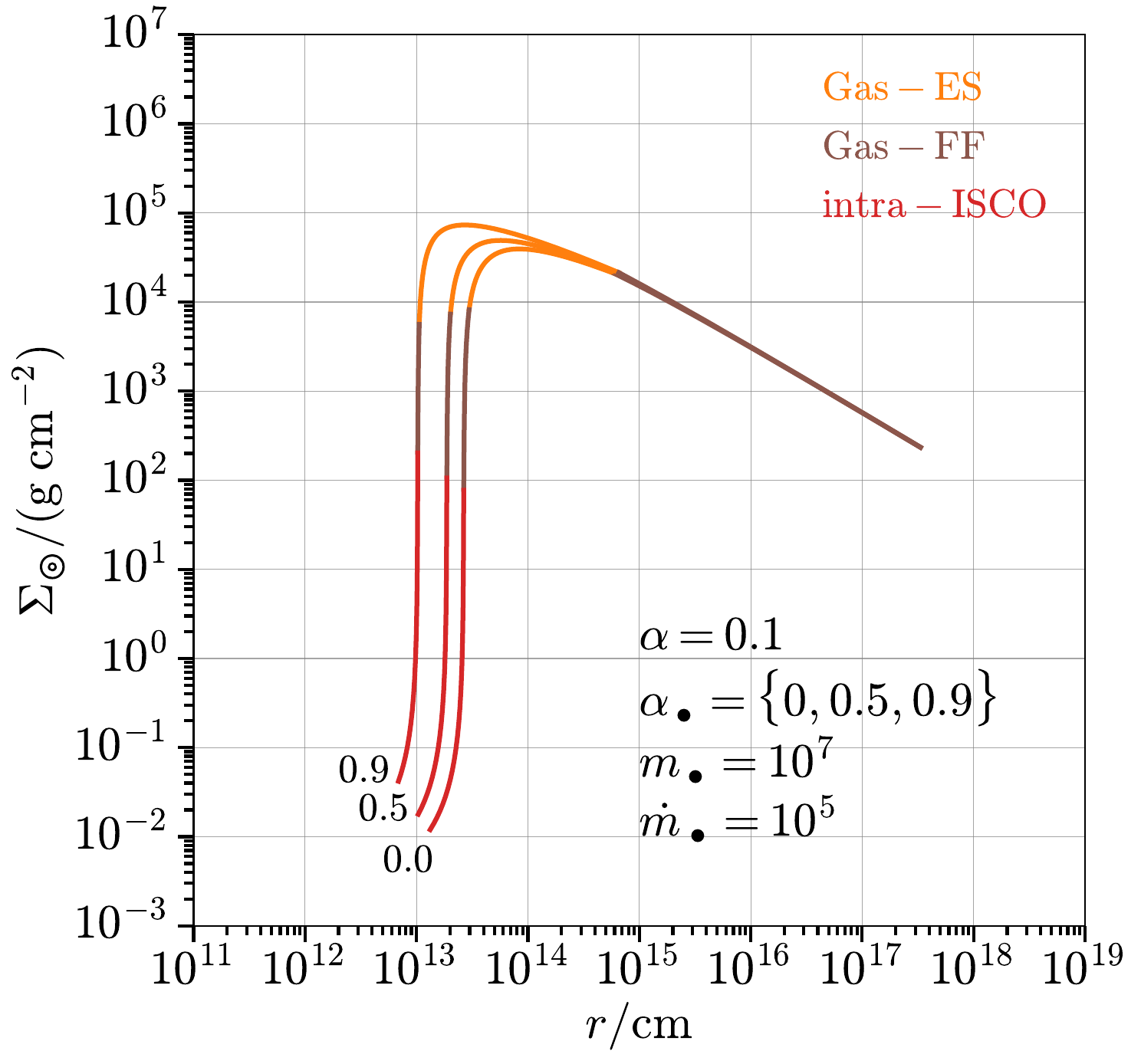} \includegraphics[width=0.32\textwidth]{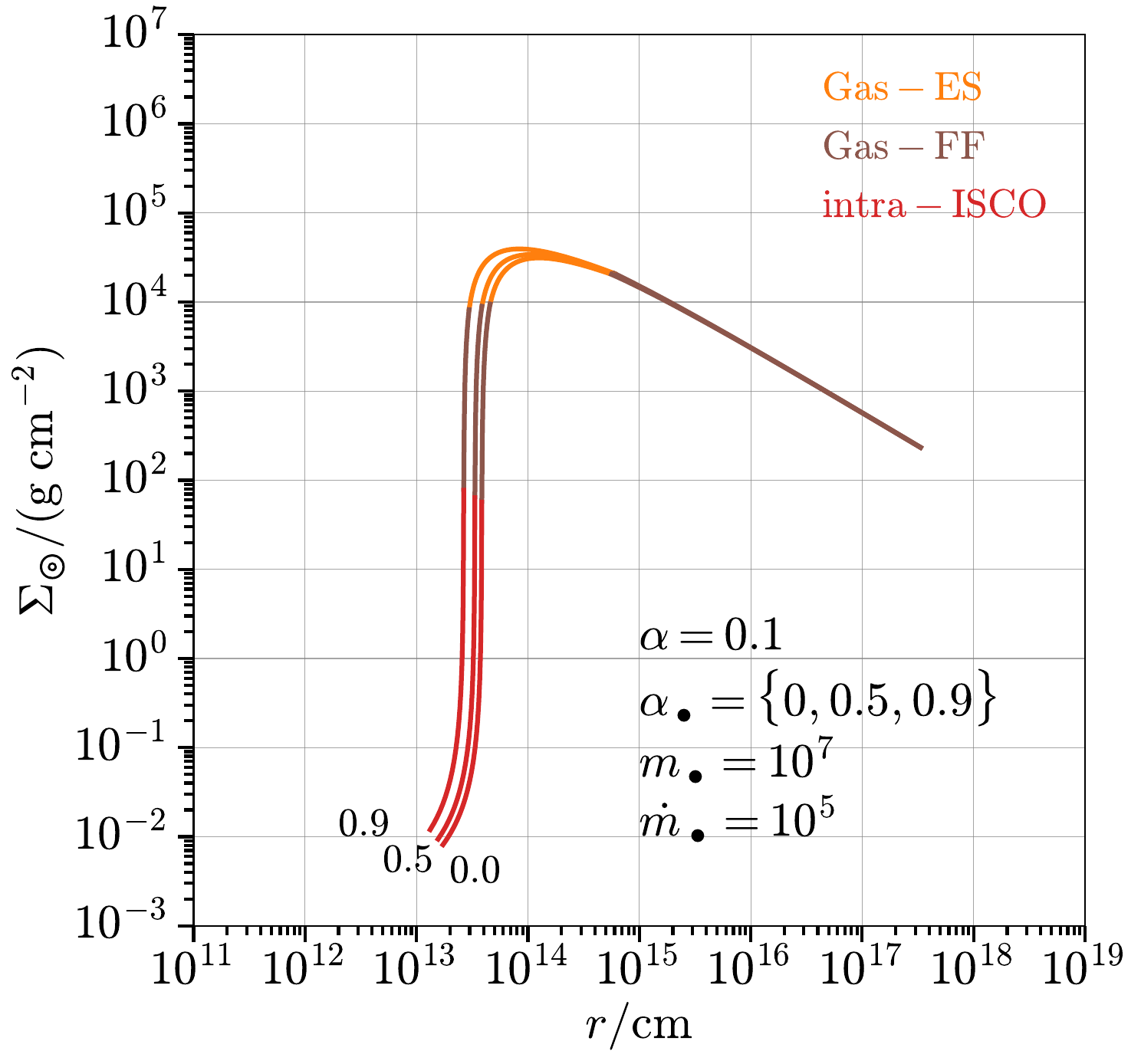} \\
\includegraphics[width=0.32\textwidth]{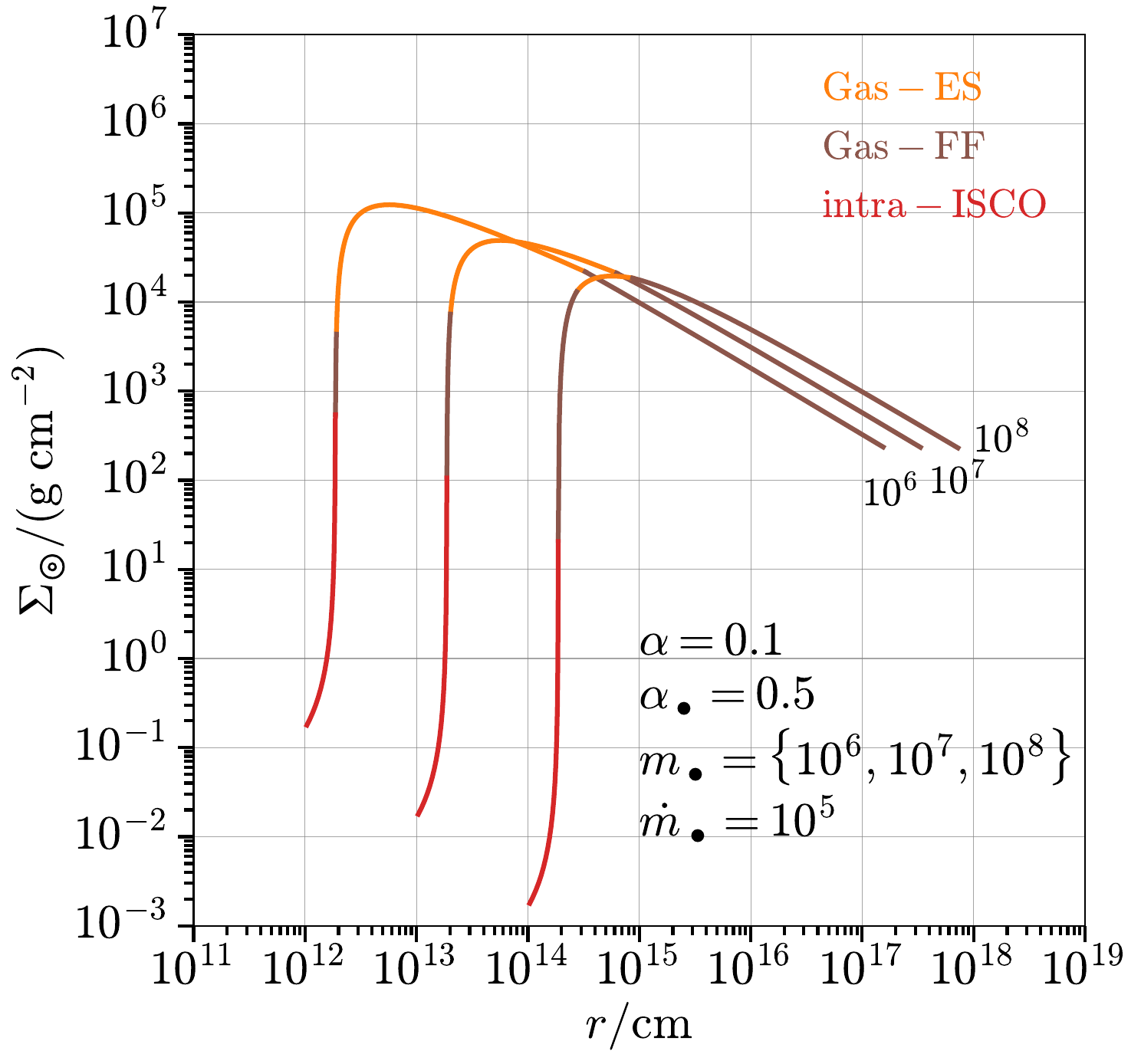} \includegraphics[width=0.32\textwidth]{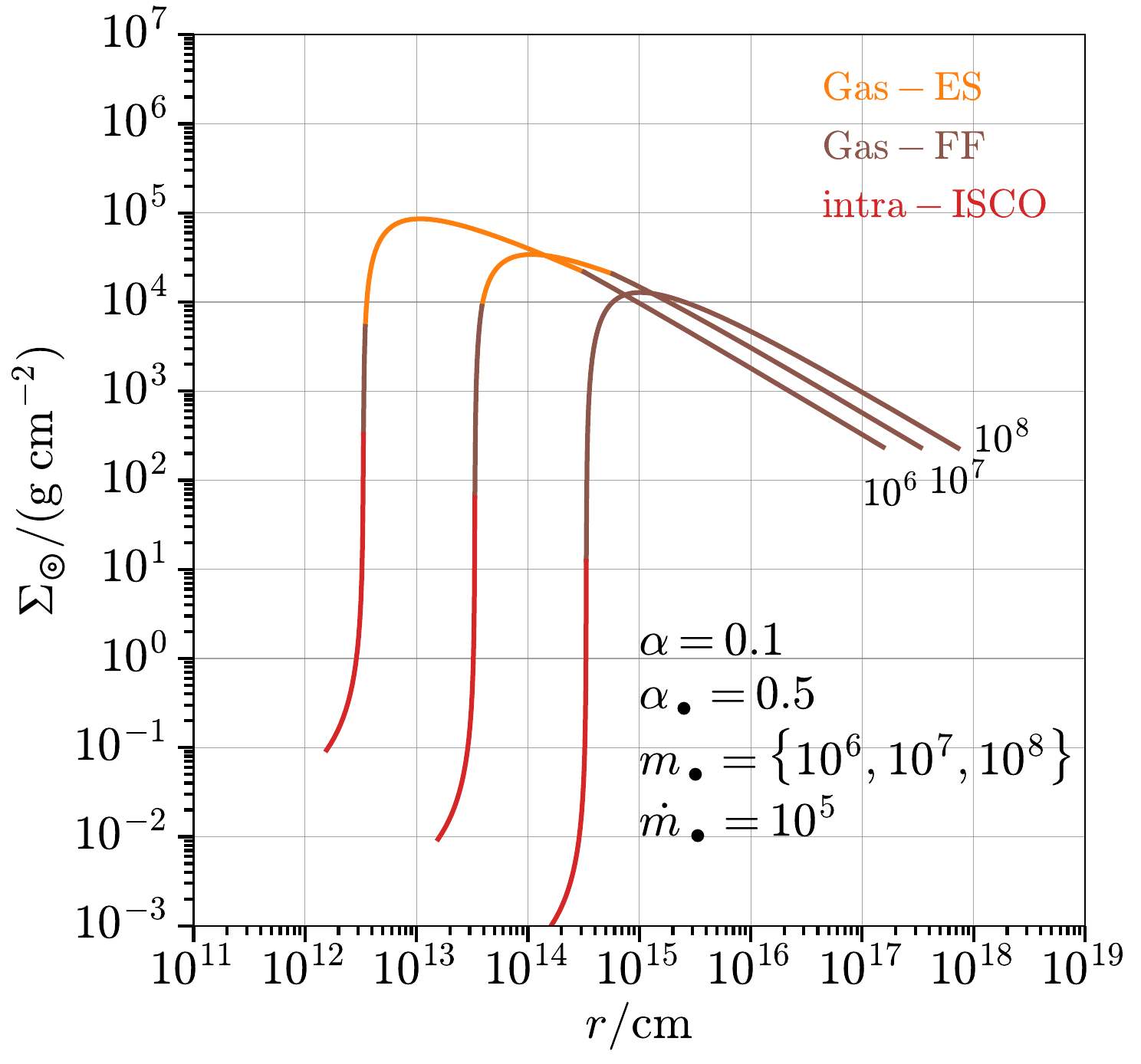} \\
\includegraphics[width=0.32\textwidth]{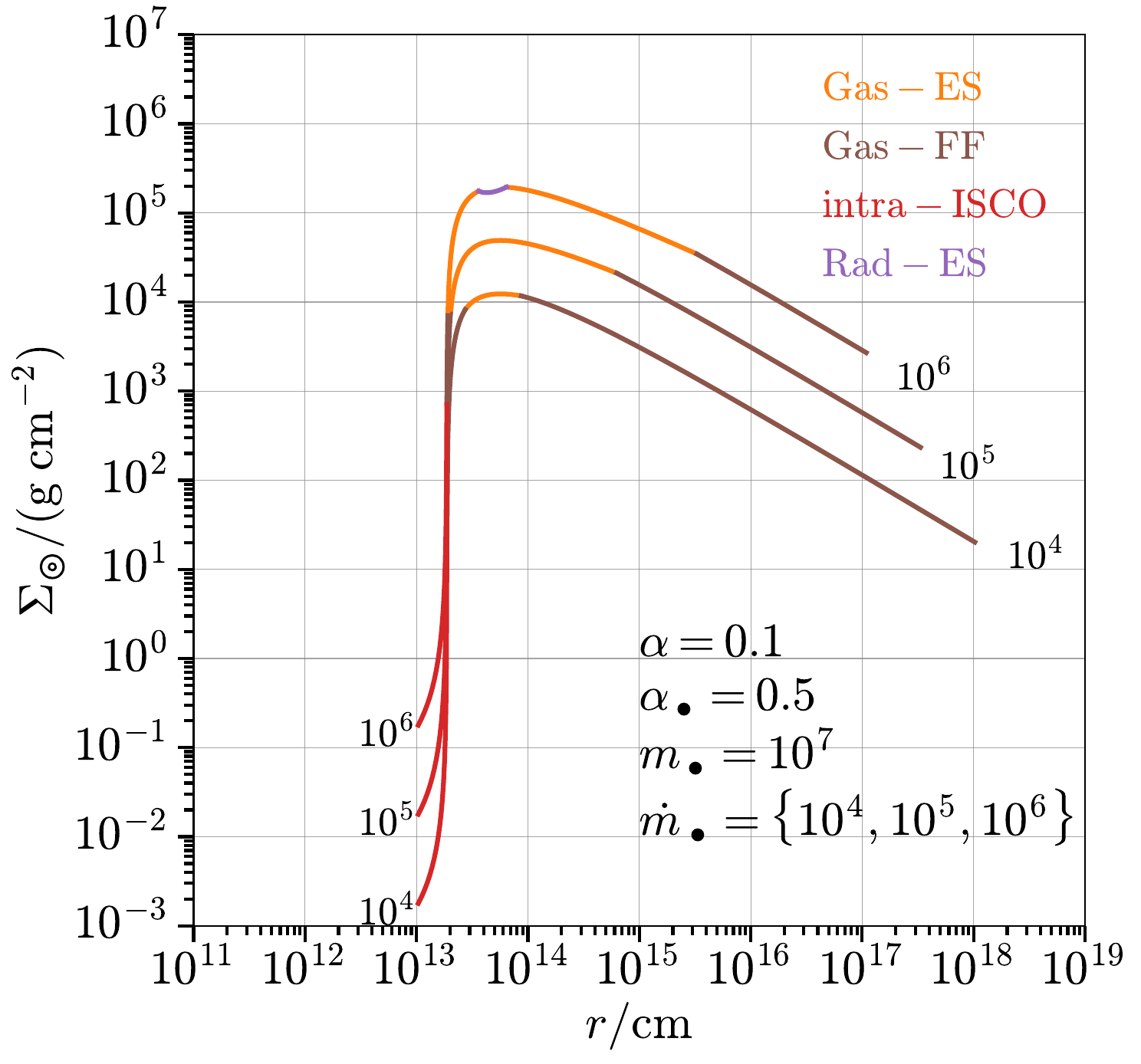} \includegraphics[width=0.32\textwidth]{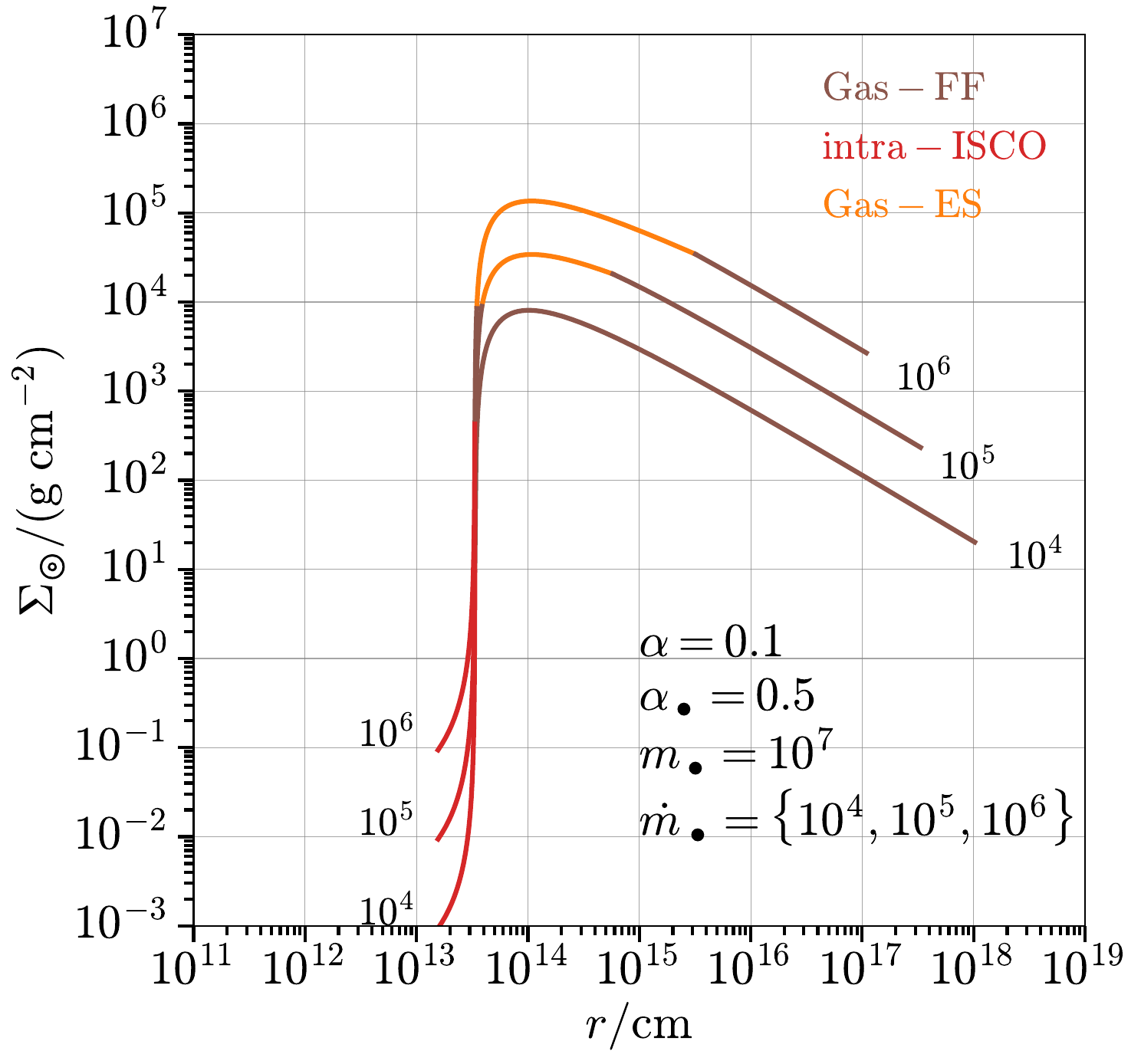}
\caption{Surface density profiles for different combinations of parameters as indicated by the text within each panel, and for two different accretion disc orientations (left--hand column: co--rotating, right--hand column: counter--rotating). In both columns, from top to bottom we explore different values of accretion disc viscosity parameter $\alpha$, black hole spin parameter $\alpha_\bullet$, dimensionless black hole mass $m_\bullet$, and dimensionless black hole accretion rate $\dot{m}_\bullet$. The text next to each line indicates which value corresponds to which line.}
\label{fig:surface_densities_all}
\end{figure*}

In this and the next subsection, we analyse further how different values of the four physical parameters that control the structure of the accretion disc and different accretion disc orientations affect its properties. We begin here this parameter space exploration by showing in \Fig{surface_densities_all} the effects on the surface density profiles. 

Even though in the left--hand column of \Fig{regional_solutions_all} we saw no significant dependence of the regions of validity on the viscosity parameter except for the extent of the accretion disc, in the top left--hand panel of \Fig{surface_densities_all} the surface density profiles [see equations (\ref{eq:SigmaGES}), (\ref{eq:SigmaRES}), and (\ref{eq:SigmaGFF})] have an inversely proportional dependence on $\alpha$ which results in the highest viscosity accretion disc (i.e. $\alpha = 1$) having the lowest surface density profile. This is the reason why this accretion disc also has the largest extent since as can be seen from \Eq{Q}, the smaller the surface density the more stable the accretion disc is. We further explore this phenomenon in the next subsection. Finally, we notice that different orientations (see top right--hand panel of \Fig{surface_densities_all} for the case of counter--rotation) have a small--scale effect on the accretion disc by slightly shifting the photon radius to larger distances and reducing the region of validity of the \textbf{Gas--ES} regime, however without affecting the outer edge of the accretion disc [see also the top two panels in \Fig{stabilities_all}].

The second to top row shows an exploration of the spin parameter where the different lines correspond to $\alpha_\bullet = \left\{ 0.0, 0.5, 0.9 \right\}$, as indicated by the text next to each line, whilst the rest of the parameters are kept fixed (i.e. $\alpha = 0.1$, $m_\bullet = 10^7$, and $\dmbullet = 10^5$). Higher spins result in higher peaks of the surface density profiles, although spins do not affect the overall extent of the accretion disc [only its starting radius through \Eq{Rph}]. This latter effect (i.e. \Router\ being independent of $\alpha_\bullet$) has been discussed above in e.g. \Fig{regional_solutions} where a spin parameter of 0.5 was used to calculate the extent of the accretion disc. Furthermore, the variations in the overall structure of the accretion disc are small among the three lines in both the co--rotating and counter--rotating scenarios -- particularly for the latter case in the right--hand side panel -- indicating that how fast a black hole is spinning is not the dominant parameter (out of the four explored in \Fig{surface_densities_all}) affecting the structure of the accretion disc. This behaviour arises from the relativistic corrections for the two cases because, as can be seen in \Fig{correction_functions_all}, the range of values for these corrections is wider for the co--rotating (blue colour map) than the counter--rotating (red colour map) case, resulting in more similar accretion disc structure in the latter than the former scenario.

The second to bottom left--hand panel shows an exploration of the dimensionless black hole mass on a co--rotating accretion disc, where the different lines correspond to $m_\bullet = \left\{ 10^6, 10^7, 10^8 \right\}$, as indicated by the text next to each line, whilst the rest of the parameters are kept fixed (i.e. $\alpha = 0.1$, $\alpha_\bullet = 0.5$, and $\dmbullet = 10^5$). The effect of different black hole masses is multifaceted as they not only affect the starting and stopping radii of the accretion disc, they also determine the regions of validity of the \textbf{intra--ISCO}, \textbf{Gas--ES}, \textbf{Rad--ES}, and \textbf{Gas--FF} regimes (as discussed in \Fig{regional_solutions_all}). As a general trend, higher mass black holes have accretion discs that start further away but also extent longer, and they have lower temperatures as their pressure is dominated by the gas pressure and free--free is the main opacity mechanism (i.e. the \textbf{Gas--FF} regime covers increasingly higher proportion of the accretion disc's structure). The same trend holds for counter--rotating accretion discs (third row, right--hand panel), however it is more enhanced than in the co--rotating case and leads to a complete disappearance of the \textbf{Gas--ES} regime for $m_\bullet = 10^8$.

The bottom row shows an exploration of the dimensionless black hole accretion rate where the different lines correspond to $\dmbullet = \left\{ 10^4, 10^5, 10^6 \right\}$, as indicated by the text next to each line, whilst the rest of the parameters are kept fixed (i.e. $\alpha = 0.1$, $\alpha_\bullet = 0.5$, and $m_\bullet = 10^7$). Apart from vertically scaling the surface density profile, higher accretion rates also increase the temperature of the accretion disc leading to the expression of the \textbf{Rad--ES} regime near the black hole. In this regime, radiation pressure dominates over gas hence AGN winds/outflows are expected to develop in these regions. This is the reason why the \textbf{Rad--ES} surface density profile (i.e. purple region) dips. Finally, as in the second to bottom panels, counter--rotation results in slightly cooler and less dense accretion discs which are more dominated by gas pressure and free--free absorption.

\subsection{Stability profiles} \label{sec:Results:Stability}

\begin{figure*}
\includegraphics[width=0.32\textwidth]{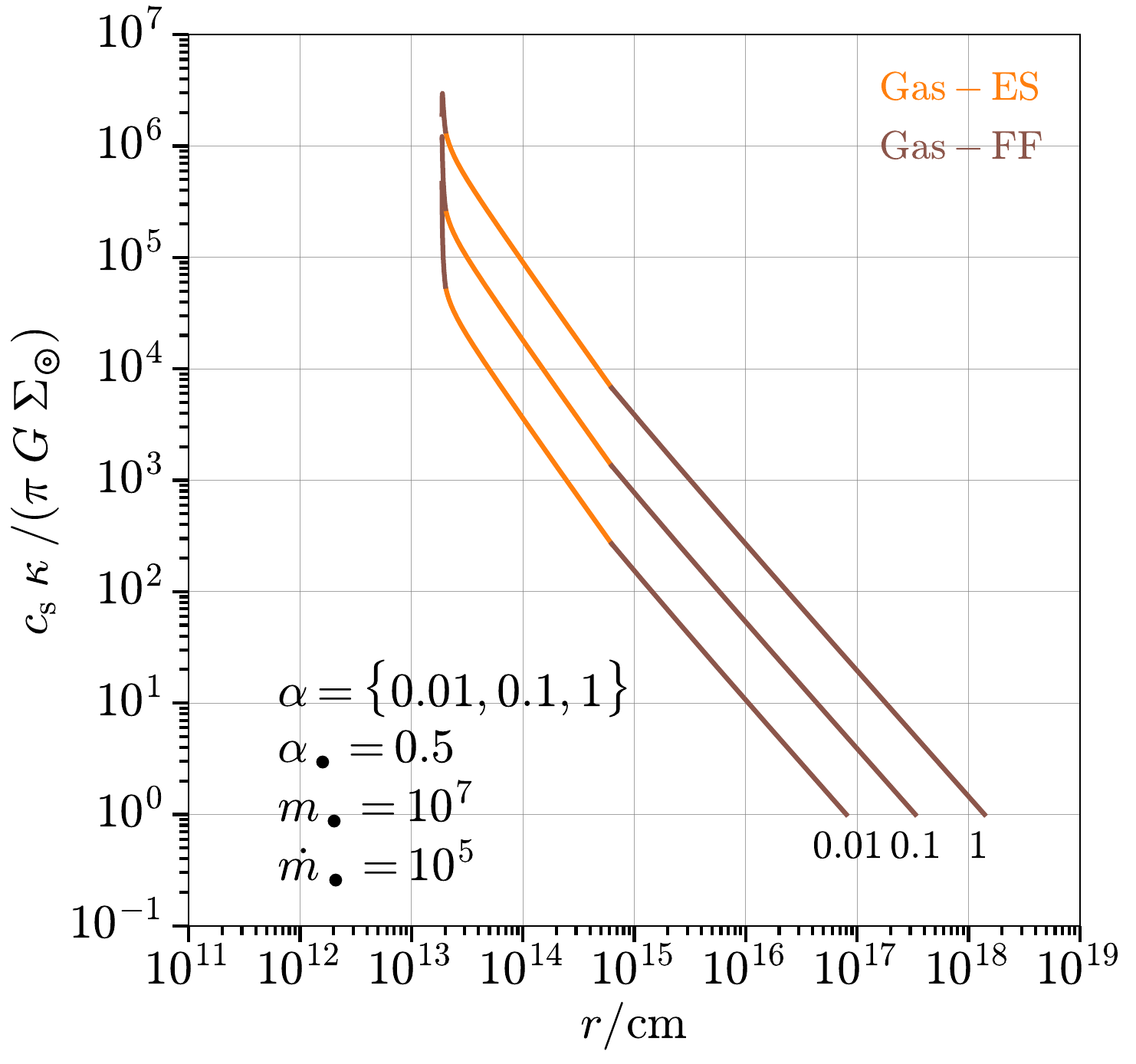} \includegraphics[width=0.32\textwidth]{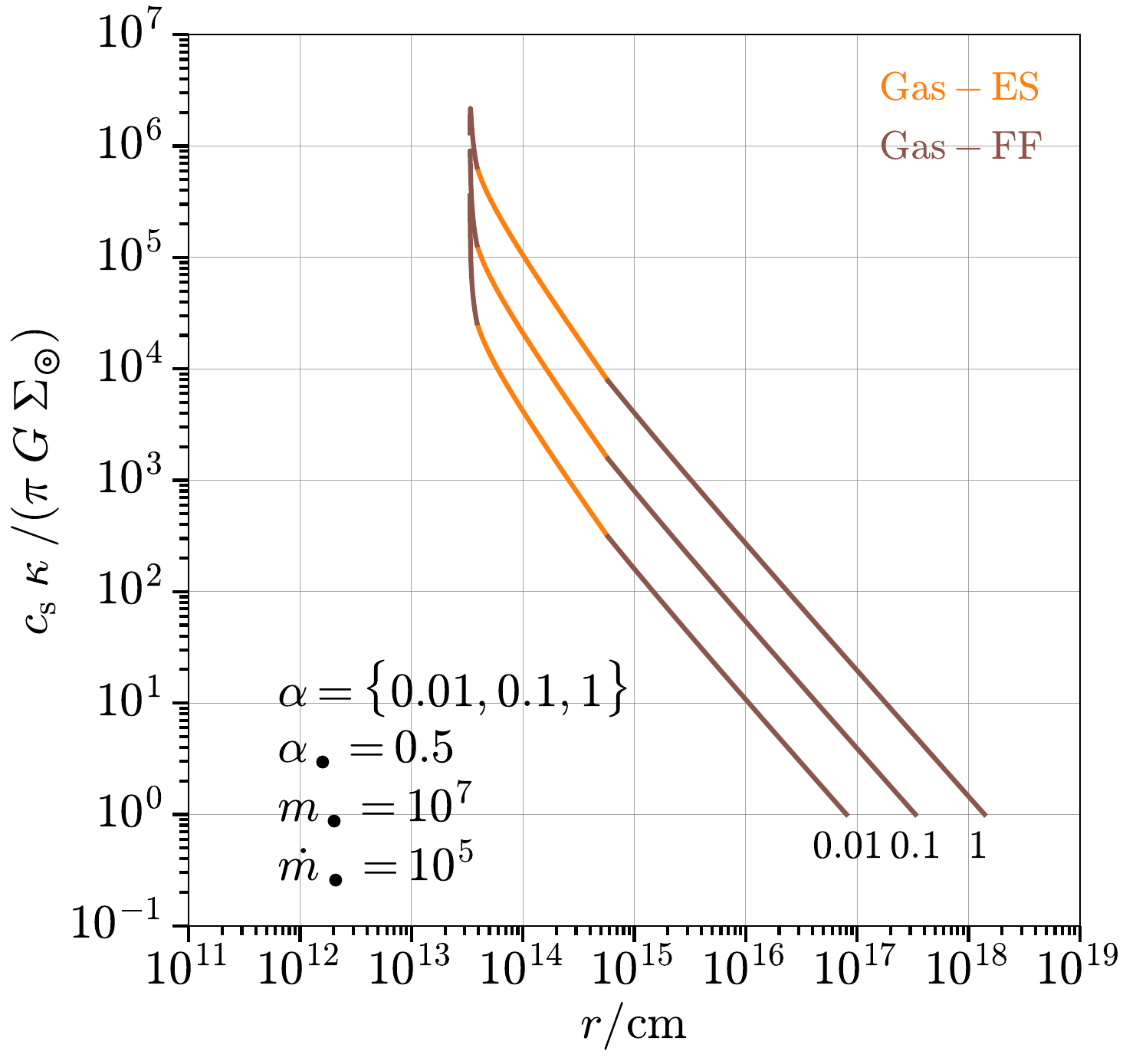} \\
\includegraphics[width=0.32\textwidth]{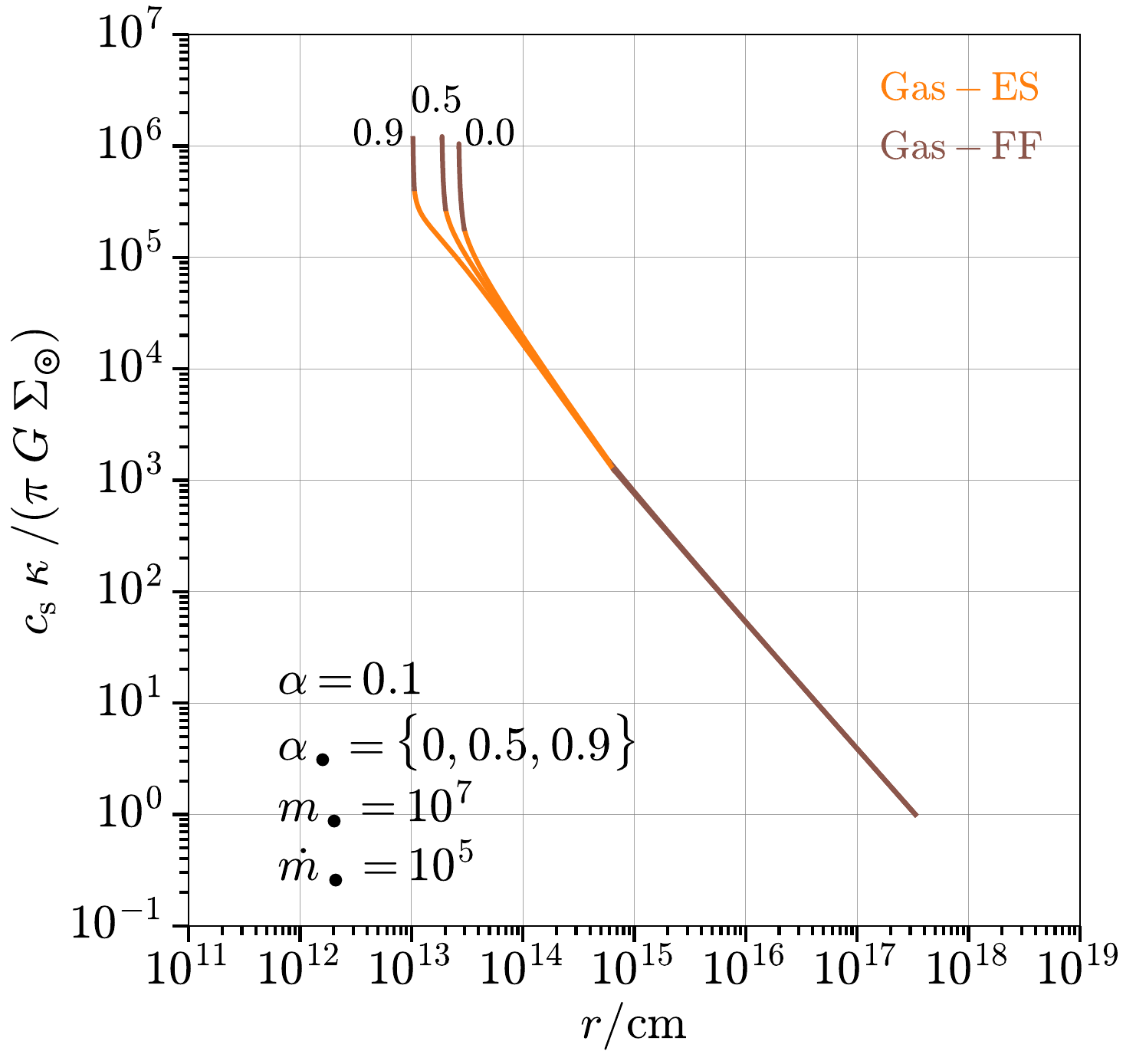} \includegraphics[width=0.32\textwidth]{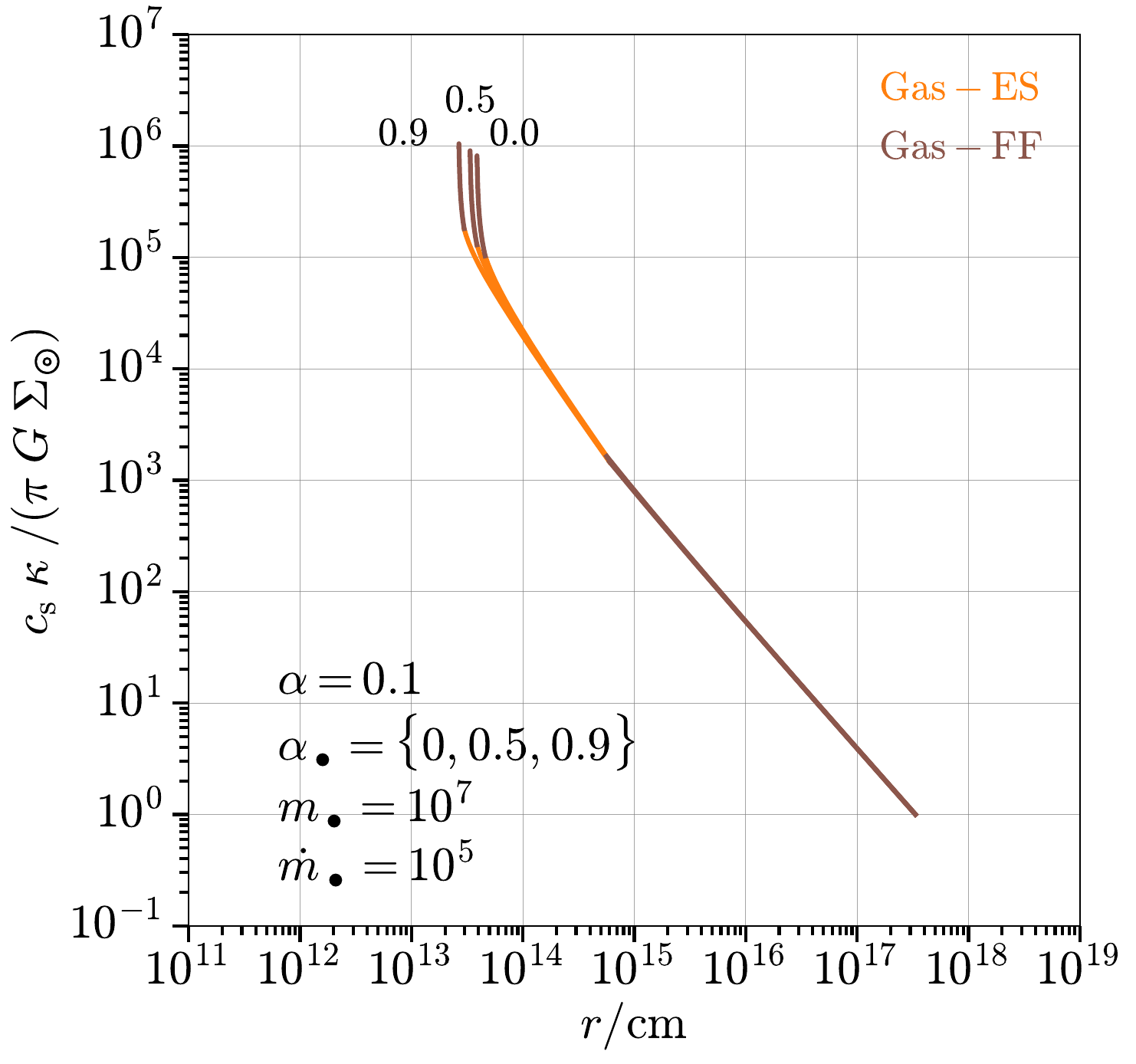} \\
\includegraphics[width=0.32\textwidth]{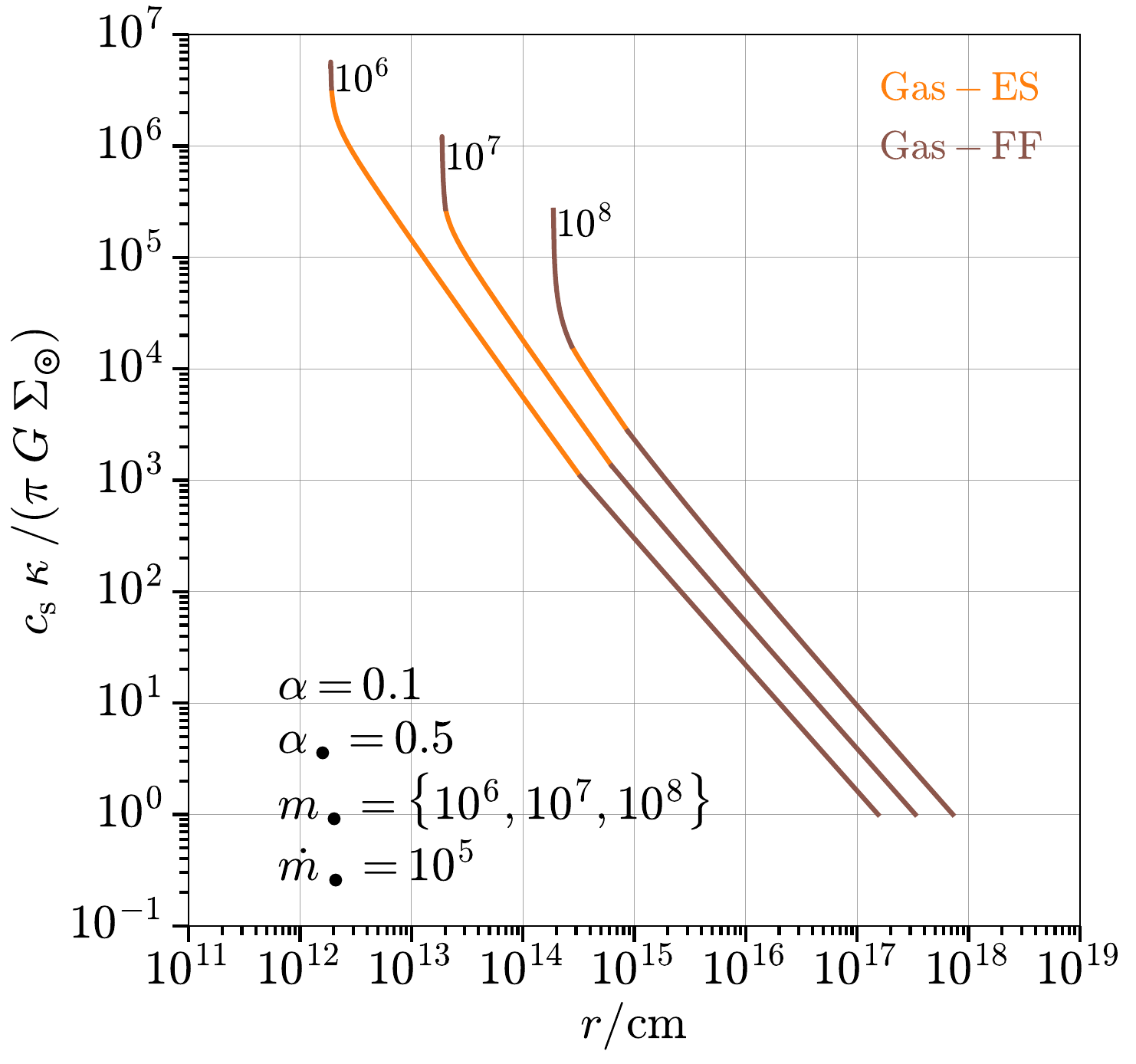} \includegraphics[width=0.32\textwidth]{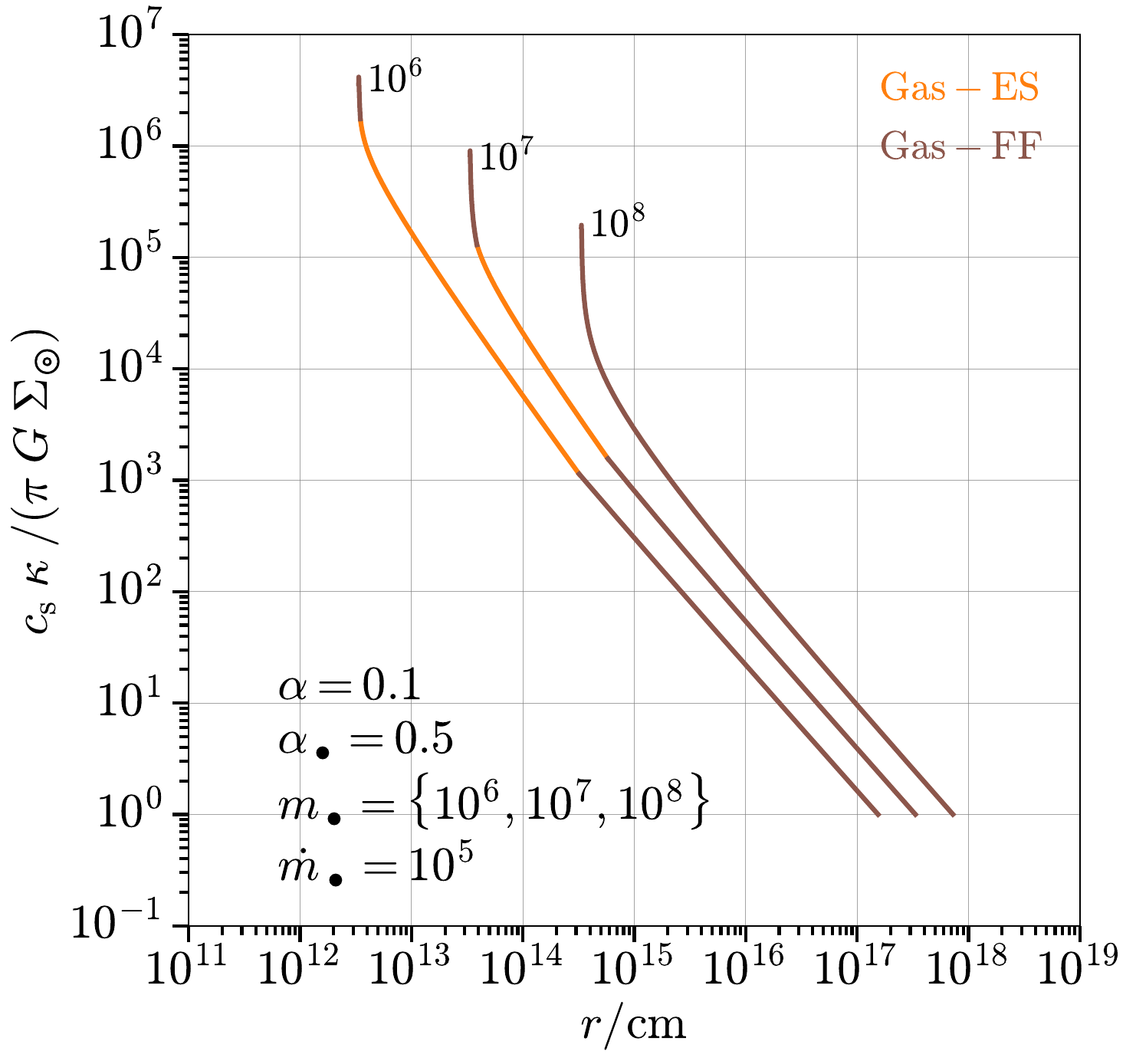} \\
\includegraphics[width=0.32\textwidth]{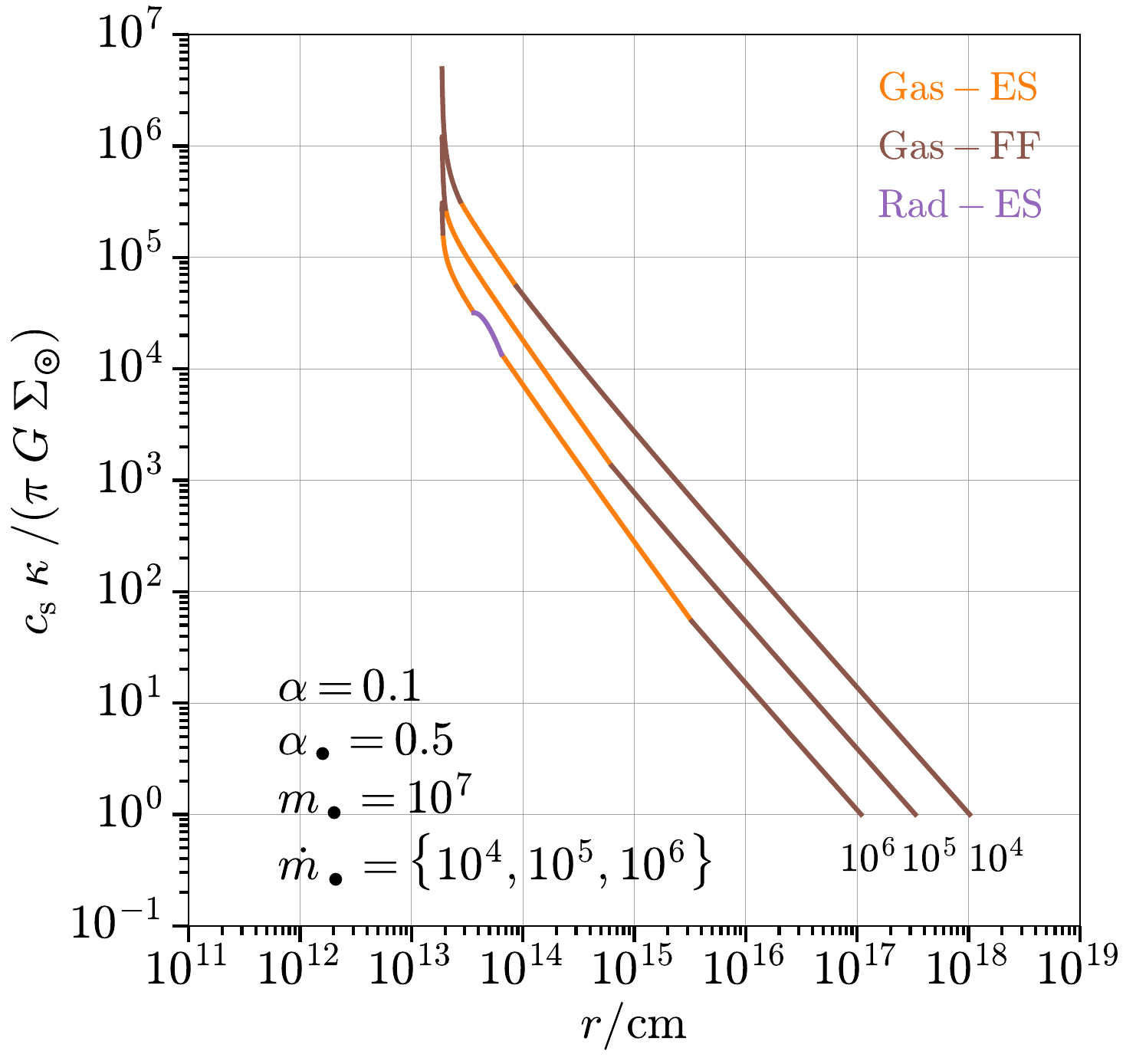} \includegraphics[width=0.32\textwidth]{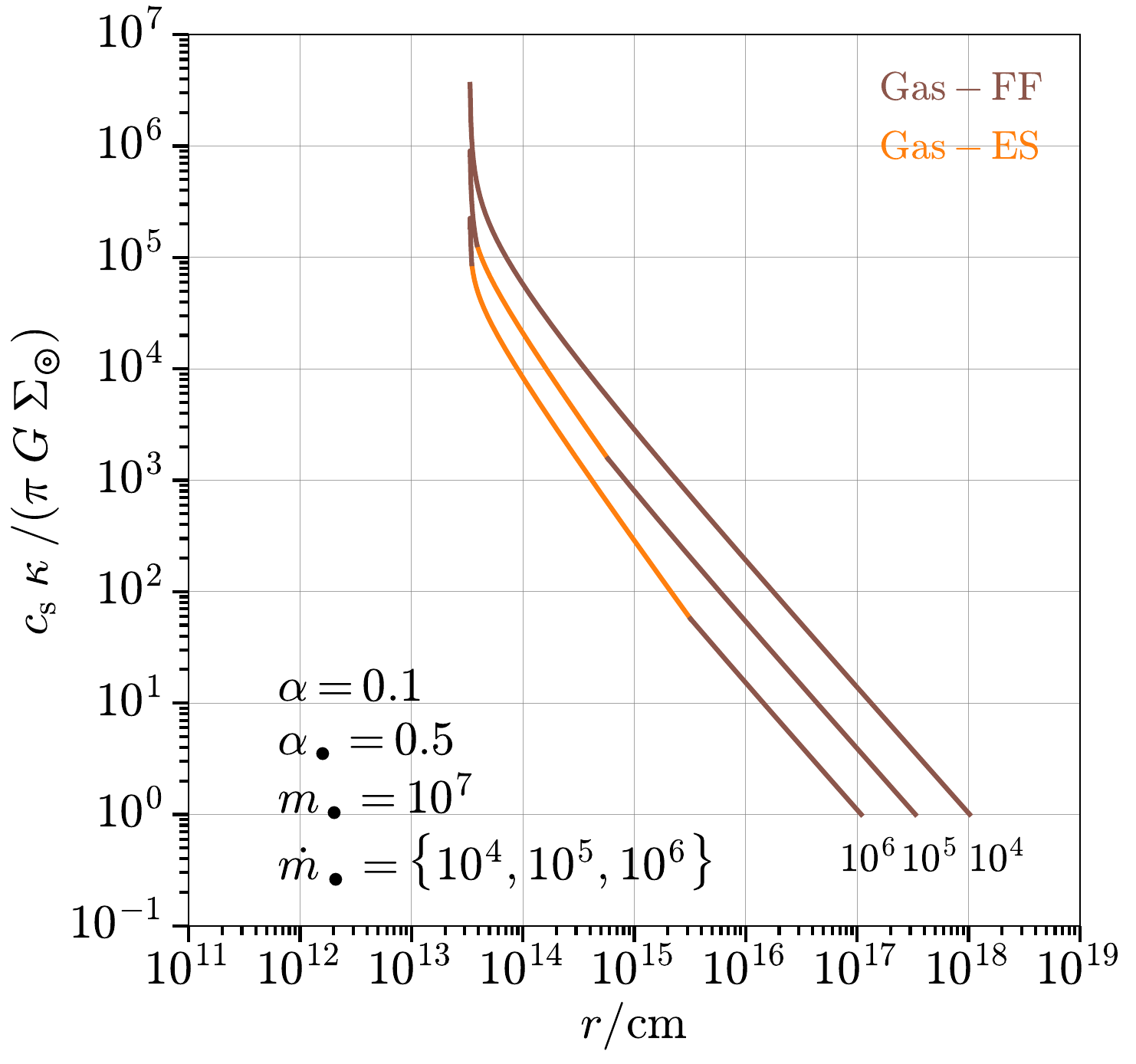}
\caption{Toomre Q stability profiles for different combinations of parameters as indicated by the text within each panel, and for two different accretion disc orientations (left--hand column: co--rotating, right--hand column: counter--rotating). In both columns, from top to bottom we explore different values of accretion disc viscosity parameter $\alpha$, black hole spin parameter $\alpha_\bullet$, dimensionless black hole mass $m_\bullet$, and dimensionless black hole accretion rate $\dot{m}_\bullet$. The text next to each line indicates which value corresponds to which line.}
\label{fig:stabilities_all}
\end{figure*}

In this final subsection, we explore how the accretion disc's self--gravity, and in turn its extent, is affected by the BH--AD properties. We do that by plotting the `relativistic' Toomre Q parameter of \Eq{Q} as a function of radius for different orientations and for different combinations of parameters. In contrast to all of the above plots, when evaluating and plotting the stability of the accretion disc we exclude the \textbf{intra--ISCO} regime from the calculations, as it is not relevant for this process.

In \Fig{stabilities_all}, we see that all the differences highlighted in the previous subsection -- concerning the dependencies of the accretion disc structure on $\alpha$, $\alpha_\bullet$, $m_\bullet$, and $\dot{m}_\bullet$ -- are present. However, by comparing each panel from the left--hand side column (co--rotating case) with its corresponding panel in the right--hand side column (counter--rotating case) we see that the self--gravitating radius is independent of the accretion disc's orientation. Furthermore, even though the extent of the accretion disc depends on different values of viscosity, black hole mass and accretion rate, it is practically independent of the spin of the black hole (second to top panels).

\section{Applications} \label{sec:Applications}

In this Section, we implement the model described in \Sec{Model} in an SPH galaxy formation simulation, and present applications relevant for the galaxy formation and black hole astrophysics communities; more in--depth analyses will be performed in future studies of the \textsc{Rabbits} series. 

We use the galaxy formation simulation of the \textsc{Rabbits} suite, last described in the \textsc{Rabbits--I} and \textsc{II} papers \citep[][respectively]{LIJ24a,LIJ24b}. Briefly, the simulations are performed with a version of the \textsc{Gadget-3} code \citep[see][for relevant information]{S05,SPZ21} which incorporates the \textsc{Mstar} integrator \citep{RPM20} and includes post--Newtonian terms in the black hole equations of motions to resolve the small--scale SMBH dynamics \citep{RPJ17,MRJ23}. Hydrodynamical calculations are performed with the \textsc{Sphgal} code \citep{HNW14}, and astrophysical processes -- including gas cooling, star formation, supernova Type Ia and Type II feedback and asymptotic giant branch star winds -- are captured following the models introduced in \cite{STW05,STW06,AWN13} and \cite{NON17}. 

\subsection{Isolated galaxy formation simulation} \label{sec:Applications:Isolated}

\subsubsection{Initial conditions} \label{sec:Applications:Isolated:ICs}

\begin{table}
	\centering
	\caption{Initial parameters of the isolated disc galaxy simulation used in this work. See also discussion in \Sec{Applications:Predictions:Radii} and \App{Testing} regarding testing our accretion disc model's sensitivity to different number of simulation particles (i.e. different resolutions) and black hole masses.}
	\label{tab:Isolated ICs}
	\begin{tabular}{lcc}
		\hline
		Halo properties & Values & Units \\
		\hline
		Virial mass & 55.20 $\times\ 10^{10}$ & \Msun \\
		Virial radius & 168.00 $\times\ 10^{3}$ & pc \\
		Halo concentration parameter & 15.00 & -- \\
		Spin parameter & 0.1 & -- \\
		\hline
		Galaxy properties & Values & Units \\
		\hline
		Dark matter mass & 49.68 $\times\ 10^{10}$ & \Msun \\
		Stellar disc mass & 3.31 $\times\ 10^{10}$ & \Msun \\
		Stellar bulge mass & 1.38 $\times\ 10^{10}$ & \Msun \\
		Gas disc mass & 0.82 $\times\ 10^{10}$ & \Msun \\
		Black hole mass & 1.00 $\times 10^{7}$ & \Msun \\
		Stellar disc scale length & 6.28 $\times 10^{3}$ & pc \\
		Gas disc scale length & 6.28 $\times 10^{3}$ & pc \\
		Stellar disc scale height & 1.26 $\times 10^{3}$ & pc \\
		Stellar bulge scale radius & 1.26 $\times 10^{3}$ & pc \\
		\hline
		Simulations properties & Values & Units \\
		\hline
		Stellar disc particles & 4.80 $\times\ 10^5$ & -- \\
		Dark matter particles & 4.00 $\times\ 10^5$ & -- \\
		Stellar bulge particles & 2.00 $\times\ 10^5$ & -- \\		
		Gas disc particles & 1.20 $\times\ 10^5$ & -- \\
		\hline
	\end{tabular}
\end{table}

To generate the initial conditions of a disc galaxy in isolation, we follow the technique described in \cite{SDH05}. Our setup is composed of a \cite{H90} dark matter halo within which we place two rotationally--supported discs (one stellar and one gaseous), a \cite{H90} stellar bulge, and a black hole. The properties of these components are shown in \Tab{Isolated ICs} and as detailed in \cite{LJR18}, these values (excluding the black hole) correspond to galaxy NGC 4038 -- one of the Antennae galaxies -- which we use as a testbed and simulate for 3 Gyr in isolation.

Given the component masses and particle numbers quoted in \Tab{Isolated ICs}, our baryonic (i.e. stellar or gas) components consist of particles each having an initial mass of 7 $\times\ 10^4$ \Msun, whilst each dark matter particle has a mass of 5 $\times\ 10^6 $ \Msun. Gravitational interactions are softened using a Plummer equivalent softening length of 281.70 pc for dark matter and 28.17 pc for baryonic and black hole particles. \footnote{We have additionally tested the scaling of the accretion disc model in simulations with 2 times and 0.5 times the particle masses and scale lengths, and the results are consistent with the behaviours reported in this section.}

\subsubsection{Initial black hole--accretion disc properties} \label{sec:Applications:Isolated:Initial setup}

We begin by initialising the dimensionless spin parameter $\alpha_\bullet$ with a value randomly drawn uniformly between 0 and 0.998,\footnote{\cite{T74} showed that isolated black holes which have been spun up only by a thin, radiative accretion disc cannot have spin values higher than 0.998, since they preferentially accrete negative angular momentum photons (i.e. counter--rotating). However, thick and/or magnetised accretion discs can lead to black holes having different spin equilibrium values \citep[e.g.][]{AJS78,GSM04,SBA11}.} and an initial direction which is also taken to be uniformly random in space. The next step is building an accretion disc for the given black hole spin and mass (set to $10^{7}$ \Msun\ in this simulation). 

Since in our simulations the formation phase of an accretion disc is not resolved \citep[as is in e.g.][]{HGS24,HSS24}, we opt for an analytic prescription to model this process (see Appendix \ref{app:Implementation:Evolve}). In our model, gas particles that (i) have, with respect to the black hole, specific angular momentum $j_\circledast$ less than that of a particle at a circular orbit with velocity $\sqrt{G\ M_{\bullet + \circledcirc}/ r_\circledast}$ and (ii) relative distance $r_\circledast$ smaller than the \cite{HL39} radius $r_{\mathrm{HL}} \equiv 2G\ M_{\bullet + \circledcirc} / (u_\circledast^2 + \ccircledastsqr)$, where $c_\circledast$ is the sound speed of the surrounding medium; will be captured by the BH--AD system and fuel the accretion disc with mass.

The next step is assigning to the accretion disc an angular momentum vector with magnitude equal to the angular momentum of the captured mass evaluated at the outskirts of the accretion disc. For its direction, we incorporate two different methods. Following an approach similar to the chaotic accretion scheme \citep[e.g.][]{KP06,KP07,NKP12a,GRO13}, we assume that the initial angular momentum of the accretion disc has a random direction in space. In addition, we also incorporate an alternative, more coherent accretion scheme \citep[e.g.][]{DVS14,BS19} in which we assume that -- since we identify gas particles with angular momenta lower than that corresponding to a circular orbit -- the gas particles will form an accretion disc perpendicular to their orbital angular momentum vector (i.e. the accretion disc aligns with the gas particle's orbital plane). This allows us to more directly study the interplay between the SPH--resolved scales and our BH--AD analytic model.

In order to finalise the initial accretion disc, a mass accretion rate is required (see Sections \ref{sec:Model} and \ref{sec:Results}). In this first formation episode (or any future episode after a complete depletion of the accretion disc), we use as an estimate of the accretion rate of mass onto the BH--AD system the captured mass (discussed above) divided by the Bondi timescale given by $G\ M_\bullet\ c_\bullet^{-3}/ (1 + \mathcal{M}^2)^{-3/2}$, where $\mathcal{M}$ is the Mach number. Note that this accretion rate is used to build an accretion disc instead of immediately being used to increase the black hole's mass, as is traditionally done in galaxy formation simulations. In all simulations presented in this work, the viscosity parameter of the accretion disc is initialised as a constant and kept fixed to 0.1.

\subsubsection{Results} \label{sec:Applications:Isolated:Results}

\begin{figure*}
\includegraphics[width=0.48\textwidth]{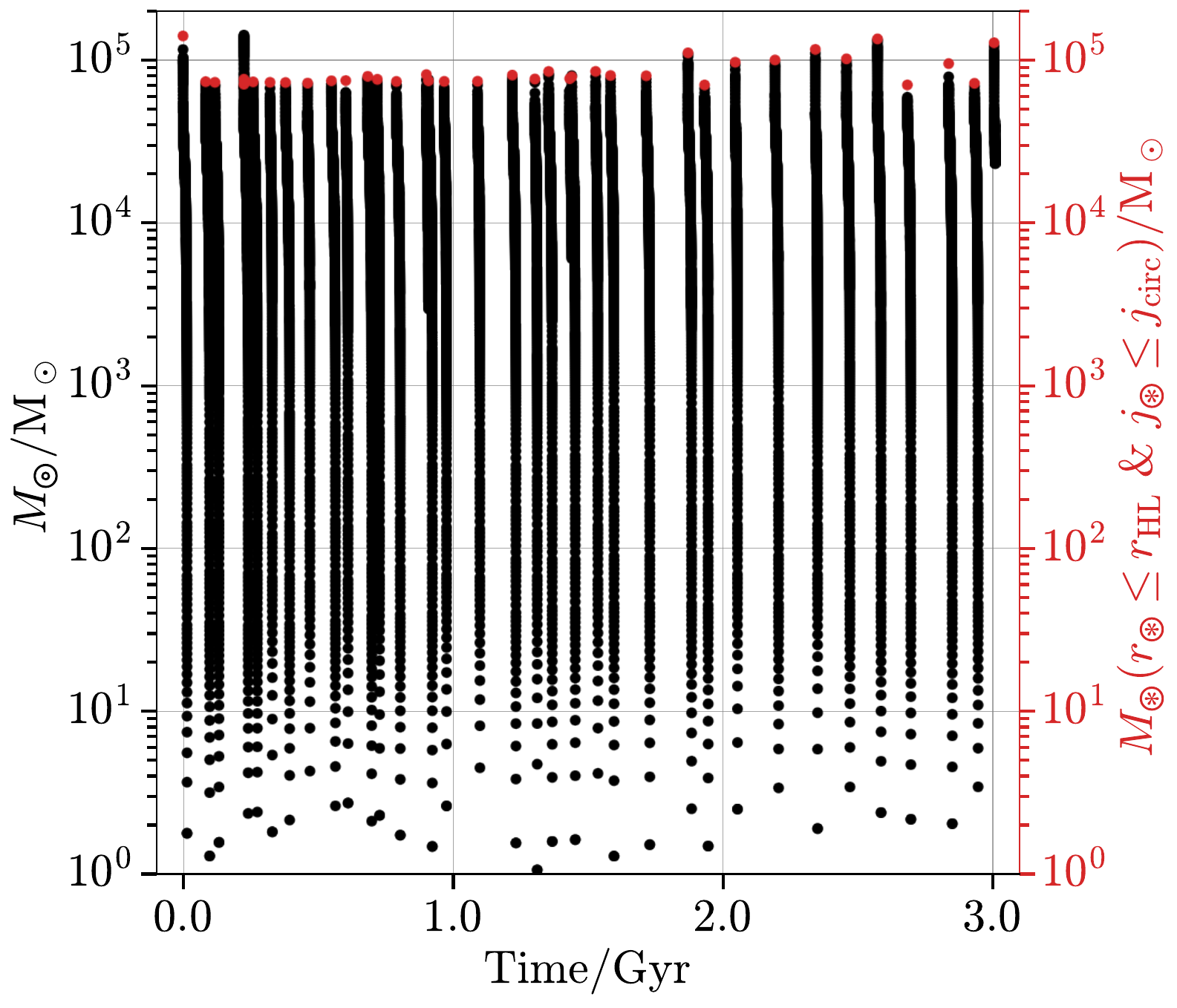} \includegraphics[width=0.48\textwidth]{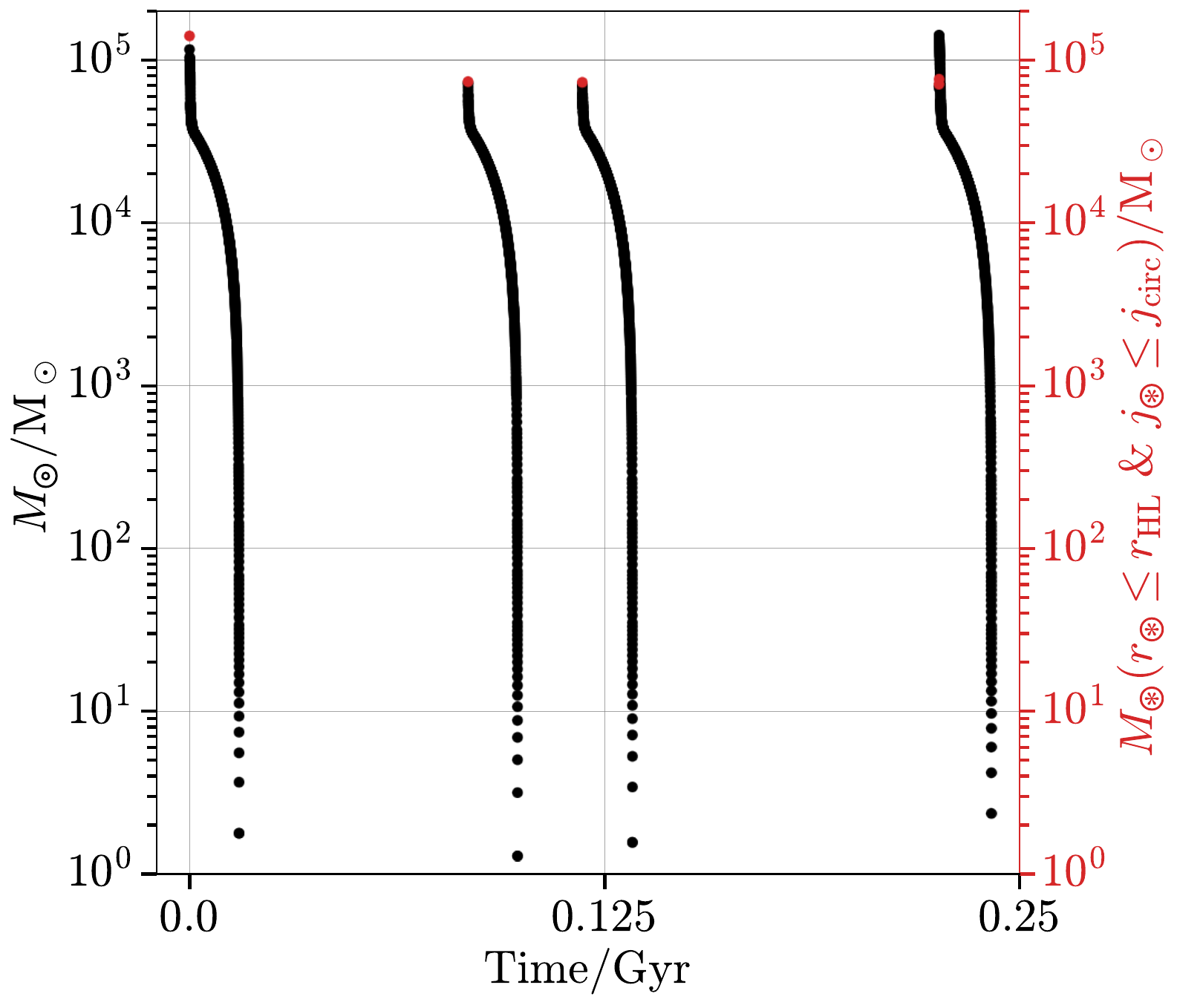} \\
\caption{Evolution of the accretion disc mass (black points) over the duration of the simulation (left--hand panel) and a zoom in of the first 250 Myr of the first four accretion episodes (right--hand panel). Red points represent the mass of gas particle(s) captured by the BH--AD system that increase the accretion disc's mass.}
\label{fig:ad_masses}
\end{figure*}

In \Fig{ad_masses}, we show the evolution of the accretion disc mass (black points) over the duration of the simulation (left--hand panel) and a zoom in of the first 250 Myr of the first four accretion episodes (right--hand panel). Red points represent the mass of gas particle(s) captured by the BH--AD system that increase the accretion disc's mass.

\begin{figure*}
\includegraphics[width=0.48\textwidth]{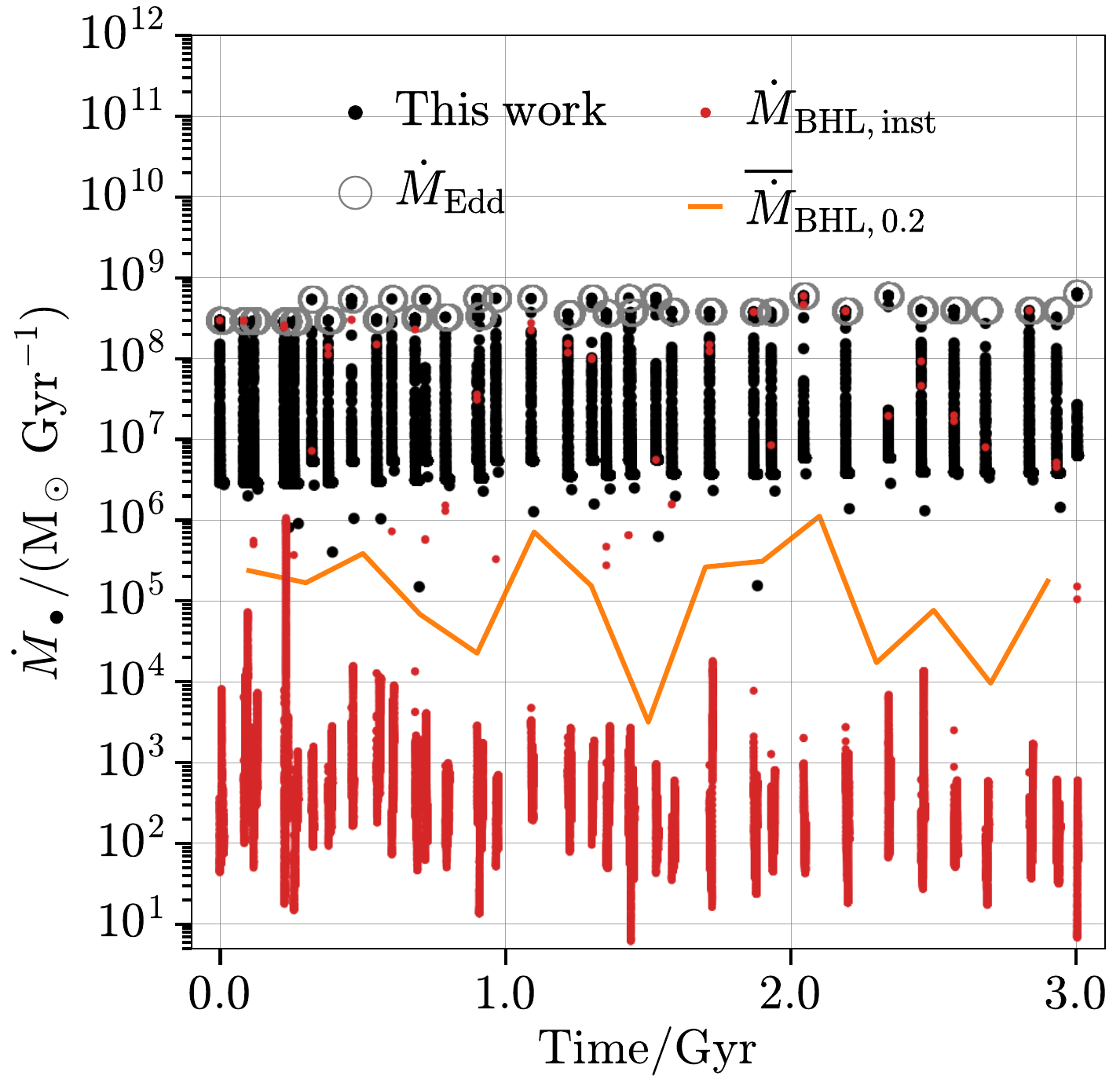} \includegraphics[width=0.48\textwidth]{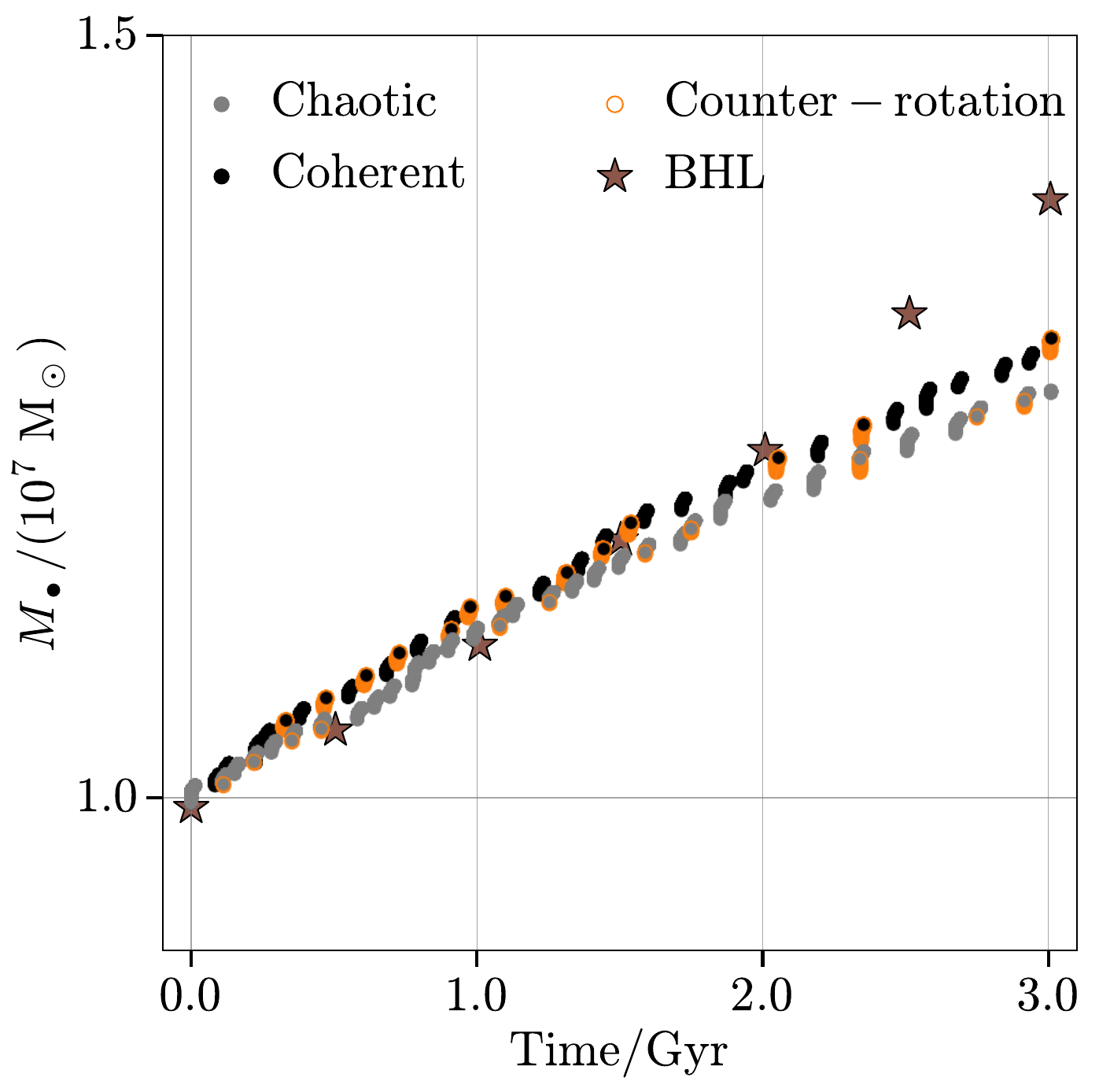}
\caption{Evolution of the black hole accretion rate (left--hand panel) and mass (right--hand panel) over the duration of the simulation. In the left--hand plot, we show in grey circles the Eddington accretion rate, in red points the instantaneous Bondi--Hoyle--Lyttleton accretion rate, and the orange line shows the 0.2 Gyr time--averaged Bondi--Hoyle--Lyttleton accretion rates (see text for more details). In the right--hand plot, we show in black (grey) points a simulation with coherent (chaotic) accretion model, orange circles around either back or grey points indicate counter--rotating accretion episodes, and brown stars show the evolution of the black hole mass every 0.5 Gyr of a simulation identical to the one described in \Sec{Applications:Isolated:Initial setup}, that was instead run with a Bondi--Hoyle--Lyttleton accretion model (see text for more details).}
\label{fig:bh_masses}
\end{figure*}

From \Fig{ad_masses}, it becomes evident how our model follows the processes of (i) capturing gas particles; (ii) using them to build/feed an accretion disc; and (iii) gradually migrating mass to the black hole through the accretion disc (see also \Fig{bh_masses}). Regarding the capture of gas particles, three distinct features -- relevant to the numerical implementation of the model -- arise that warrant commenting on.
\begin{enumerate}[wide=0pt,labelindent=10pt,labelwidth=10pt]
\item The first feature in \Fig{ad_masses} -- more clearly visible in the zoomed--in right--hand plot -- is that in the first accretion episode, two gas particles meet the capturing criteria discussed in \Sec{Applications:Isolated:Initial setup} and \App{Implementation:Evolve}, hence the accretion disc starts with a mass of $\sim 1.4 \times\ 10^5$ \Msun. This capture of two gas particles at the same time--step happens two more times in this simulation, which makes it a rare but possible event ($\sim$8$\%$ of all capturing events).
\item The second feature appears during the fourth accretion episode, roughly around 0.25 Gyr where, in contrast to the rest of the accretion episodes, the accretion disc captures another gas particle before it has been completely depleted of mass, resulting in a sudden increase in its mass. 
\item Finally, the third feature is the clear evolution of the captured gas particles' masses as the simulation progresses. This increase in mass is a consequence of feedback from the surrounding stellar particles which apart from energy also deposit mass/metals to their neighbouring gas particles, thus supplementing their mass over time. Note that to maintain a relatively uniform mass resolution, a gas particle splitting routine is in place in the simulation where gas particles that acquire two times their initial mass are split into two individual gas particles.
\end{enumerate}
The above three points translate to accretion discs being roughly twice as massive when either two particles get captured at the same or in adjacent time--steps; or when the captured particles' mass has increased due to stellar feedback. In practise, both how effectively and how much mass is captured by the BH--AD systems reflect physical assumptions of the model and numerical parameters of the simulation.

Finally, in \Fig{ad_masses}, we also demonstrate that the apparently smooth draining of accretion disc mass seen in the left--hand panel, in reality has a non--linear evolution with time in the right-hand plot. The nature of this evolution has both physical and numerical aspects (e.g. time--step limiters), which we will disentangle in a future work. Additionally, in a subsequent paper, we will further investigate the physical characteristics of individual accretion episodes. Parameters such as their depletion timescale (i.e. how long it takes for complete consumption) and the time intervals between distinct capture episodes (i.e. how frequently the accretion disc gets replenished) leave imprints on the radiated luminosities, black hole variability, and properties such as their spin up/down rate, which have direct observational implications.

In \Fig{bh_masses}, we show the evolution of the black hole accretion rate (left--hand panel) and mass (right--hand panel) over the duration of the simulation. In the left--hand plot, we show in grey circles the Eddington accretion rate, in red points the instantaneous Bondi--Hoyle--Lyttleton (BHL) accretion rate, and the orange line shows the 0.2 Gyr time--averaged BHL accretion rate. In the right--hand plot, we show in black (grey) points a simulation with coherent (chaotic) accretion model (see \Sec{Applications:Isolated:Initial setup}), orange circles around either back or grey points indicate counter--rotating accretion episodes. Finally, brown stars show the evolution of the black hole mass every 0.5 Gyr of a simulation identical to the one described in \Sec{Applications:Isolated:Initial setup}, that was instead run with a BHL accretion model boosted by a factor of 25 \citep[e.g. equation 30 in][]{SDH05}, with a radiative efficiency of 0.1 and a coupling efficiency of 0.05. 

It is worth noting that in the left--hand panel, the accretion rates calculated by the simulation at a given time--step are multiplied by the time--step and the corresponding efficiency when converted into mass increments (i.e. $\delta M_\bullet = (1 - \epsilonr)\dot{M}_\bullet \times dt$). Hence, the plotted values do not directly correspond to the growth rate of the black hole seen in the right--hand panel.

An advantage of our model compared to the standard BHL prescription can be seen from the red data points in the left--hand panel of \Fig{bh_masses}. It is evident that the BHL accretion rates exhibit a significantly larger scatter, arising from the intrinsically stochastic evaluation of local gas properties in particle--based simulations. This necessitates computing time--averaged values of the instantaneous accretion rates \citep[e.g.][]{MBH17,WKI25} in simulations that use BHL. {By contrast, fluctuations in our accretion disc model reflect physical variability in the gas supply and disc evolution, rather than numerical noise. Hence our work} naturally alleviates stochasticity as at every time--step the accretion rate calculated based on the properties of the accretion disc provides a meaningful value of the accretion rate.\\

In the right--hand panel of \Fig{bh_masses}, we see that for almost 1.5 Gyr all three simulations (i.e. the coherent accretion, the chaotic accretion, and the BHL simulation) result in very similar black hole masses. The three black hole growth tracks start to diverge around 2 Gyr and continue to separate until the end of the simulation, where the BHL simulation results in the most massive black hole (even though the difference is relatively small). A direct comparison between the BHL method and the model developed in this work is beyond the scope of this work as numerical parameters (e.g. different time--stepping, BHL boosting factor, spin--dependent feedback efficiency) need to be marginalised first before the two methods can be fairly compared in a controlled environment. What is more important to address here is why the coherent accretion model (where the accretion disc's angular momentum direction is inherited from the captured gas particles' orbital angular momentum) results in relatively higher black hole masses compared to the chaotic (where the accretion disc's angular momentum direction is random at every capture episode).

\begin{figure*}
\includegraphics[width=0.48\textwidth]{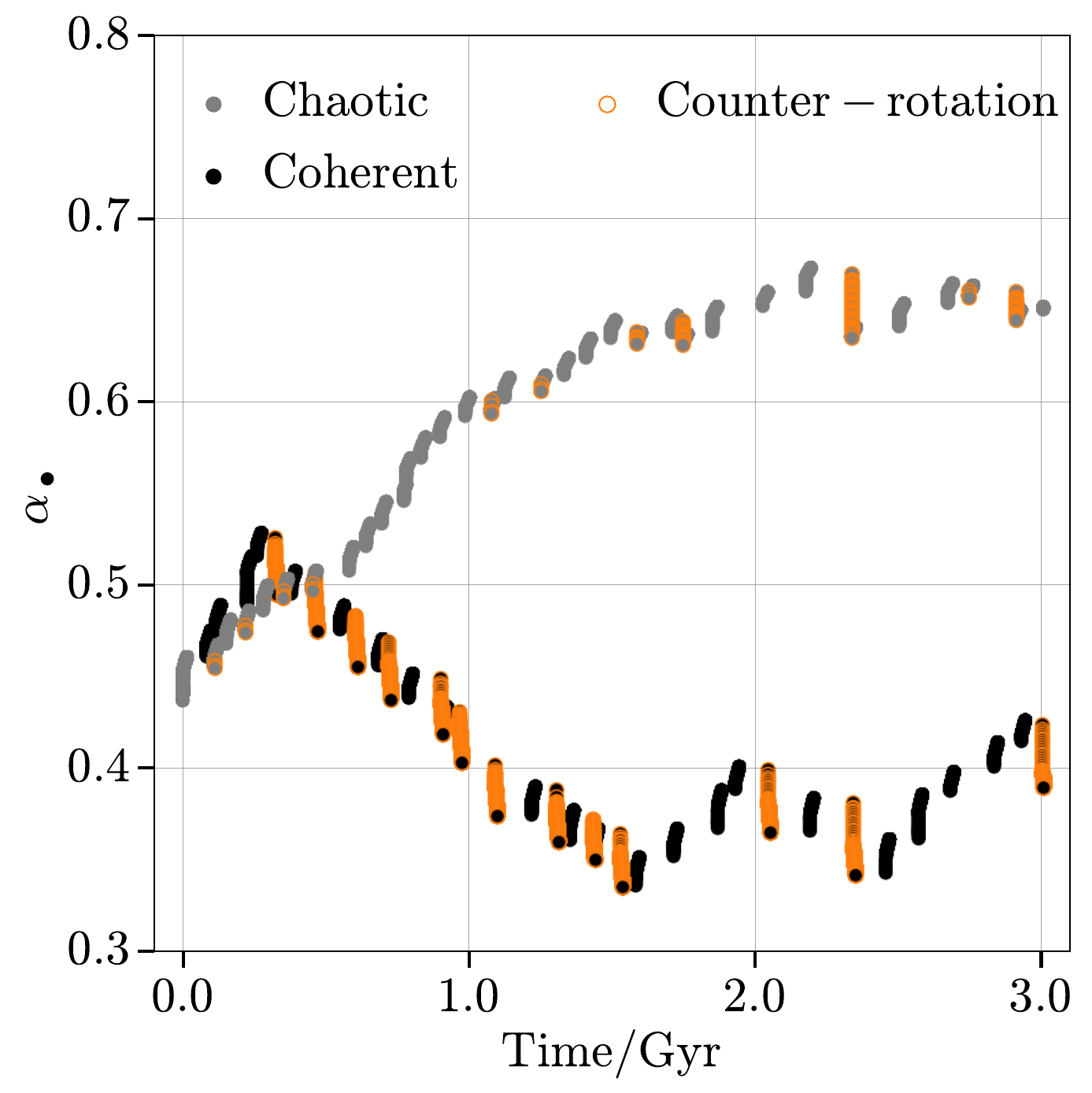} \includegraphics[width=0.48\textwidth]{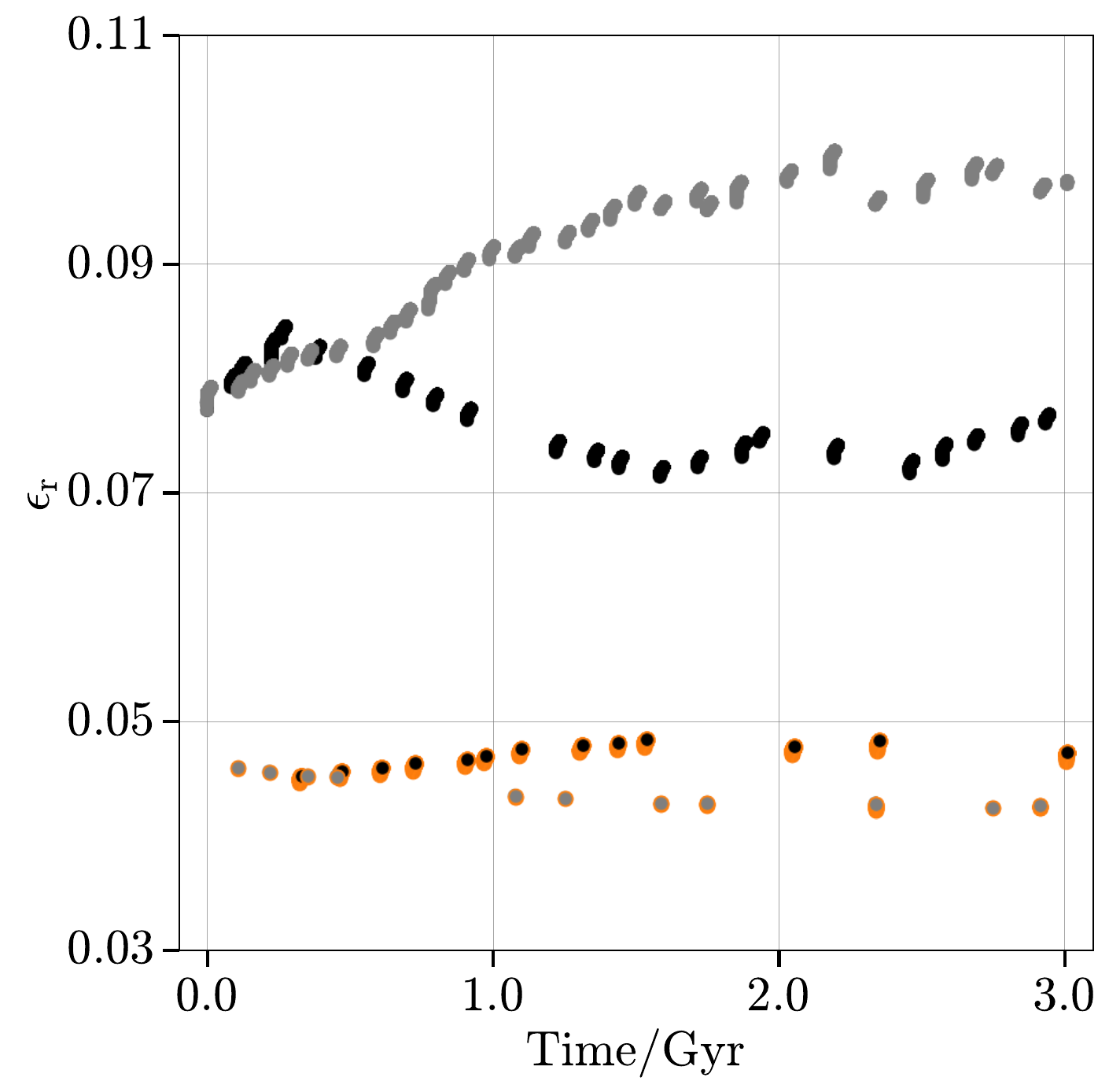}
\caption{Evolution of the black hole spin parameter (left--hand panel) and feedback efficiency (right--hand panel). In both panels, we show in black (grey) points a simulation with coherent (chaotic) accretion model and orange circles around either back or grey points indicate counter--rotating accretion episodes.}
\label{fig:bh_spins}
\end{figure*}

To understand why the coherent simulation is more effective at growing the black hole compared to the chaotic simulation, in \Fig{bh_spins} we show the evolution of the spin parameter (left--hand panel) and feedback efficiency (right--hand panel). In both panels, we show in black (grey) points a simulation with coherent (chaotic) accretion model and orange circles around either back or grey points indicate counter--rotating accretion episodes.

In the left--hand panel, it is evident that the chaotic run (grey points) results in a higher spinning black hole than the coherent run (black points). The reason behind this behaviour is that the chaotic run has less counter--rotating accretion episodes (orange circles) compared to the coherent run, which are the natural culprit of black hole spin--down. Even though one would (statistically speaking) expect that the random orientation in the chaotic run will result in half the accretion episodes being through a co--rotating and half through a counter--rotating accretion disc \citep[e.g.][]{WC95,MS96,VMQ05}, this is not the case here. Furthermore, even if slightly more than half of the accretion episodes are from an aligned accretion disc, one would still expect the spin parameter to decrease over time, since a counter--aligned accretion disc is more efficient at spinning down a black hole than an aligned at spinning it up \citep{KP06}. However, what we observe in \Fig{bh_spins} is an abundance of co--rotating accretion episodes in the chaotic accretion simulation that spin up the black hole. A question that naturally arises is: what is the reason behind this phenomenon in the chaotic simulation? 

Recall that following \cite{KLO05} we use \Eq{costheta} to evaluate if the accretion disc is aligned or not with respect to the black hole spin. This inequality does not simply require $\sbullethat \cdot \jcircledcirchat$ to be positive (negative) for co--(counter--) rotation, but it takes into account the total angular momentum of the accretion disc. Since in our model we calculate the angular momentum of the accretion disc via \Eq{Jcircledcirc}, we can explicitly quantify this property. 

Given that the angular momentum increases monotonically as a function of radius, the extent of the accretion disc plays a crucial role in determining the alignment via integrating \Eq{Jcircledcirc} for \Eq{costheta}. In this work, we use the self--gravitating radius based on \Eq{Q} as an estimate of the extent of the accretion disc, which even though is not infinite like e.g in \cite{SF96} \cite[see discussion in][]{KLO05} it results in angular momenta high enough to make the term $- J_\circledcirc /2S_\bullet$ in the right--hand side of \Eq{costheta} large (i.e. $\theta$ just above 90$\degree$ is not enough for counter--rotation). In other words, large accretion discs are in `disadvantage' when their alignment is classified via integrating \Eq{Jcircledcirc}, which translates to more co--rotating than counter--rotating accretion discs in a chaotic accretion scheme. This direct evaluation of the \cite{KLO05} alignment criterion is an additional advantage of our approach to model the structure of the accretion disc, which leads to behaviours that diverge from previous findings \citep[e.g.][]{VMQ05}. To summarise, when the structure of the accretion discs is taken into account, assigning random orientation to the accretion disc does not necessarily result in black hole spin--down since for fixed black hole spin: the larger the angular momentum of the accretion disc is the harder it is for the inequality in \Eq{costheta} to be satisfied.

An immediate implication of the above discussed spin up (down) due to co-- (counter--) rotation, is demonstrated in the right--hand panel of \Fig{bh_spins}. Since in our model the accretion--related radiative efficiency is spin--dependent [see \Eq{Lobs}], the higher spinning black hole of the chaotic simulation results in higher feedback efficiencies (i.e. stronger heat up of the surrounding ISM for fixed accretion rate) compared to the coherent run. This has the direct consequence of making the BH--AD system in the chaotic run on average less capable of attracting and acquiring gas particles (see \App{Implementation:Evolve}) than the coherent run, hence the former black hole ends up being less massive than the latter, as discussed in \Fig{bh_masses}. In a future work (Liao \& Irodotou in prep.), we additionally investigate how faster spinning black holes affect the evolution of SMBH binaries due to the impact their enhanced feedback has on the surrounding star formation compared to slower spinning black holes. As we showed in \cite{LIJ24a,LIJ24b}, the evolution and merger timescale of a SMBH binary is affected by the surrounding stellar density since three--body interaction with stellar particles extract energy from the binary and accelerate its coalescence \citep{BBR80,MV92,Q96,SHM06}.

Finally, regarding the right--hand panel of \Fig{bh_spins}, it is worth noting that the total radiative efficiency in our model incorporates an extra, accretion--related term [see \Eq{Lobs}]. In  the presented simulations however, this term is of the order of $10^{-4}-10^{-3}$ so it does not significantly contribute to the total radiative efficiency which is of the order of $10^{-2}-10^{-1}$.

\begin{figure*}
\includegraphics[width=0.48\textwidth]{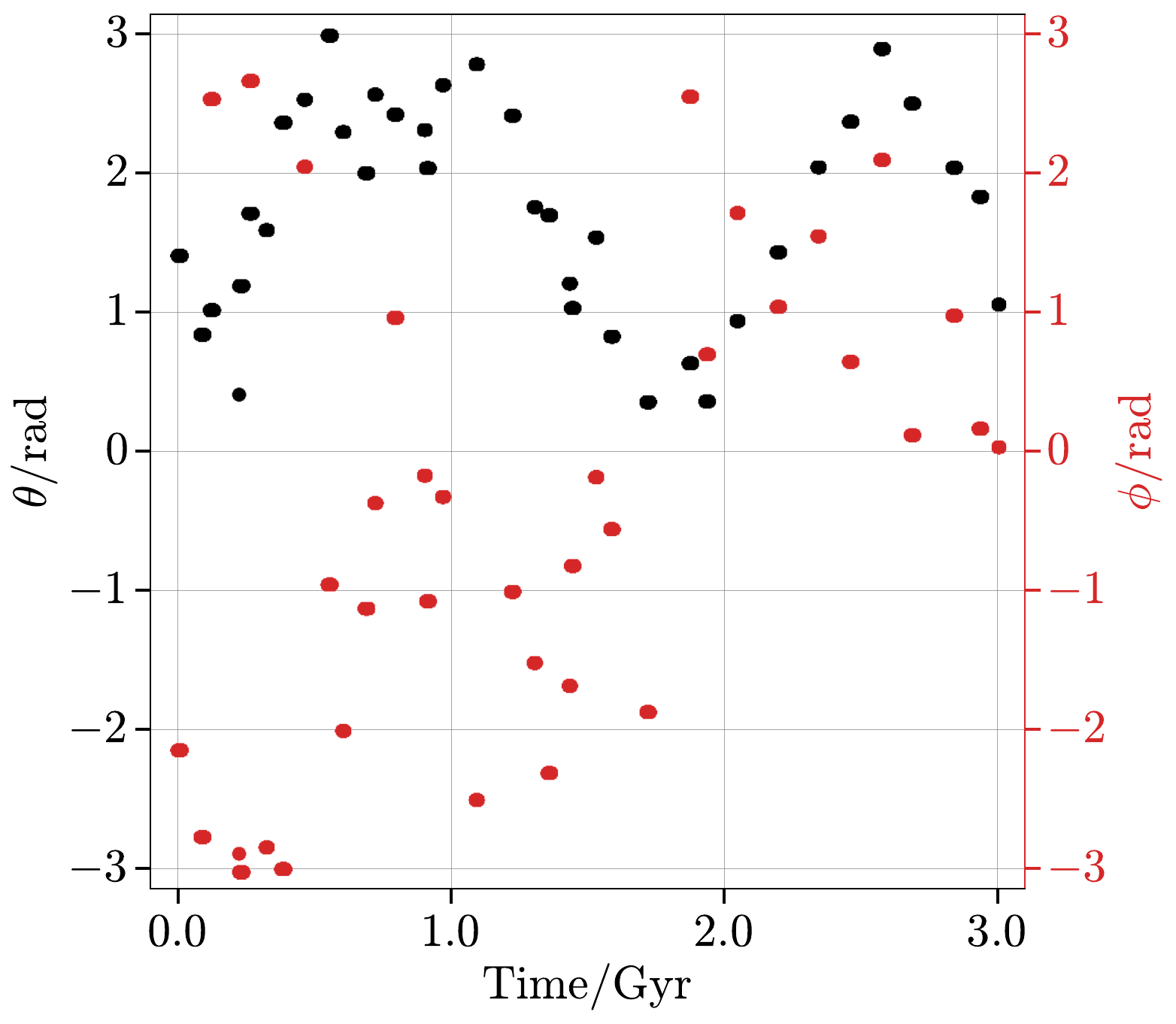} \includegraphics[width=0.48\textwidth]{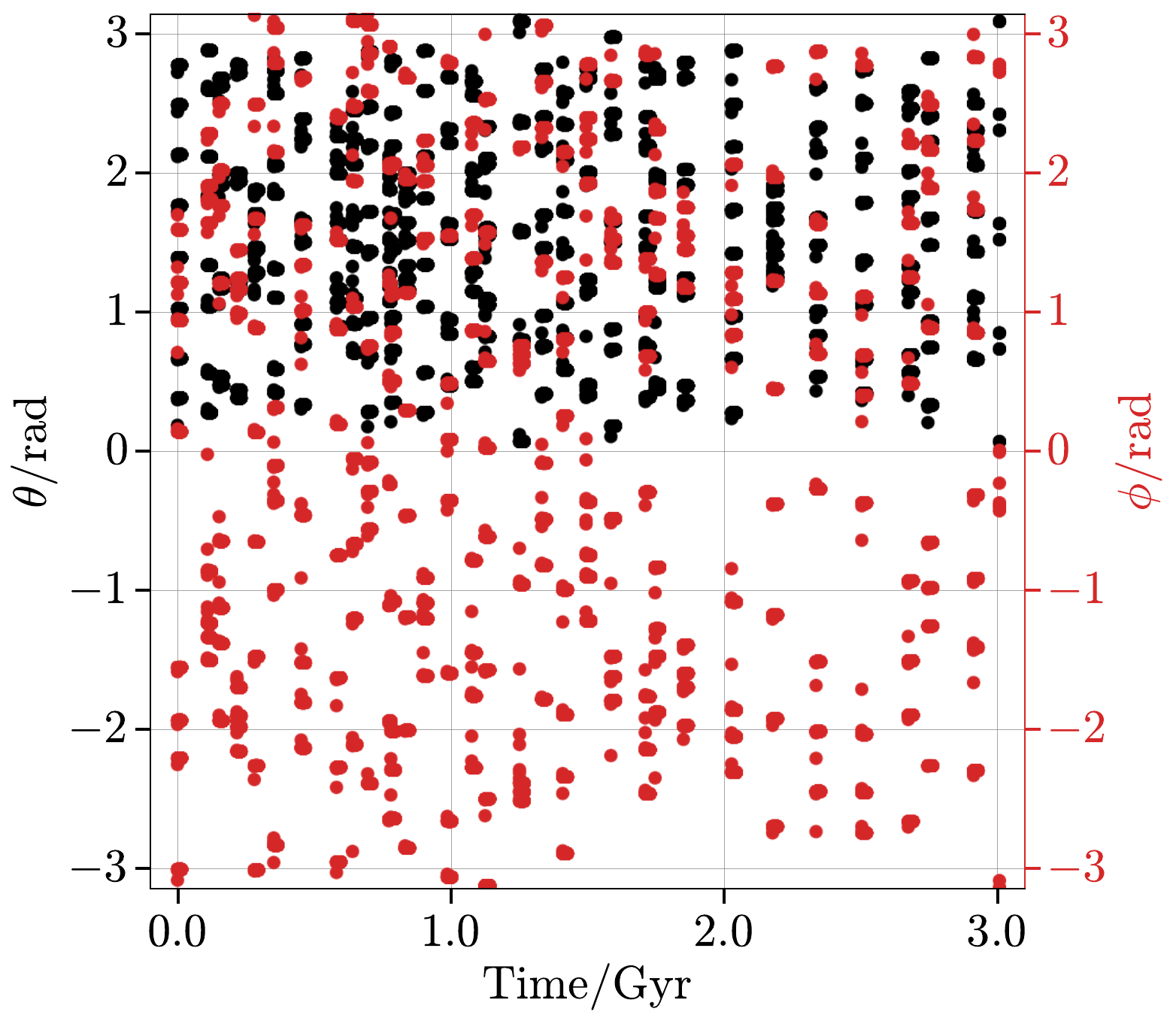}
\caption{Evolution of the black hole's spin vector direction in a coherent (left--hand panel) and chaotic (right--hand panel) accretion simulation.}
\label{fig:vectors_all}
\end{figure*}

As a final demonstration of the behaviour of our model, in \Fig{vectors_all}, we plot the evolution of the black hole's spin vector direction for the coherent (left--hand plot) and chaotic (right--hand plot) simulations.

Even though visually the coherent simulation appears to be taking fewer time--steps than the chaotic, the two simulations have a similar number of data points ($\sim$35,000 versus $\sim$44,000). The visual discrepancy arises because in the coherent case on the left--hand panel the black hole spin evolves more smoothly on short time--scales, instead of being random. Consequently, the changes in the direction of the black hole spin are more gradual than the abrupt shifts observed in the right--hand panel. Over longer time--scales, the coherent accretion leads to oscillations in the polar angle $\theta$ with a period of $\sim$2 Gyr, whilst the azimuthal angle $\phi$ systematically increases.

\subsection{Novel predictions} \label{sec:Applications:Predictions}

Our approach of modelling the structure of accretion discs around spinning black holes allows us to make predictions from galaxy formation simulations that -- to the best of our knowledge -- were not possible before.

\subsubsection{Accretion disc half--light radii} \label{sec:Applications:Predictions:Radii}

\begin{figure*}
\includegraphics[width=\textwidth]{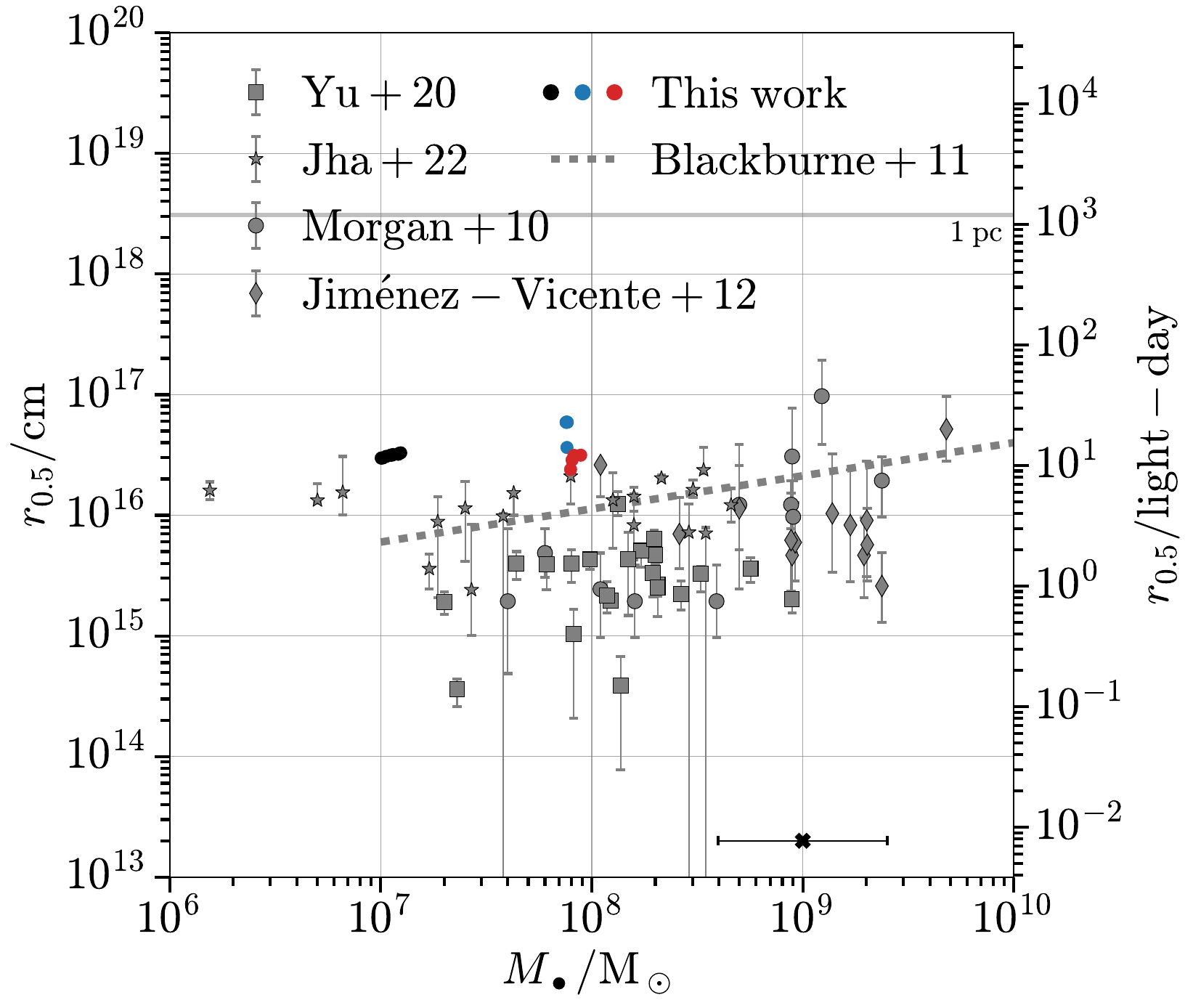}
\caption{Accretion disc half--light radius as a function of black hole mass. Black data points show results from the simulation described in \Sec{Applications:Isolated:ICs}. Blue and red data points correspond to the disc galaxy setup of the \textsc{RABBITS} series \protect\citep{LIJ24a,LIJ24b} (see text for more details), where for the former/latter (blue/red points) the accretion disc has been forced to co--/counter--rotate with respect to the black hole spin. We compare these three simulations of this work with the observational datasets of \protect\cite{MKM10,BPR11,JMM12,YMD20} and \protect\cite{JJC22} (see text for more details). The typical uncertainty of the observed black hole masses is about 0.4 dex, shown as the black cross symbol with errorbars in the bottom right corner.}
\label{fig:ad_size}
\end{figure*}

In \Fig{ad_size}, we show the accretion disc half--light radius as a function of black hole mass for the isolated disc galaxy of \Sec{Applications:Isolated} (black points) and for two additional disc galaxy simulations of the \textsc{RABBITS} series \protect\citep{LIJ24a,LIJ24b} with $M_\bullet = 7.5 \times 10^7 \Msun$ and $M_\mathrm{DM} = 2.5 \times 10^{12} \Msun$, $M_\star = 1.2 \times 10^{11} \Msun$, and $\Mgas = 2.5\times 10^{10} \Msun$ (which correspond to particle resolution of $1.6 \times 10^6 \Msun$, $10^5 \Msun$, and $10^5 \Msun$, respectively). The blue/red points correspond to a simulation where the accretion disc is forced to be co--/counter--rotating with respect to the black hole spin for the duration of the simulation (1 Gyr). These scaled up setups with fixed accretion disc orientation allow us to test two extreme cases of accretion and radiative feedback; and are part of our resolution testing we discuss in \App{Testing}.

We compare the aforementioned three simulations with the following observational datasets
\begin{enumerate}[wide=0pt,labelindent=10pt,labelwidth=10pt]
\item \textbf{Morgan+10:} a sample of eleven gravitationally lensed quasars \citep{KMM06} that were modelled using the \textsc{Gravlens} package of \cite{K01}. Black hole masses were estimated with $\sim0.3$ dex uncertainty following \cite{YD91,MJ02,AAA06,PIR06}. As noted by \cite{MKM10}, the microlensing size can be converted to half--light radii by multiplying with a factor of 2.44 \citep{MSW05}.
\item \textbf{Blackburne+11:} a sample of twelve quadruply lensed quasars from the literature with flux ratios in eight bands covering from 0.36 to 2.2 $\upmu$m. Accretion disc sizes were derived by combining optical/IR and X--ray flux ratios and using a Bayesian method to constrain the half-light radii of each quasar in each filter. Black hole masses are calculated following the virial method of \cite{PIR06} or estimated using the bolometric luminosity \citep{PBR07}. We extract from their Fig. 10 the best fit line for half--light radii at a rest wavelength of 2500\textup{~\AA}, similar to \cite{MKM10}.
\item \textbf{Jim{\'e}nez-Vicente+12:} a sample of nineteen lensed quasars from the \cite{MMF09} sample. Half--light sizes were calculated at 1736\textup{~\AA} from microlensing magnifications of image pairs, whereas black hole masses were extracted from \cite{MK11}.
\item \textbf{Yu+20}: a sample of twenty-two quasars from the Dark Energy Survey \citep[DES][]{AAA18}. Accretion disc sizes at rest-frame 25000\textup{~\AA} were calculated using the \textsc{Javelin} thin disk model as in \cite{MMZ18}, and black hole masses with 0.4 dex uncertainty were inferred by \cite{SRS11} using the single--epoch method.
\item \textbf{Jha+22:} a sample of nineteen reverberation mapped AGN from the Zwicky Transient Facility survey \citep{GKB19}. Accretion disc sizes were derived via fitting multi--band light curves with the \textsc{Javelin} thin disk model \citep{MMZ18}. Black hole masses were obtained from the AGN Black Hole Mass database \citep{BK15} and \cite{GTS17}.
\end{enumerate}

Observationally, two techniques can be used to directly measure accretion disc sizes in AGN. In micro--lensing observations \citep[e.g.][]{K04a}, accretion disk sizes are estimated by analysing the flux variability in images of gravitationally lensed quasars. In continuum reverberation mapping observations, the accretion disc sizes are inferred by measuring the inter--band time delays of continuum light curves, since the delays correspond to the time light takes to travel between different wavelength emitting regions. In our simulations, we perform a much simpler analysis. 

For given black hole mass, spin and accretion rate, we start by building the structure of the accretion disc as discussed in \Sec{Model}. Then, we calculate the total (bolometric) luminosity of the accretion disc by integrating over the accretion disc the relativistic flux profile of \cite{CO17}
\begin{flalign} \label{eq:flux}
F = \left( 5.50\ \mathrm{erg\ cm^{-2}\ s^{-1}} \right) \left( m_\bullet\ \dmbullet^{-2}\ x^{-7} \right) \times \mathcal{C}^{-1}\ \mathcal{P} \;,
\end{flalign}
which is the same for all three regimes (\textbf{Gas--ES}, \textbf{Rad--ES}, and \textbf{Gas--FF}). Once the total luminosity has been calculated, we can infer the half--light radius of the accretion disc as the radius within which half of the total luminosity was emitted.

By following this method, we expect our half--light radii to be systematically above the observed as we integrate the flux profile over the whole extent of the accretion disc, and not restrict to a particular area in which specific wavelengths are emitted \citep[see for example][for the effects of multiwavelength observations on microlensed accretion disk sizes]{BPR11}. As can be seen from our data in \Fig{ad_size}, all simulated accretion discs are a factor of a few higher than the observed. In addition, our counter--rotating merger setup has on average lower accretion rates compared to the co--rotating setup, which is the reason why the former accretion discs (red data points) are more compact than the latter (blue data points).

It is worth noting that \cite{MKM10,BPR11,JMM12,JJC22} have, respectively, found a factor of 4, 10, 5, and 3.9 on average larger observed half--light radii compared to the predictions of the standard accretion disc theory of \cite{SS73}. Contrary, \cite{HTG19} analysed ninety-six galaxies from the Sloan Digital Sky Survey Reverberation Mapping (SDSS-RM) project \citep{SBD15} and found their accretion disc sizes to be consistent with this theory. More recently, \cite{GLZ22} found the median ratio of observed to theoretical accretion disk sizes to be 1.24, further complicating the picture.

In addition, both numerical resolution and the absence of a fully cosmological environment may influence the detailed gas supply to the central black hole, and hence the inferred accretion disc properties. Higher-resolution simulations can better resolve the thermodynamic structure of the nuclear gas, while a cosmological context would naturally include sustained inflows, mergers, and environmental effects that may be important for reproducing the diversity of observed AGN. Addressing these aspects will be the subject of future work.

From the above discussion, it becomes apparent that a more robust comparison is essential but beyond the scope of this work. In a future study, we will investigate more systematically (i.e. larger simulation sample) and in a more similar approach (i.e. include wavelength and inclination corrections) the discrepancies between observed and simulated accretion disc sizes. This long--standing discrepancy between theory and observation \citep{KB99} remains until today an unresolved problem as was recently discussed by \cite{DT20}; and our galaxy formation simulations can contribute towards better understanding and testing current and future accretion disc theories.

\subsubsection{Spectral energy distribution} \label{sec:Applications:Predictions:SED}

\begin{figure*}
\includegraphics[width=0.48\textwidth]{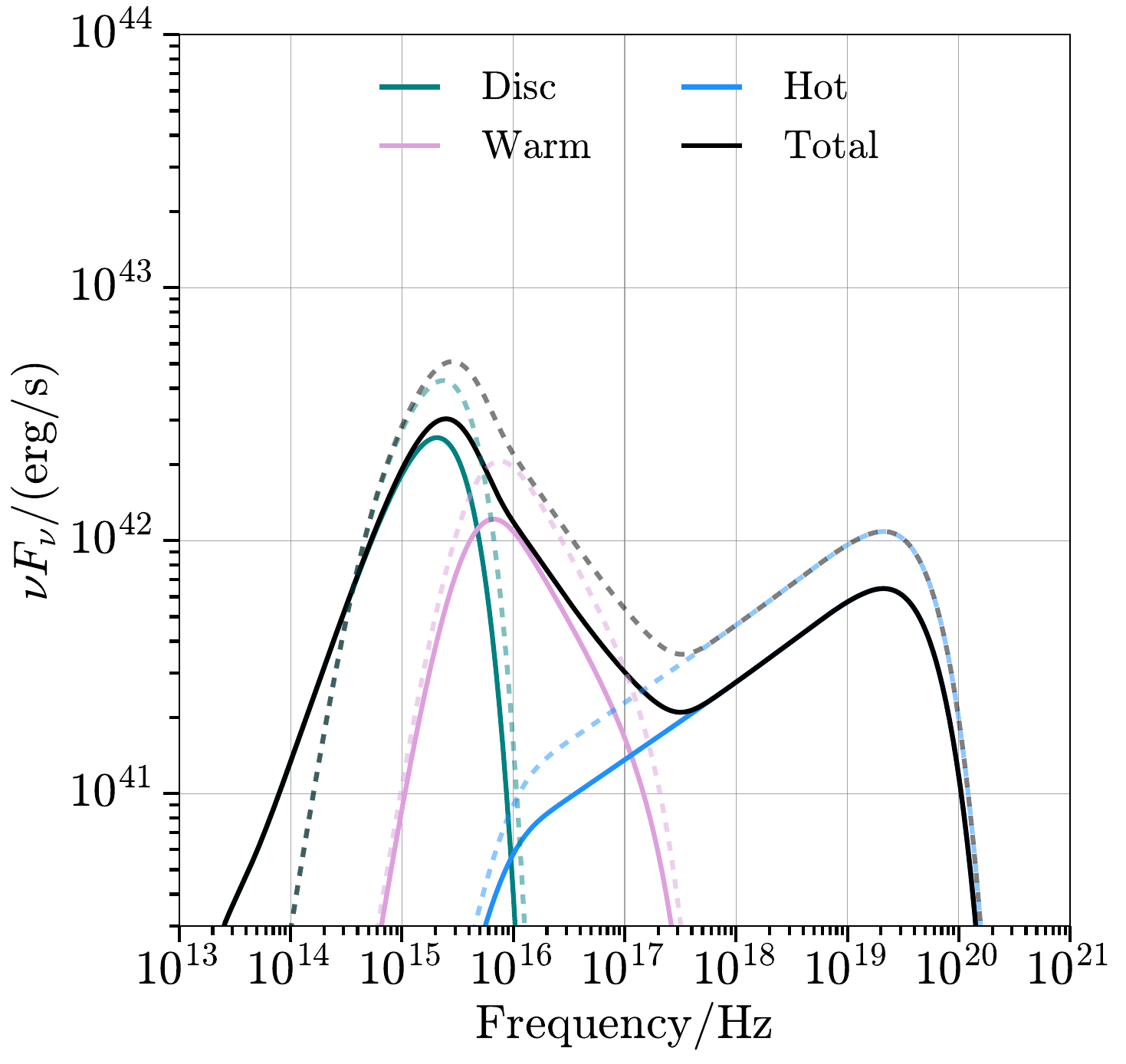} \includegraphics[width=0.48\textwidth]{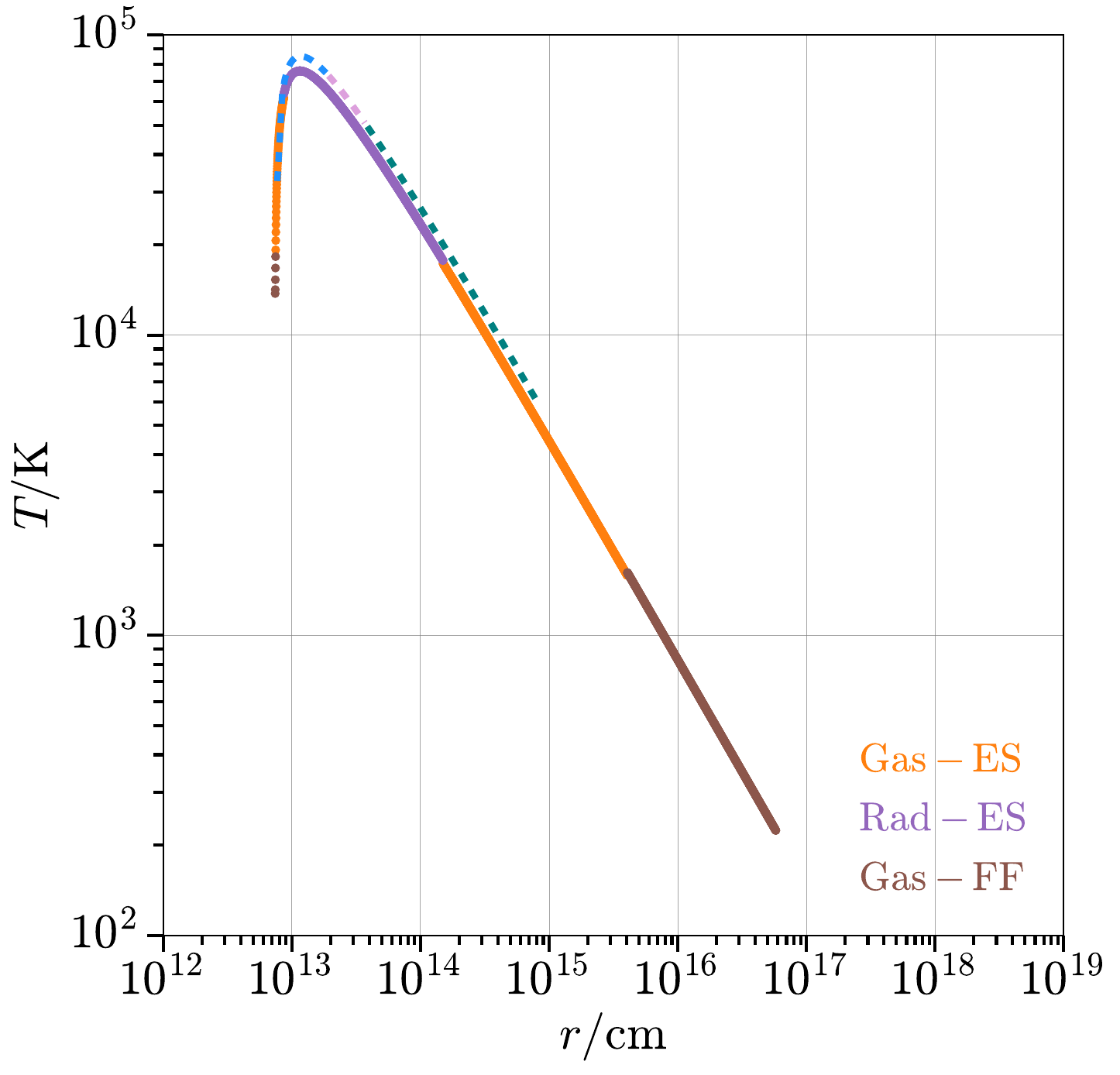}
\caption{Black hole spectral energy distribution (SED) generated by the \textsc{Relagn} package (left--hand side panel) and midplane temperature profile (right--hand panel). In the left--hand side panel, the solid lines are produced by integrating the temperature structure of the accretion disc (shown on the right--hand side panel with solid lines) based on the model presented in this work, whilst the dashed lines show the SED generated by assuming blackbody radiation (i.e. the default \textsc{Relagn} method shown on the right--hand side panel with dashed lines). The green, pink, and blue lines show the disc, warm, and hot components of the SED, and the total energy distribution is shown in black. Both models begin calculating the temperature profile at the ISCO, but for \textsc{Relagn} we plot the coordinates of the radial bin midpoints, hence the data points between the two models are not perfectly aligned. The outer edge for both models is set to be the self--gravitational radius of the accretion disc, however in our model this is calculated via relativistic equations resulting in more stable accretion discs.}
\label{fig:sed}
\end{figure*}

By using the model presented in this work, we are able to constrain five of the fifteen free parameters of the \textsc{Relagn} model \citep{HD23}, namely the black hole mass, accretion rate, spin, accretion disc inclination and outermost radius. However, the biggest advantage of modelling the accretion disc is knowing its composition, hence being able to provide more realistic temperature profiles by using, at a given distance from the black hole, the appropriate mid--plane temperature and opacity formula. 

For the opacity profiles, we use for \textbf{Gas--ES}, \textbf{Rad--ES}, and \textbf{Gas--FF} the equations \ref{eq:kappaFFkappaESGES}, \ref{eq:kappaFFkappaESRES} and \ref{eq:kappaESkappaFFGFF}, respectively. For the temperature profiles, we use the following set of equations from \cite{CO17}
\begin{flalign} \label{eq:TGES}
T \big|^\mathrm{gas}_\mathrm{es} =& \left( 2.60 \times 10^8\ \mathrm{K} \right) \left (\alpha^{-1/5}\ m_\bullet^{-3/5}\ \dmbullet^{2/5}\ x^{-11/5} \right) \nonumber \\ 
&\times \mathcal{C}^{-1/5}\ \mathcal{D}^{-1/5}\ \mathcal{P}^{2/5} \;,
\end{flalign}
\begin{flalign} \label{eq:TRES}
T \big|^\mathrm{rad}_\mathrm{es} =& \left( 3.20 \times 10^7\ \mathrm{K} \right) \left (\alpha^{-1/4}\ m_\bullet^{-1/4}\ x^{-3/4} \right) \nonumber \\ 
&\times \mathcal{C}^{1/4}\ \mathcal{D}^{-1/4}\ \mathcal{R}^{1/4} \;,
\end{flalign}
\begin{flalign} \label{eq:TGFF}
T \big|^\mathrm{gas}_\mathrm{ff} =& \left( 7.70 \times 10^7\ \mathrm{K} \right) \left (\alpha^{-1/5}\ m_\bullet^{-1/2}\ \dmbullet^{3/10}\ x^{-9/5} \right) \nonumber \\ 
&\times \mathcal{C}^{-1/10}\ \mathcal{D}^{-1/5}\ \mathcal{P}^{3/10}\ \mathcal{R}^{1/20} \;.
\end{flalign}

In \Fig{sed}, we show the black hole SED generated by the \textsc{Relagn} package based on modelling the temperature profile of the accretion disc (solid lines), and based on the original \cite{NT73} model (dashed lines). To demonstrate an example of application, we use a snapshot from the simulation presented in \Sec{Applications:Isolated} with $\alpha = 0.1$, $\alpha_\bullet = 0.56$, $M_\bullet \sim 10^9\ \Msun$, $\dot{M}_{\circledcirc \rightarrow \bullet} \sim 3.5 \times 10^{-3}\ \Msun /$yr, and a co--rotating accretion disc. By knowing these parameters, the full structure of the accretion disc is generated which allows us to use the appropriate combination of equations and calculate the temperature and opacity profiles of the material generating the observed luminosity. The global temperature profile of this accretion disc -- formed by piece--wisely combining local solutions -- is shown with solid lines on the right--hand panel of \Fig{sed}, along with green, pink, and blue dashed lines for the disc, warm, and hot components of the original \textsc{Relagn} model (i.e. the \cite{NT73} blackbody temperature profile).

The small differences in the temperature profiles are reassuring that the structure of the accretion discs generated by the model presented in \Sec{Model} is in close agreement with the \cite{NT73} theory. However, some key differences are present that differentially affect the shape of the SED, due to the non-linear relation between the temperature of the accretion disc and the emitted luminosity.

A prominent difference between the two models appears at the low frequency tail and up to $\sim$ $10^{15-16}$ Hz (i.e., at the infrared, optical and ultra--violet parts of the electromagnetic spectrum) with the original \textsc{Relagn} resulting in a higher UV peak, but less luminous tail. Since this part of the spectrum is mainly produced by the coolest regions of the disc, we conclude that for this galaxy the structure of the standard \cite{NT73} and of the \cite{CO17} model diverge the most in the outskirts of the accretion disc due to the different radial extent of the two models, as can be also seen from the temperature profiles on the right--hand panel of \Fig{sed}. At higher frequencies, the two spectra are also distinct -- with the hot component being the most divergent as a result of the difference at the peak of the temperature profile -- but converge towards the gamma-ray end of the spectrum ($>10^{20}$ Hz).

The above differences can have notable implications when comparing observed and simulated SEDs (e.g. effects on the ``big blue bump'' of quasars, time lags in reverberation mapping, etc.), highlighting the importance of knowing the structure of the accretion disc and providing a more physically--motivated prediction of fluxes. 

Unfortunately, a more systematic parameter space exploration and comparison between this work and the \textsc{Relagn} model, which uses the blackbody flux for the entire accretion disc, is beyond the scope of this study. In a future work (Irodotou et al. in prep.), we perform the aforementioned comparison in addition to updating the black hole feedback model so the energy released in the surrounding gas is calculated directly from the luminosity emitted by the accretion disc.

\section{Discussion \& Conclusions} \label{sec:Discussion}

In this third paper of the \textsc{Rabbits} series, we have developed and implemented a relativistic accretion disc model that self--consistently evolves accretion disc and black hole properties. This model incorporates a set of local relativistic solutions \citep{CO17} that are combined to represent the global structure of accretion discs. The public release of the model offers a tool for the broader astrophysical and black hole physics communities and allows for hydrodynamical galaxy formation simulations to incorporate accretion discs on-the-fly (C version), as well as for post-processing studies to generate mock accretion discs and observable quantities (Python version).

Our model’s ability to simulate and constrain accretion disc sizes is a critical step forward in connecting black hole accretion physics with observable properties of quasars. In galaxy formation simulations, this enables a direct link between the resolved gas supply on galactic scales and accretion disc--scale quantities such as flux and temperature radial profiles. These profiles naturally facilitate the construction of spectral energy distributions (SEDs), enabling comparisons with observations and providing new avenues to test accretion physics within a galaxy evolution context.

A number of recent works have explored the role of accretion-disc-regulated feedback in galaxy formation simulations. For example, \cite{SCC21} implement a subgrid disc model to regulate quasar feedback energetics and duty cycles, focusing primarily on the impact of disc physics on large-scale outflows. While complementary in spirit, our approach differs in that we explicitly evolve the relativistic disc structure and black hole spin self-consistently, allowing us to directly connect accretion geometry, radiative efficiency, and angular momentum evolution. This makes our framework particularly well suited for studying the coupled evolution of black hole growth, spin, and feedback in a physically motivated manner.

Alternative numerical strategies have also been proposed to bridge the gap between accretion disc and galaxy scales, such as cyclic zoom techniques that repeatedly re--simulate the nuclear region at higher resolution \citep{GSQ25}. \newpage These approaches offer valuable insights into the detailed gas dynamics near the black hole, but are computationally expensive and difficult to integrate self--consistently into long--term galaxy evolution simulations. By contrast, our method is designed to operate efficiently within standard galaxy formation frameworks, capturing the essential physics of relativistic accretion discs while remaining tractable for multi-gigayear simulations and large parameter studies.

We note that our model adopts the standard geometrically thin, radiatively efficient accretion disc paradigm, which is known to break down at very high or very low accretion rates. Global radiation magnetohydrodynamic simulations have demonstrated that effects such as disc thickening, magnetic pressure support, and strong outflows can modify the disc structure and emission properties \citep[e.g.][]{BB99,FEA18,JSD19}. While incorporating such effects directly into galaxy formation simulations remains challenging, our framework provides a flexible foundation upon which additional accretion states (e.g. radiatively inefficient or super-Eddington regimes) can be implemented in future work.

In addition, as part of the \textsc{Rabbits} series which investigates BH merger time--scales and event rates, this model will allow us to study how gas accretion versus black hole mergers contribute to black hole spin distributions, which has implications for gravitational wave predictions.

Future improvements and additions to the model will include: i) a more detailed feedback mechanism where the luminosity will be extracted from the accretion disc; ii) the addition of accretion states other than radiatively efficient and the inclusion of a jet--mode feedback that can have significant effects on the black hole spin evolution; iii) a property--conserving transition from our (single black hole) accretion model to the binary accretion model of the \textsc{Rabbits} project \citep{LJM23,LIJ24a,LIJ24b} to follow the complete evolution of SMBHs from formation to merger.

\section*{Acknowledgements}
DI would like to dedicate this work to the memory of Prof. Peter Thomas, whose wisdom and mentorship will always be remembered.

We list here the roles and contributions of the authors according to the Contributor Roles Taxonomy (CRediT)\footnote{\href{https://credit.niso.org}{https://credit.niso.org}}. \textbf{DI:} Conceptualization, Data curation, Formal analysis, Methodology, Software, Visualization, Writing – original draft. \textbf{SL:} Conceptualization, Methodology, Software, Validation, Writing – review \& editing. \textbf{TN, GC, RO:} Formal analysis, Resources, Validation, Writing – review \& editing.
\textbf{JMH, AR, SS, APV:} Validation, Writing – review \& editing.

We thank the developers of \astropy\ \citep{AC13,A18}, \matplotlib\ \citep{H07}, \numpy\ \citep{VCG11}, and \scipy\ \citep{VGO20}.

The work of DI, JMH, and SS was partially supported by the European Research Council via ERC Consolidator Grant KETJU (no. 818930), we acknowledge Peter H. Johansson for this funding acquisition.
SL acknowledges the support of the National Natural Science Foundation of China (No. 12588202, 12473015).
TN is supported by IBS under the project code IBS-R018-D3.
GC is the Research Director of the FNRS.
AR acknowledges the support by the University of Helsinki Research Foundation and the European Research Council via ERC Consolidator Grant KETJU (no. 818930). 
APV acknowledges support from the Sussex Astronomy Centre STFC Consolidated Grant
(ST/X001040/1).

\section*{Data Availability} \label{sec:Data}

The accretion model and simulation data associated with the article are publicly available at \href{https://github.com/DimitriosIrodotou/RABBITS-III}{https://github.com/DimitriosIrodotou/RABBITS-III}.


\bibliographystyle{mnras}
\bibliography{paper}

\appendix


\section{Correction functions} \label{app:Corrections}

\begin{figure*}
\includegraphics[width=0.44\textwidth]{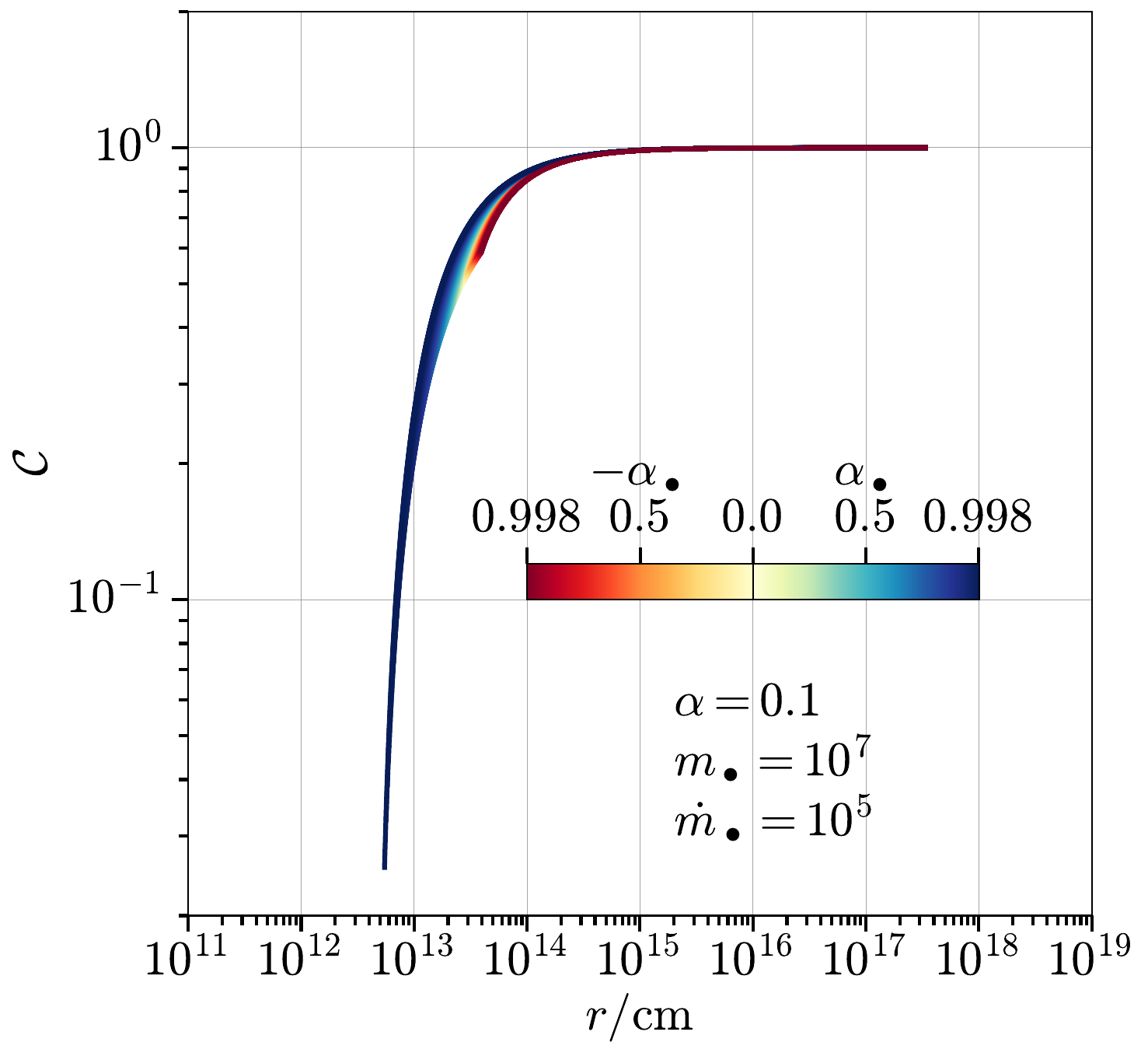} \includegraphics[width=0.44\textwidth]{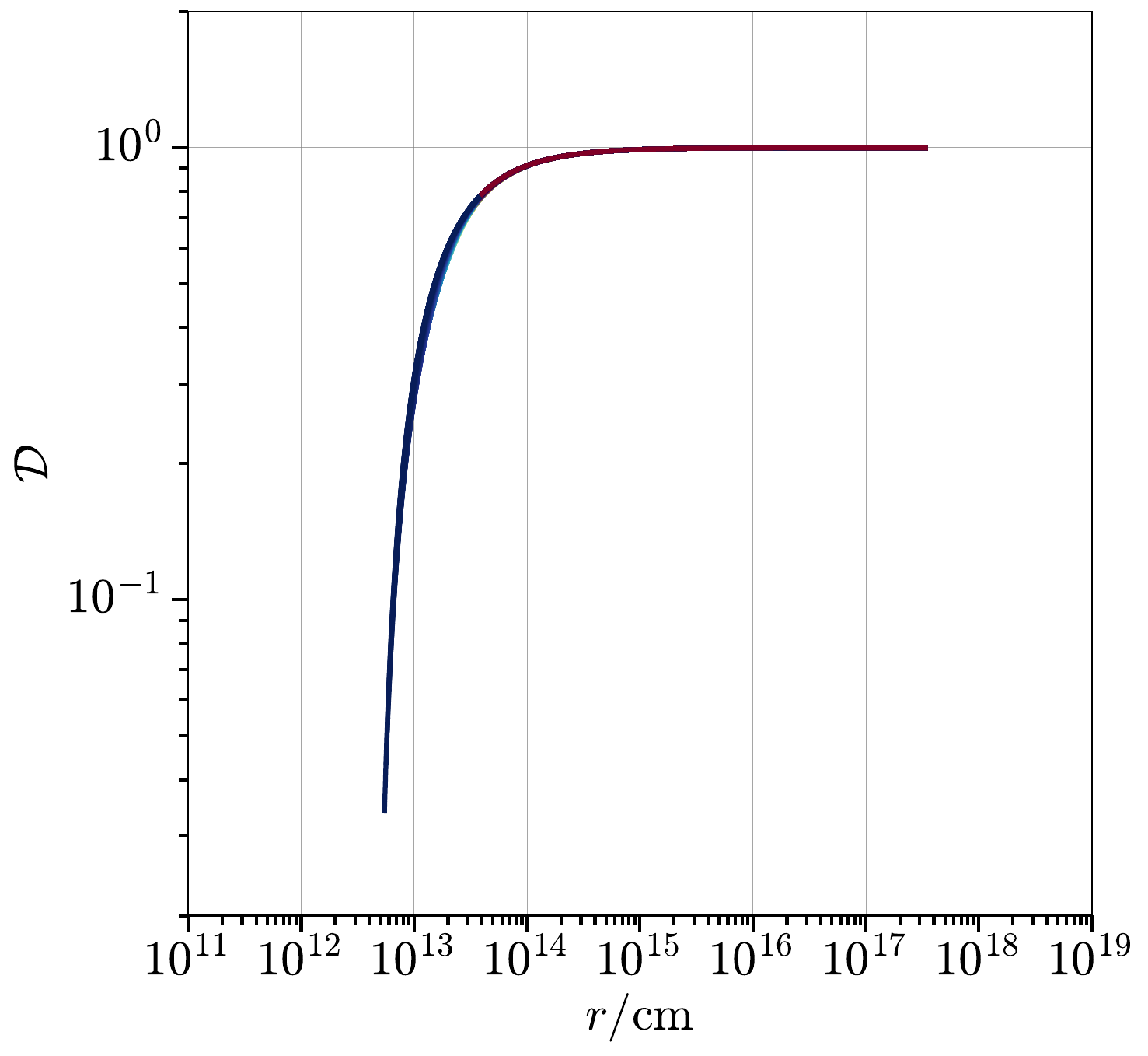} \\
\includegraphics[width=0.44\textwidth]{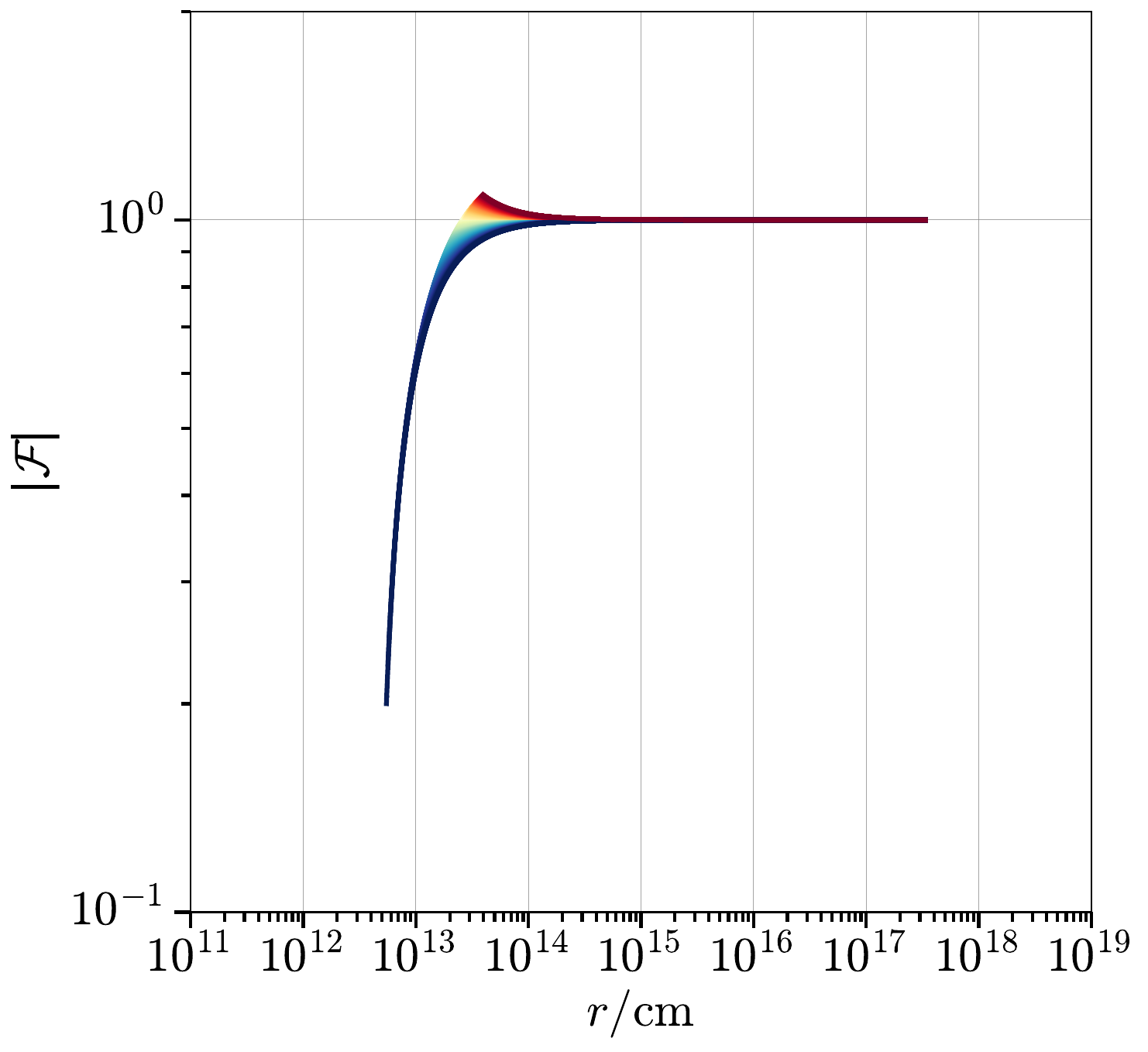} \includegraphics[width=0.44\textwidth]{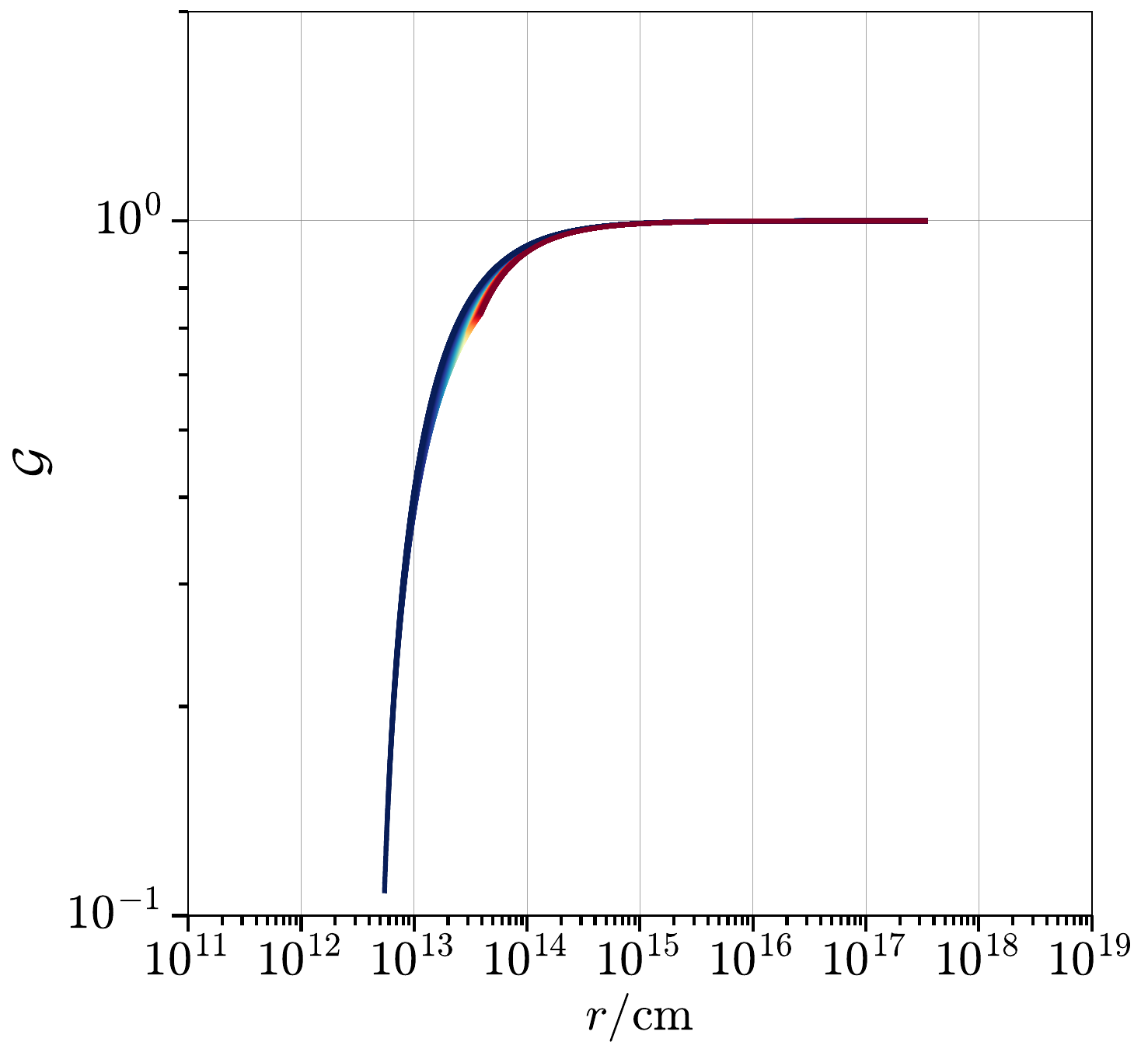} \\
\includegraphics[width=0.44\textwidth]{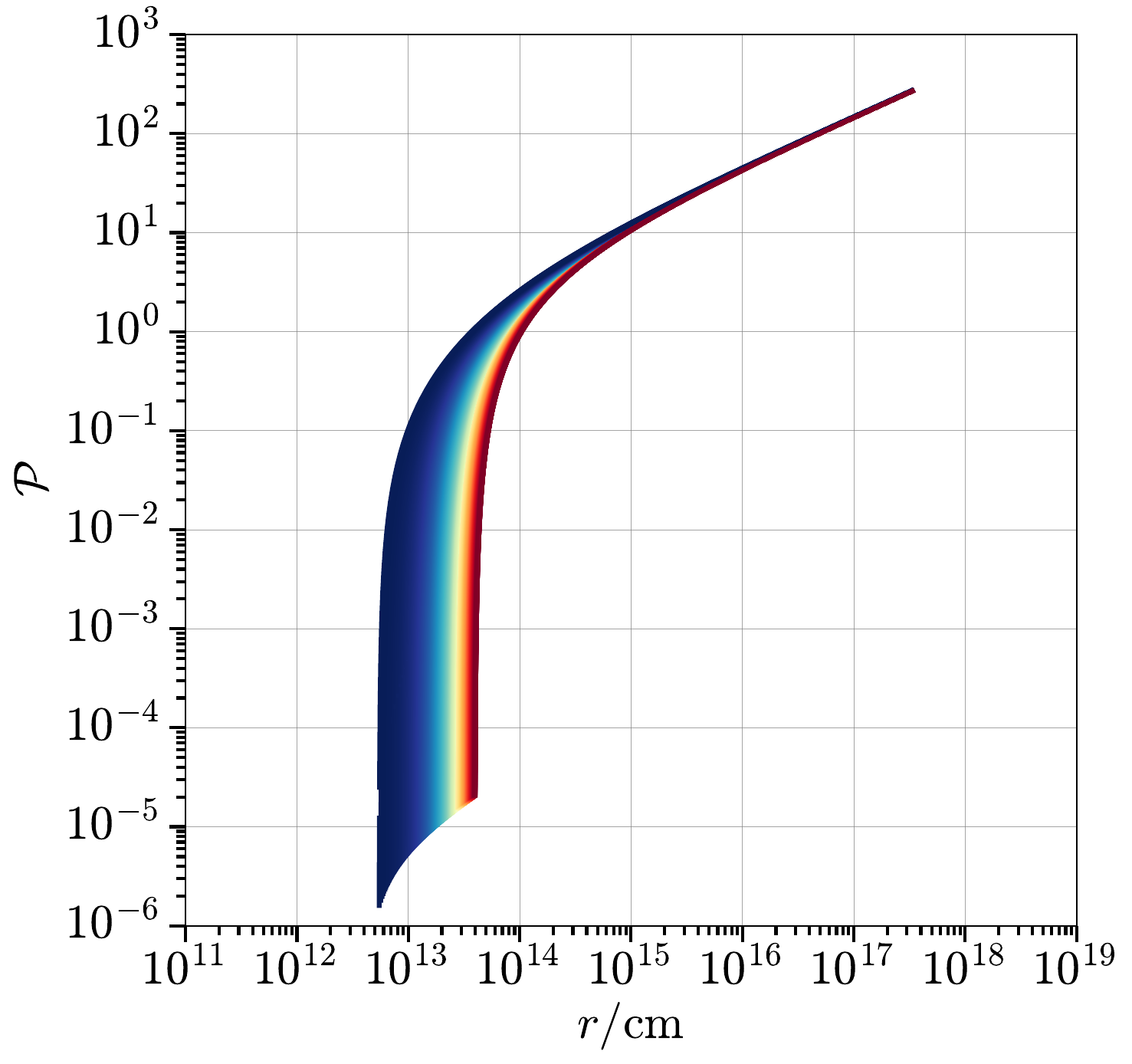} \includegraphics[width=0.44\textwidth]{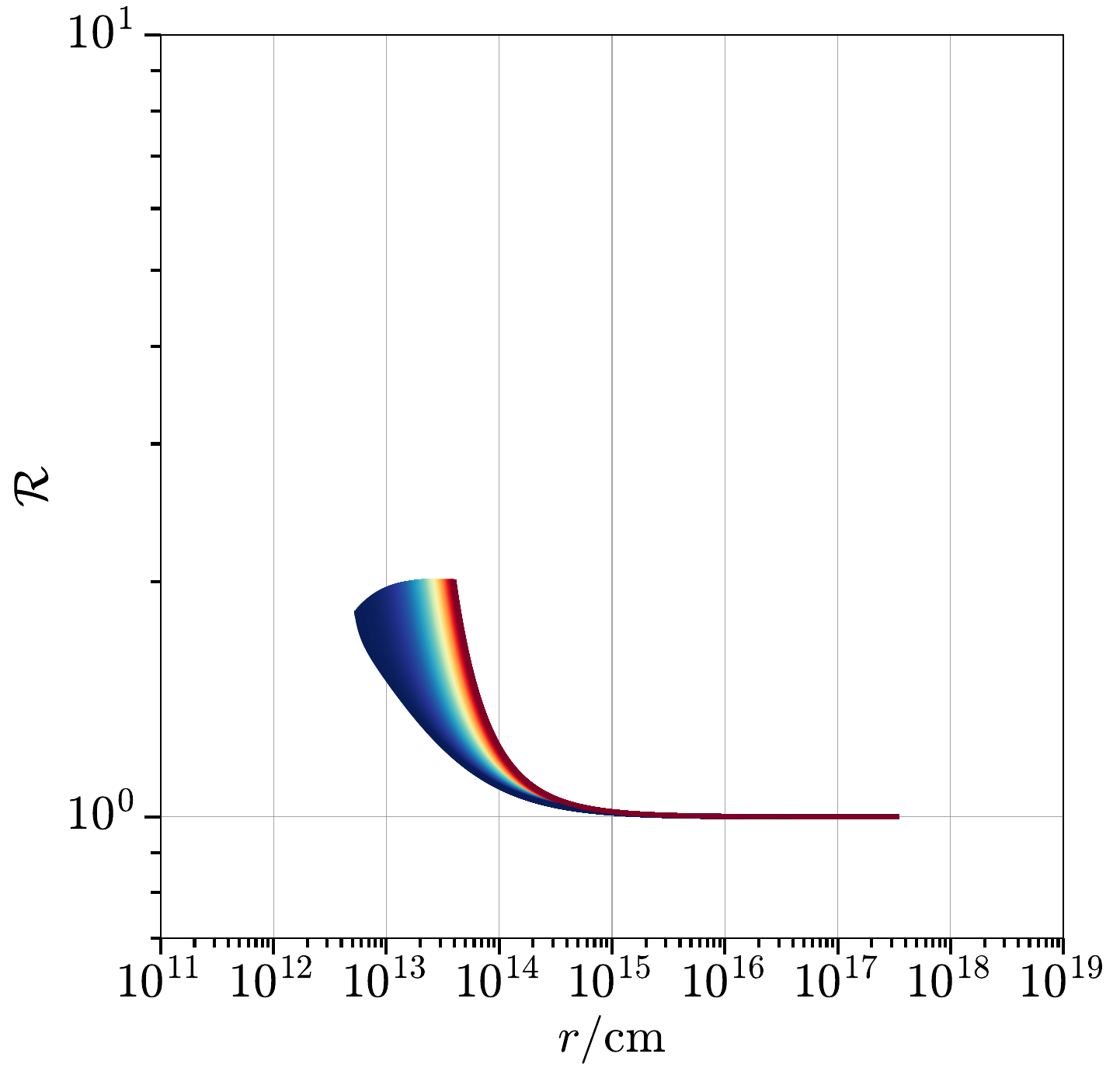}
\caption{Correction functions for the parameters used in \Sec{Model:Properties:Example}: $\alpha = 0.1$, $\alpha_\bullet = [0.000-0.998]$, $m_\bullet = 10^7$, and $\dmbullet = 10^5$, and two different orientations (positive/negative spin parameters correspond to co--/counter--rotation and blue/red colour maps). Not that for $\mathcal{F}$ we plot the absolute value since in the counter--rotation case this correction function is negative, hence cannot be plotted in logarithmic space. The gap in $\mathcal{P}$ for $\alpha_\bullet = 0.998$ is purely due to the radial discretisation of the accretion disc (we use 5000 logarithmically spaced bins for these plots).}
\label{fig:correction_functions_all}
\end{figure*}

Below we provide the dimensionless correction functions as initially defined by \cite{NT73} and updated by \cite{CO17}, following the convention of e.g. \cite{KGB11}. All functions represent the sum of the Newtonian limit plus relativistic corrections which approach unity far from the black hole (apart from $\mathcal{P}$, as can be seen in \Fig{correction_functions_all})
\begin{flalign}
\mathcal{C} &= 1 - 3x^{-2} \pm 2\alpha_\bullet\ x^{-3} \;, \label{eq:C} \\
\mathcal{D} &= 1 - 2x^{-2} + \alpha_\bullet^2\ x^{-4} \;, \label{eq:D} \\
\mathcal{F} &= \pm (1 \mp 2 \alpha_\bullet\ x^{-3} + \alpha_\bullet^2\ x^{-4}) \; \label{eq:F} \\
\mathcal{G} &= 1 - 2x^{-2} \pm \alpha_\bullet\ x^{-3} \;, \label{eq:G} \\
\mathcal{R} &= \mathcal{C}^{-1}\ \mathcal{F}^{2} - \alpha_\bullet^2\ x^{-2} \left( \mathcal{C}^{-1/2}\ \mathcal{G} - 1 \right)   \label{eq:R} \;,
\end{flalign}
where $x \equiv (r / \Rgrav)^{1/2}$ is a dimensionless radial coordinate, $0 \leq \alpha_\bullet \leq 0.998$ is the dimensionless spin parameter, and upper/lower signs refer to co-/counter--rotating orbits with respect to \Sbulletvec. An additional function was introduced by \cite{CO17} as a result of the additional torque at the ISCO \citep{PSM12}
\begin{flalign} \label{eq:P}
\mathcal{P} = \Pisco + x - \xisco \mp\frac{3\alpha_\bullet}{2} &\mathrm{ln}\left(\frac{x}{\xisco} \right) \nonumber \; \\
-\frac{3(x_1 \mp \alpha_\bullet)^2}{x_1(x_1 - x_2)(x_1 - x_3)} &\mathrm{ln}\left(\frac{x - x_1}{\xisco - x_1} \right) \nonumber \; \\
-\frac{3(x_2 \mp \alpha_\bullet)^2}{x_2(x_2 - x_3)(x_2 - x_1)} &\mathrm{ln}\left(\frac{x - x_2}{\xisco - x_2} \right) \nonumber \; \\
-\frac{3(x_3 \mp \alpha_\bullet)^2}{x_3(x_3 - x_1)(x_3 - x_2)} &\mathrm{ln}\left(\frac{x - x_3}{\xisco - x_3} \right) \;,
\end{flalign}
where
\begin{flalign} \label{eq:Pcirc}
\Pisco = 2^{-1/2}\alpha\ \xisco \ \hisco\ \Disco^{1/2}\ \Rcalisco^{1/2} \;,
\end{flalign}
where $\alpha$ is the viscosity parameter, $\xisco = (\Risco / \Rgrav)^{1/2}$ is the dimensionless radius of the ISCO, and \hisco\ is the accretion disc opening angle [see \Eq{hGES}, (\ref{eq:hRES}) and (\ref{eq:hGFF}) ] at the ISCO given by 
\begin{flalign} \label{eq:hcircGES}
\hisco \big|^\mathrm{gas}_\mathrm{es} =&\; 1.80 \times 10^{-3} \left( \alpha^{1/8}\ m_\bullet^{-3/8}\ \dmbullet^{1/4} \right) \nonumber \\
&\times\ \xisco^{1/8}\ \Cisco^{-1/8}\ \Rcalisco^{-1/2} \;,
\end{flalign}
\begin{flalign} \label{eq:hcircRES}
\hisco \big|^\mathrm{rad}_\mathrm{es} \equiv 2.00 \times 10^{-3} \;,
\end{flalign}
\begin{flalign} \label{eq:hcircGFF}
\hisco \big|^\mathrm{gas}_\mathrm{ff} =&\; 1.30 \times 10^{-3} \left( \alpha^{1/17}\ m_\bullet^{-5/17}\ \dmbullet^{3/17} \right) \xisco^{5/17} \nonumber \\ 
&\times\ \Cisco^{-1/17}\ \Disco^{-1/34}\ \Rcalisco^{-8/17} \;,
\end{flalign}
where the upper equation corresponds to the case when the ISCO lies inside the \textbf{Gas--ES} regime and the lower case when in the \textbf{Gas--FF} regime. When the ISCO lies in the \textbf{Rad--ES} regime the opening angle of the accretion disc cannot be constrained, since \Eq{hRES} linearly depends on $\mathcal{P}$, which in turn linearly depends on $\hisco \big|^\mathrm{rad}_\mathrm{es}$, hence both opening angle terms cancel out. Thus, we follow \cite{CO17} and fix $\hisco \big|^\mathrm{rad}_\mathrm{es}$ to 0.002. Finally,  
\begin{flalign} \label{eq:x}
x_1 &= 2\mathrm{cos}\left[ \mathrm{arccos}(\pm \alpha_\bullet)/3 - \uppi/3 \right] \; \\
x_2 &= 2\mathrm{cos}\left[ \mathrm{arccos}(\pm \alpha_\bullet)/3 + \uppi/3 \right] \; \\
x_3 &= -2\mathrm{cos}\left[ \mathrm{arccos}(\pm \alpha_\bullet)/3 \right] \;
\end{flalign} 
are the three roots of $\mathcal{C} = 0$ calculated in \cite{PT74}.

\section{Numerical implementation} \label{app:Implementation}

\begin{figure*}
\includegraphics[trim=2cm 0cm 2cm 0cm, width=\textwidth, clip]{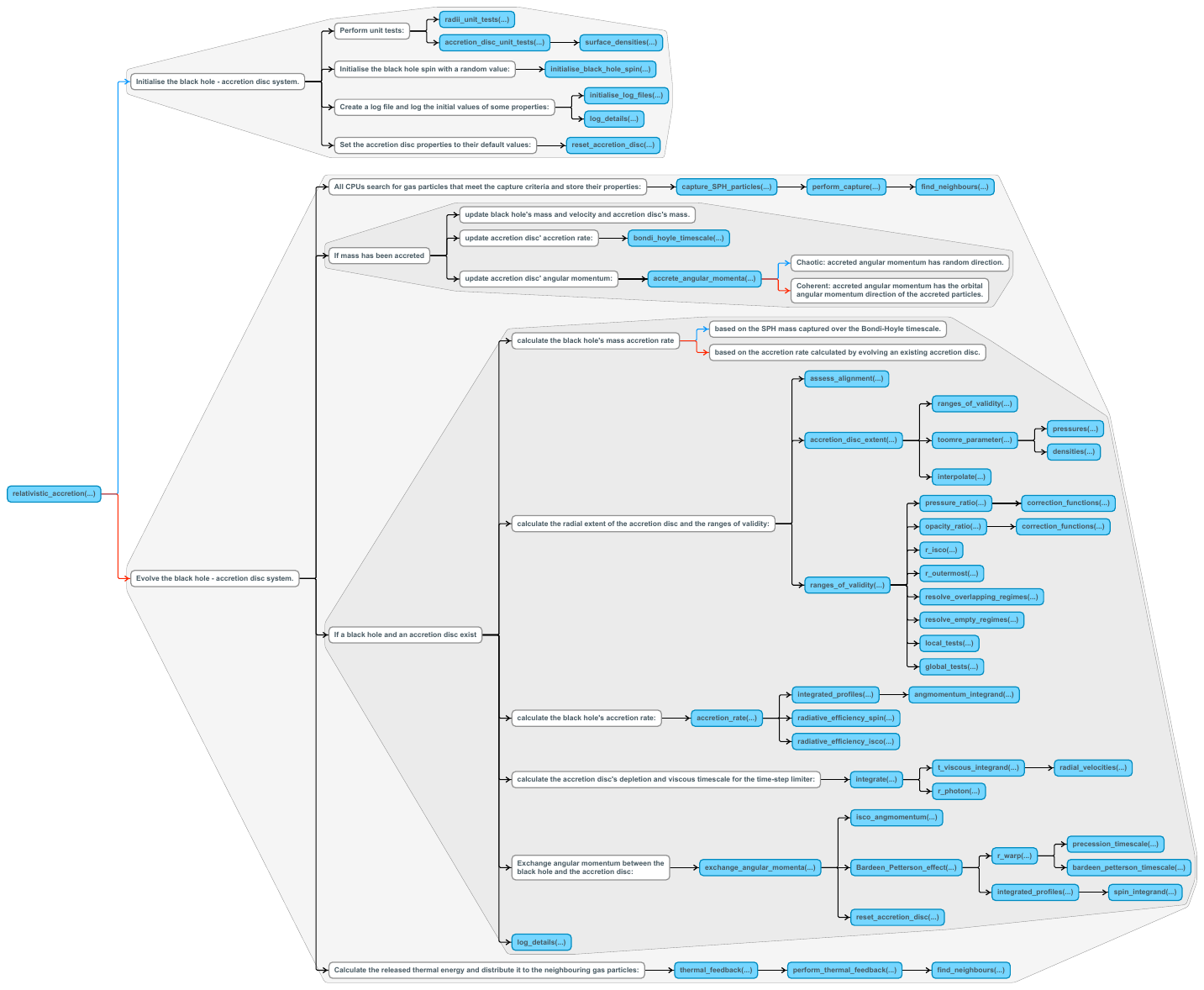}
\caption{Structure of the whole model. Light blue boxes represent core functions in the code and red and blue lines correspond to conditional statements in the code.}
\label{fig:chart}
\end{figure*}

The model developed in this work -- presented in Sections \ref{sec:Model} and \ref{sec:Results}, and applied in \Sec{Applications} -- can be attached to (in principal) any C--based galaxy formation simulation as a module, after making it appropriate to one's specific setup (e.g. we follow the pressure--entropy formulation to compute SPH properties, we use \textsc{Gadget}--type of routines to search for neighbouring particles etc. Theses lines/methods might need to be altered). Implementing the presented model can be done by bypassing one's existing black hole physics routines (by for example introducing \texttt{\#ifdef} directives that control which parts of the code will be included in the compilation process) and instead invoking the function \texttt{relativistic\_accretion(...)}. 

In \Fig{chart}, we provide the structure of the whole code where light blue boxes represent core functions in the code and red and blue lines correspond to conditional statements in the code. This figure summarises the steps the model takes and which functions are called. Below, we provide more information for each of these steps.

\subsection{Initialise the black hole--accretion disc system} \label{app:Implementation:Initialise}

\begin{figure*}
\includegraphics[trim=0cm 6cm 0cm 6cm, width=\textwidth, clip]{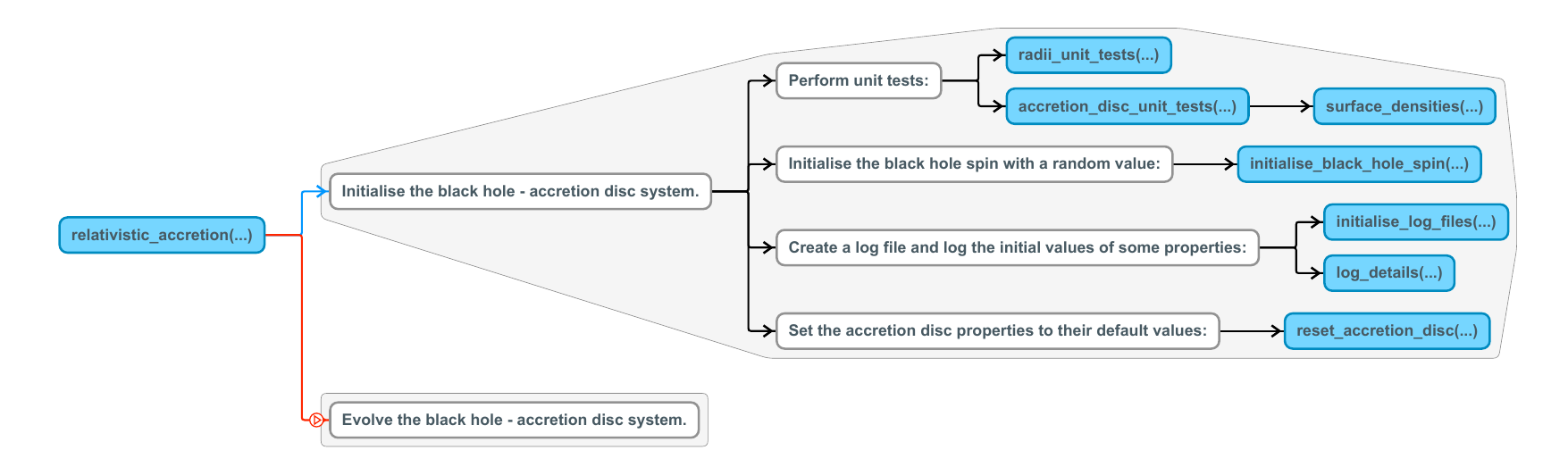}
\caption{View of the processes described in \App{Implementation:Initialise}, same as \Fig{chart} but the rest of the structure of the code has been collapsed. Light blue boxes represent functions in the code and red and blue lines correspond to conditional statements in the code.}
\label{fig:chart1}
\end{figure*}

In \Fig{chart1}, the first step once the function \texttt{relativistic\_accretion(...)} is called for the first time (i.e. the first time--step of the simulation), is to initialise the BH--AD system in \texttt{initialise\_setup(...)}. 

The code starts by performing unit test in functions \texttt{radii\_unit\_tests(...)} and \texttt{accretion\_disc\_unit\_tests(...)} for each active black hole by building a mock accretion disc and comparing it with known solutions. These unit tests ensure that a valid structure of an accretion disc can be created and no unexpected behaviours will occur in the simulation.

Then, a black hole spin is initialised with a random spin parameter value and direction in \texttt{initialise\_black\_hole\_spin(...)}. Furthermore, log files are generated and parameters that represent the values from the initial conditions get printed\footnote{Each subsequent time--step a black hole is active, \texttt{log\_details(...)} is called at the beginning and at the end of \texttt{relativistic\_accretion(...)}, which even though doubles the size of the log file, it allows for detailed quantification of how the model operates.} inside them in \texttt{initialise\_log\_files(...)} and \texttt{log\_details(...)}, respectively. Finally, accretion disc properties get initialised in \texttt{reset\_accretion\_disc(...)}.

\subsection{Evolve the black hole--accretion disc system} \label{app:Implementation:Evolve}

We split this subsection into six steps: (i) the capture of ambient ISM gas by the BH--AD system; (ii) the transfer of the captured gas to the accretion disc; (iii) the effects of accretion on the accretion disc and (iv),(v) on the black hole; and finally (vi) the effects of accretion--related feedback on the surrounding ISM.

\begin{figure*}
\includegraphics[trim=0cm 6cm 0cm 6cm, width=\textwidth, clip]{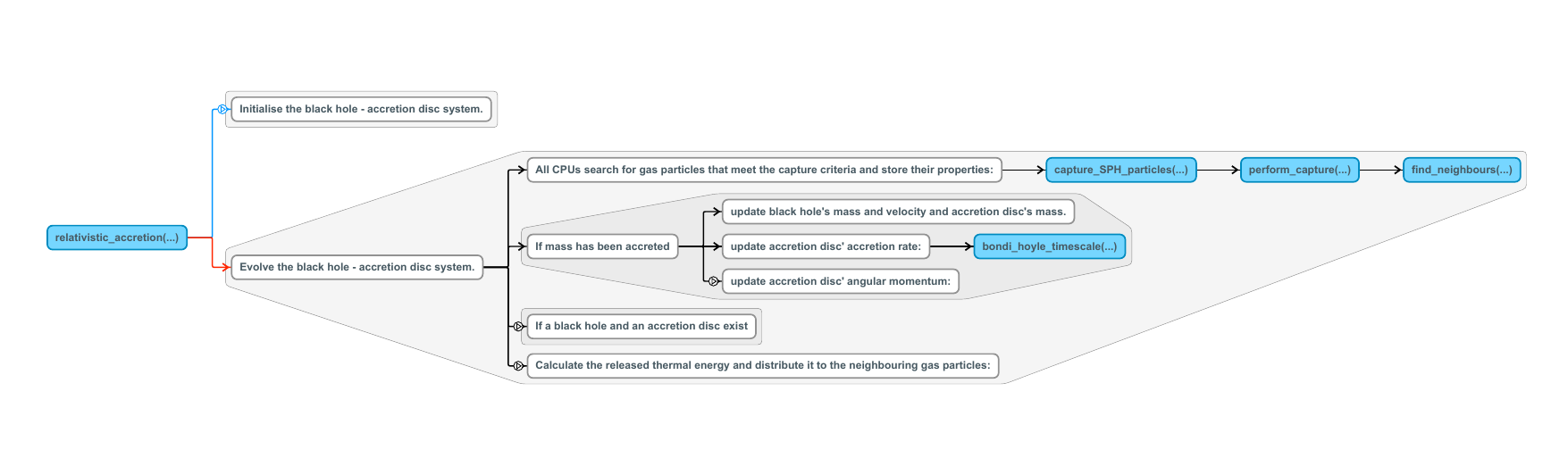}
\caption{View of the processes described in the steps (i) and (ii) of \App{Implementation:Evolve}, same as \Fig{chart} but the rest of the structure of the code has been collapsed. Light blue boxes represent functions in the code and red and blue lines correspond to conditional statements in the code.}
\label{fig:chart2}
\end{figure*}

\begin{enumerate}[wide=0pt,labelindent=10pt,labelwidth=10pt]
\item As shown in \Fig{chart2}, the initial step towards forming and/or evolving an accretion disc is identifying which gas particle(s) will provide the material. To this end, we use in \texttt{capture\_SPH\_particles(...)} an SPH particle capturing scheme \citep[e.g.][]{PNK11,NK13} in which gas particle(s) that form the accretion disc must meet both of the following conditions
\begin{flalign} \label{eq:rcircledast}
r_\circledast \leq \rcapt \equiv \mathrm{max}(\varepsilon_\bullet, \RHL) \;,
\end{flalign}
and
\begin{flalign} \label{eq:jcircledast}
u_\circledast \leq \ucirc \equiv \sqrt{G\ M_{\bullet + \circledcirc}/ r_\circledast} \;,
\end{flalign}
where $r_\circledast$ and $u_\circledast$ are the position and velocity, respectively, of a gas particle with respect to the black hole particle, $\varepsilon_\bullet$ is the black hole's gravitational softening length, $M_{\bullet + \circledcirc} = M_\bullet + {M_\circledcirc}$ is the BH--AD system's total mass, and \RHL\ is the \cite{HL39} \citep[see also][]{SMS85} radius given by
\begin{flalign} \label{eq:RHL}
\RHL = \frac{2G\ M_{\bullet + \circledcirc}}{u_\circledast^2 + \ccircledastsqr} \;,
\end{flalign}
where $c_\circledast$ is the sound speed of the surrounding medium given by
\begin{flalign} \label{eq:cssqr}
\ccircledastsqr = \gamma\ A\ \rho^{\gamma - 1} \;,
\end{flalign}
where $\gamma = 5/3$ is the adiabatic index and $A \equiv p / \rho^{\gamma}$ is the entropy of the gas. The entropy associates the pressure $p$ to the density $\rho$ and based on the \cite{H13} pressure--entropy formulation for each particle $i$ these parameter are related as
\begin{flalign} \label{eq:pressure}
\bar{p_i} = \left[ \sum_{j=1}^{N} m_j\ A_j^{1/\gamma}\ W_{ij}(h_j) \right]^\gamma \;,
\end{flalign}
where the sum runs over $N=100$ neighbouring gas particles, $m_j$ is the neighbouring gas particle's mass, $W_{ij}$ is the Wendland $C^4$ smoothing kernel \citep{DA12}, and $h_j$ is the smoothing length.

The radius of \Eq{RHL} defines a sphere within which gas particles are gravitationally bound to the BH--AD system and their thermal gas pressure cannot prevent them from being drawn in and accreted (i.e. at \RHL\ the gas escape velocity equals the sound speed of the surrounding medium). Hence, \Eq{rcircledast} and \Eqno{jcircledast} ensure that only gas particles that are within the smallest, resolvable length \citep[in cases where \RHL\ is not resolved we use $\varepsilon_\bullet$ as in][]{CON12} and have velocity smaller than the velocity of a circular orbit will form an accretion disc around the black hole.

\begin{figure*}
\includegraphics[width=\textwidth]{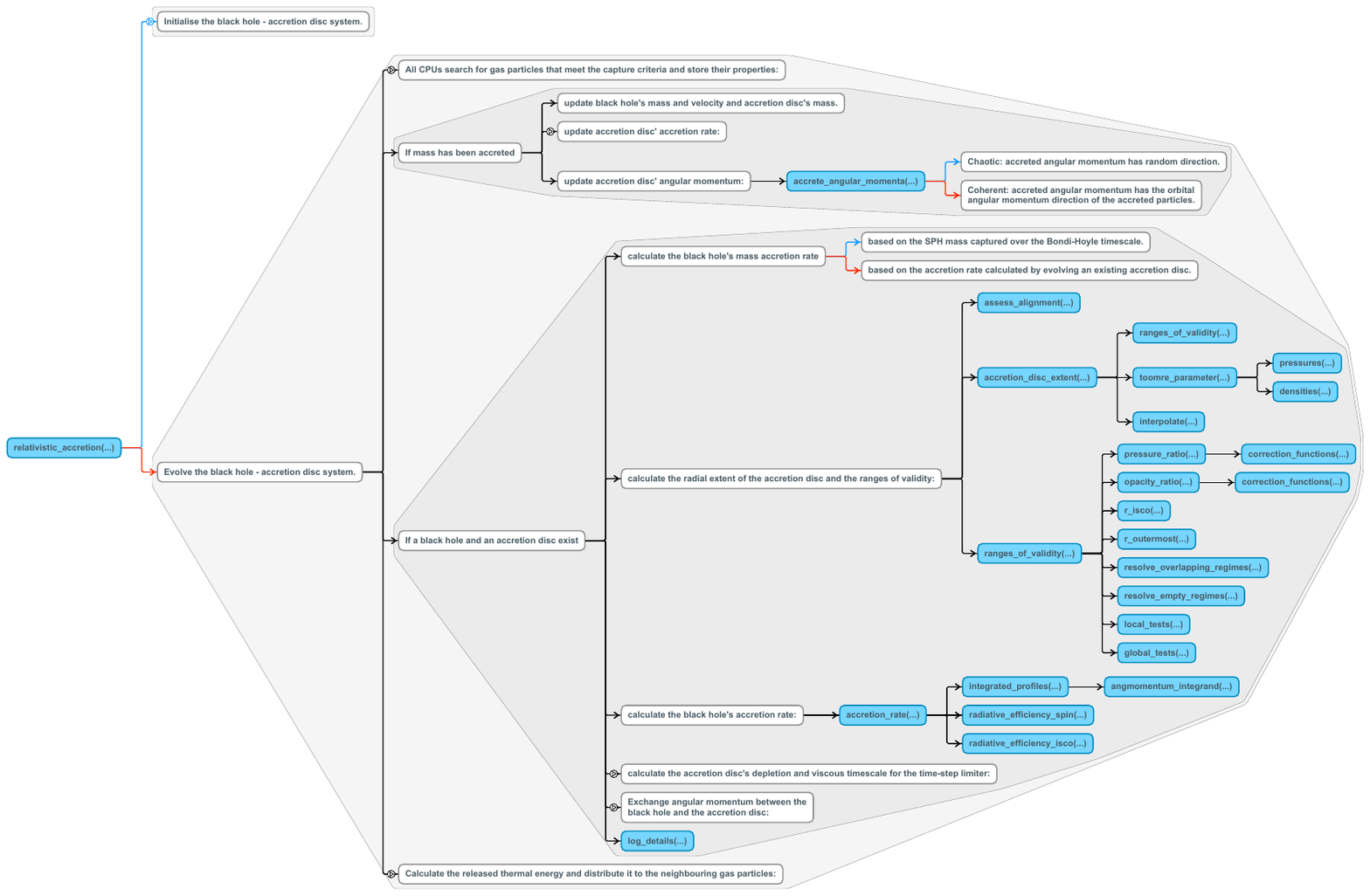}
\caption{View of the processes described in the steps (iii) and (iv) of \App{Implementation:Evolve}, same as \Fig{chart} but the rest of the structure of the code has been collapsed. Light blue boxes represent functions in the code and red and blue lines correspond to conditional statements in the code.}
\label{fig:chart3}
\end{figure*}

\item In most SPH--based black hole accretion models, the mass of the black hole is split into dynamical and sub--grid for force and accretion/feedback calculations, respectively. In our model, the dynamical black hole mass represents the BH--AD system mass. Hence, gas particle(s) that meet the radial and velocity conditions given by \Eq{rcircledast} and \Eqno{jcircledast}, respectively, are dissociated from the hydrodynamical and gravity solvers and their mass and momentum are immediately given to the dynamical black hole properties (in order to conserve mass and linear momentum in the simulation). The evolution of the sub--grid black hole mass is handled analytically via the modelling of the accretion disc. 

At each time--step, if the capture conditions are met gas flows from the ISM to the accretion disc and an amount defined by \Eq{dMcircledcircarrowbullet} migrates through the accretion disc and feeds the black hole (if $M_\circledcirc > 0$). If an accretion disc already exists, the captured gas particles are instantly transferred to the accretion disc. However, if an accretion disc forms for the first time (either because a newly formed black hole appeared in the simulation or a complete depletion due to accretion has occurred in the previous time--step) once the particle(s) that meet the above criteria have been identified their mass is assumed to be added to the accretion disc at a rate
\begin{flalign} \label{eq:dMcircledastarrowcircledcirc}
\dMcircledastarrowcircledcirc = \frac{M_\circledast(r_\circledast \leq \rcapt\ \&\ j_\circledast \leq \jcirc)}{\tBondi} \;,
\end{flalign}
where \tBondi\ is the Bondi timescale defined as
\begin{flalign} \label{eq:tBondi}
\tBondi = \frac{\rBondi}{c_\bullet\sqrt{1 + \mathcal{M}^2}} = \frac{G\ M_\bullet}{c_\bullet^3 (1 + \mathcal{M}^2)^{3/2}} \;,
\end{flalign}
where $\mathcal{M} \equiv u_\bullet / c_\bullet$ is the Mach number which is the ratio of the black hole's relative velocity with respect to the surrounding gas particles and the local sound speed calculated using gas particles inside the black hole's kernel.

Calculating \Eq{dMcircledastarrowcircledcirc} for a newly formed accretion disc is necessary because, as discussed in \Sec{Model}, the accretion rate of the black hole is one of the four parameters (along with spin, mass, and viscosity) that dictate the structure of the accretion disc. Whenever an accretion disc exists the accretion rate is explicitly calculated by \Eq{dMcircledcircarrowbullet} and used to evolve the accretion disc in time. However, when an accretion disc is formed for the first time, the rate given by \Eq{dMcircledastarrowcircledcirc} is derived utilising \texttt{bondi\_hoyle\_timescale(...)} and used to provide an estimate\footnote{Note that the accretion rate of \Eq{dMcircledastarrowcircledcirc} represents an estimate in the sense that we have not implemented a gas particle slicing scheme \citep[e.g.][]{HSB14} where captured mass is gradually transferred to the system. However, in a future work we intent to improve both the capturing and the transfer of mass mechanisms of the model.} of the mass accretion rate of the BH--AD system.

\item Finally, as can be seen in \Fig{chart3}, in \texttt{accrete\_angular\_momenta(...)} the accretion disc's total angular momentum vector can be affected by the accretion of angular momentum from the ISM which has a magnitude equal to that of a circular orbit's at the outskirts of the accretion disc and  -- depending on the method -- a direction that is either random or is governed by the gas particles' orbital angular momentum direction. Hence, in every time--step that mass is captured by the accretion disc, the magnitude of its angular momentum is altered by an amount
\begin{flalign} \label{eq:dJcircledastarrowcircledcircvec}
\sqrt{G\ M_{\bullet + \circledcirc}\ R_\circledcirc}\ \mathcal{C_\circledcirc}^{-1/2}\ \mathcal{F_\circledcirc} \;,
\end{flalign}
which is equal to evaluating \Eq{tildeJcircledcirc} at the outskirts of the accretion disc (i.e. for $r = R_\circledcirc$)\footnote{In theory this would not conserve the total angular momentum of a galaxy, but that is not a problem as the simulation is agnostic to the sub--grid modelling of the accretion disc. What is important is conserving the angular momentum of the BH--AD system -- that technically exists in isolation -- which our model does.}.

Once the above steps have finished, invoking \texttt{assess\_alignment(...)}, which uses \Eq{costheta} to measure the relative BH--AD alignment, concludes all information required to build a new or evolve an existing accretion disc: the black hole mass and spin, the accretion rate of either the black hole or the BH--AD system, and their relative orientation. 

Hence, in \texttt{accretion\_disc\_extent(...)} the extent of the accretion disc is evaluated by using the equations in \Sec{Model:Properties:AD boundaries} and following the process mentioned  in \Sec{Model:Properties:Example}. Briefly, once the \texttt{ranges\_of\_validity(...)} have been evaluated for each of the three different regimes (\textbf{Gas--ES}, \textbf{Rad--ES}, and \textbf{Gas--FF}) based on their \texttt{pressure\_ratio(...)} and \texttt{opacity\_ratio(...)}, the relativistic \texttt{toomre\_parameter(...)} of \Eq{Q} is iteratively calculated in order to identify the outermost edge of the accretion disc (i.e. its self--gravitational radius).

Knowing now the extent of the accretion disc, the final structure of the accretion disc is re--evaluated one final time by re--calculating the \texttt{ranges\_of\_validity(...)}. In addition to the unit tests performed at the beginning of the simulation (see \App{Implementation:Initialise}), inside \texttt{ranges\_of\_validity(...)} we make sure that we \texttt{resolve\_overlapping\_regimes(...)}, \texttt{resolve\_empty\_regimes(...)}, and \texttt{resolve\_lone\_regimes(...)}, meaning that once the accretion disc is split into radial bins we do not allow for the same bin to, respectively, have more than one regimes, be empty or be single (i.e. at least two adjacent bins of the same regime need to exist for any meaningful calculation to be performed).

\item The next process involves the effects of accretion on the black hole. Once a robust accretion disc exists, integrating \Eq{Jcircledcirc} over its structure and calculating the amount of mass that migrates through it and falls past the photon radius of \Eq{Rph} is done based on \Eq{dMcircledcircarrowbullet} in \texttt{accretion\_rate(...)}, as shown in \Fig{chart3}. In the same function, the \texttt{radiative\_efficiency\_spin(...)} and \texttt{radiative\_efficiency\_isco} which appear in \Eq{Lobs} are calculated and stored in variables that can be accessed later, when the accretion--related thermal feedback is evaluated (step (vi)).

\begin{figure*}
\includegraphics[trim=0cm 6cm 0cm 6cm, width=\textwidth, clip]{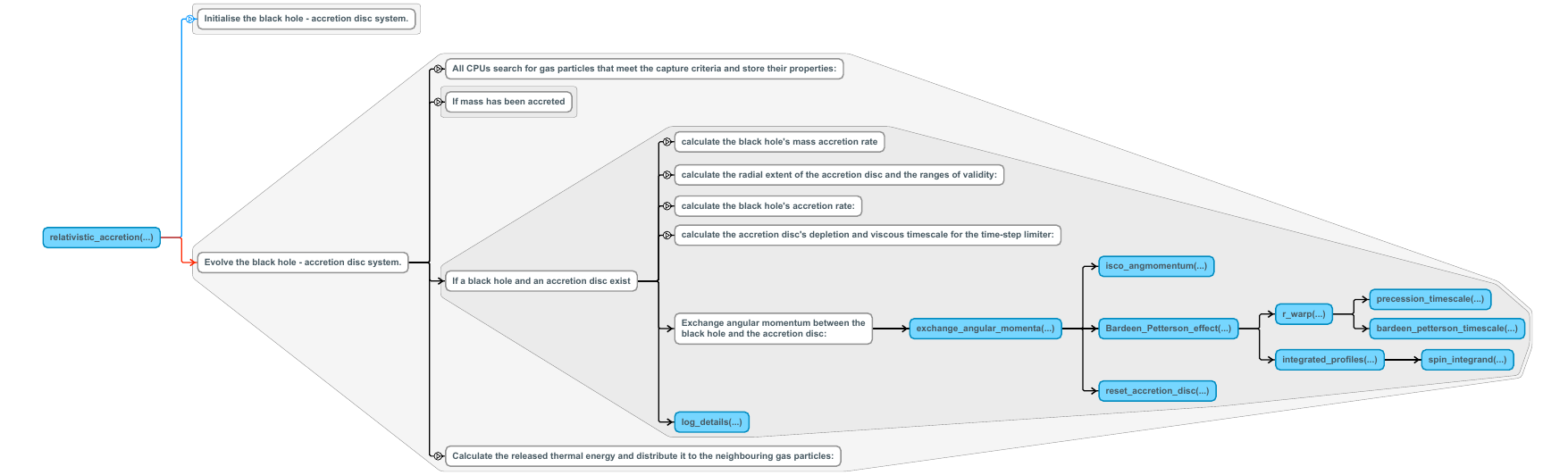}
\caption{View of the processes described in the step (v) of \App{Implementation:Evolve}, same as \Fig{chart} but the rest of the structure of the code has been collapsed. Light blue boxes represent functions in the code and red and blue lines correspond to conditional statements in the code.}
\label{fig:chart4}
\end{figure*}

\item Apart from the increase in the black hole's mass due to accretion, angular momentum is also transferred from the accretion disc to the black hole, as can be seen in \Fig{chart4}. In \texttt{exchange\_angular\_momenta(...)}, after evaluating the \texttt{isco\_angmomentum(...)} following \Eq{Jisco} we update the spin magnitude both for the black hole and the accretion disc. In addition, we update their spin direction by quantifying the \texttt{Bardeen\_Petterson\_effect(...)}.

As described in \Sec{Model:Evolve:BH}, knowing the structure of the accretion disc allows us to resolve (to some extent) the \cite{BP75} effect. In each time--step, we keep track of the portion of the accretion disc that has aligned (or anti--aligned) with the black hole by evaluating the warp radius \citep{SF96}, which represents the characteristic radius within which alignment has occurred and outside of which the accretion disc is still misaligned with respect to the black hole's spin axis. The \texttt{warp\_radius(...)} can be calculated by comparing the precession timescale to the Bardeen--Petterson timescale \citep[e.g.][]{NA99,FBB11} which are respectively given by
\begin{flalign} \label{eq:tprec}
\tprec = \frac{2\uppi}{\omega(r)} \;,
\end{flalign}
where $\omega$ is the \cite{LT18} frequency given by \Eq{omega}. The Bardeen--Petterson timescale is given by \citep{O99,LP07,PDC09}
\begin{flalign} \label{eq:tBP}
\tBP = \frac{4(1 +7\alpha^2)}{2\alpha^2(4 + \alpha^2)} \tvisc \;,
\end{flalign}
where $\alpha$ is the viscosity parameter and the viscous timescale \tvisc\ is integrated over the structure of the accretion disc as indicated by \Eq{tvisc}. 

Hence, at all times we evaluate \tprec\ and \tBP\ whose comparison (i.e. whether \tprec\  < \tBP) indicates if the radial diffusion of the warp (i.e. the gravitomagnetic perturbation) has taken longer to reach radius $r$ than the local precession timescale at $r$; in other words how far out the Lense--Thirring torque had time to align (or anti--align) the inner accretion disc with the black hole's spin axis \citep[e.g][]{MPT07}. The above resolved evaluation of the Bardeen--Petterson effect prevents us from overestimating its impact since as can be seen from \Eq{dSbulletBPvec}, only the misaligned part of the accretion disc contributes to the torque between the black hole and the accretion disc, hence the aligned part (i.e. all $r$ where \tprec\  < \tBP) of the accretion disc should be excluded from the integration by using the warp radius as lower limit in the integral of \Eq{dSbulletBPvec}. 

\begin{figure*}
\includegraphics[trim=0cm 6cm 0cm 6cm, width=\textwidth, clip]{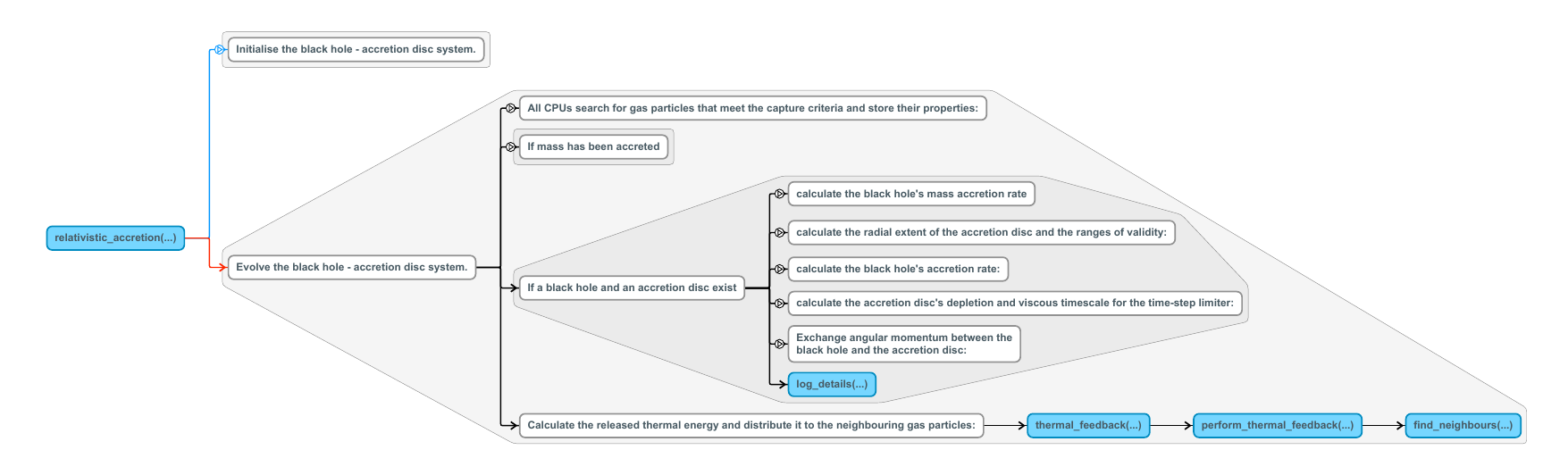}
\caption{View of the processes described in the step (vi) of \App{Implementation:Evolve}, same as \Fig{chart} but the rest of the structure of the code has been collapsed. Light blue boxes represent functions in the code and red and blue lines correspond to conditional statements in the code.}
\label{fig:chart5}
\end{figure*}

\item Finally, \texttt{thermal\_feedback(...)} is performed in a standard \textsc{Gadget}--fashion where the energy rate calculated based on \Eq{dE} is distributed to the neighbouring gas particles inside the black hole's kernel, following the steps seen in \Fig{chart5}. Note that in \Eq{Lobs} we have -- apart from the spin--dependent efficiency -- an additional mechanism that generates luminosity that corresponds to the torque at the ISCO \citep{CO17}.
\end{enumerate}

\subsection{Resolving important black hole--accretion disc processes} \label{app:Implementation:Resolving}

\begin{figure*}
\includegraphics[trim=0cm 6cm 0cm 6cm, width=\textwidth, clip]{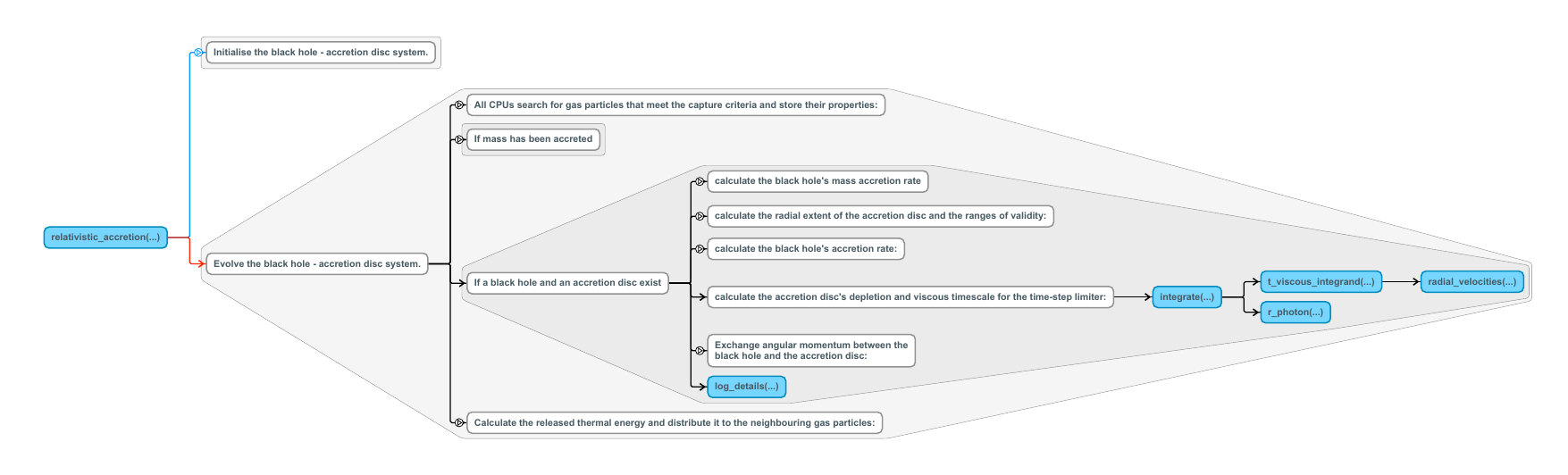}
\caption{View of the processes described in the step (vi) of \App{Implementation:Evolve}, same as \Fig{chart} but the rest of the structure of the code has been collapsed. Light blue boxes represent functions in the code and red and blue lines correspond to conditional statements in the code.}
\label{fig:chart6}
\end{figure*}

As can be seen in \Fig{chart6}, one extra calculation we perform inside \texttt{relativistic\_accretion(...)} is evaluating two timescales which is necessary for ensuring that we are capturing important process taking place in the accretion disc.

The first one is the accretion disc's depletion timescale which can be written as
\begin{flalign} \label{eq:tdep}
\tdep = \frac{M_\circledcirc}{\dMcircledcircarrowbullet} \;,
\end{flalign}
and captures how quickly the accretion disc gets depleted of its gas due to radial migration of mass from the accretion disc to the black hole. 

The second timescale is the viscous timescale of the accretion disc, which describes how long it will take for mass at a given radius to radially migrate and be accreted by the black hole. Hence, since the global structure of the accretion disc is a combination of locally valid regimes, this timescale is given by
\begin{flalign} \label{eq:tvisc}
\tvisc = \int_{\circledcirc} \frac{\mathrm{d}r}{u^r} \;,
\end{flalign}
where $u^r$ is the radial velocity. For \textbf{Gas--ES}, \textbf{Rad--ES}, and \textbf{Gas--FF}, \cite{CO17} derived the following expressions

\noindent \\
\textbf{Gas--ES}
\begin{flalign} \label{eq:urGES}
u^r \big|^\mathrm{gas}_\mathrm{es} =& \left( -7.30 \times 10^{5} \mathrm{cm\ s^{-1}} \right) \left( \alpha^{4/5}\ m_\bullet^{-3/5}\ \dmbullet^{2/5}\ x^{-1/5} \right) \nonumber \\ 
&\times \mathcal{C}^{-1/5}\ \mathcal{D}^{4/5}\ \mathcal{P}^{-3/5} \;, 
\end{flalign}

\noindent \\
\textbf{Rad--ES}
\begin{flalign} \label{eq:urRES}
u^r \big|^\mathrm{rad}_\mathrm{es} =& \left( -3.50 \times 10^{9} \mathrm{cm\ s^{-1}} \right) \left( \alpha\ m_\bullet^{-2}\ \dmbullet^{2}\ x^{-6} \right) \nonumber \\ 
&\times \mathcal{C}^{-2}\ \mathcal{D}\ \mathcal{P}\ \mathcal{R}^{-1} \;, 
\end{flalign}

\noindent \\
\textbf{Gas--FF}
\begin{flalign} \label{eq:urGFF}
u^r \big|^\mathrm{gas}_\mathrm{ff} =& \left( -2.10 \times 10^{5} \mathrm{cm\ s^{-1}} \right) \left( \alpha^{4/5}\ m_\bullet^{-1/2}\ \dmbullet^{3/10}\ x^{1/5} \right) \nonumber \\ 
&\times \mathcal{C}^{-1/10}\ \mathcal{D}^{4/5}\ \mathcal{P}^{-7/10}\ \mathcal{R}^{1/20} \;,
\end{flalign}
where the minus signs (i.e. negative radial velocities) indicate inwards migration of mass. For the \textbf{intra--ISCO} region, we follow the self--similar solutions of \cite{MB23}
\noindent \\
\textbf{intra--ISCO}
\begin{flalign} \label{eq:urintraISCO}
u^r \big|^\mathrm{intra}_\mathrm{isco} =& u_\mathrm{isco}\left[ \varepsilon^{-1} \left( \frac{\Risco}{r} - 1 \right)^{3/2} + 1 \right] \;,
\end{flalign}
where
\begin{flalign} \label{eq:epsilonISCO}
\varepsilon \equiv \frac{u_\mathrm{isco}}{c} \sqrt{\frac{3\Risco}{2\Rgrav}} \;.
\end{flalign}

Thus, the time--step limiter introduced to the simulation by our model can be written as
\begin{flalign} \label{eq:dt}
dt = \mathrm{max} \left( 10^3, 0.1 \times \mathrm{min}(\tdep, \tvisc) \right) \mathrm{yr} \;,
\end{flalign}
where a lower limit of $10^3$ years is introduced to avoid extremely small time--steps that will stall the simulation in cases when the mass of the accretion disc gets very close to complete depletion. In addition, a 0.1 factor multiplies the aforementioned two timescales to ensure that these two phenomena are at least partially captured. In reality, to fully resolve the viscous timescale everywhere in the accretion disc requires time--steps of the order of seconds, which is impossible to be achieved in a galaxy formation simulation. Finally, note that there are processes in the simulation that can restrict the time--step of black holes even further.

\section{Resolution \& Scaling testing} \label{app:Testing}

In addition to the simulation presented in \Sec{Applications}, we have generated four additional initial conditions (ICs) to test the robustness of the model. These ICs include variants with 0.5$\times$ and 2$\times$ the initial black hole mass compared to the original run (i.e. $10^7$ \Msun), and 0.5$\times$ and 2$\times$ the number of baryonic particles (i.e. modifying the resolution of the simulation). While these changes affect the process of capturing and accreting gas and the effects of feedback, we find that they do not fundamentally alter the physical framework of our model.

Reproducing the analysis presented in \Sec{Results} and \Sec{Applications} for the aforementioned four additional simulations would generate a wealth of additional plots. In addition, testing the model in a single isolated galaxy simulation even at different resolutions does not provide a statistically significant sample from which to draw robust conclusions about resolution effects. Hence, rather than introducing multiple additional plots of the same galaxy at varying resolutions, we instead make the accretion disc log files and ICs files of the aforementioned four simulations publicly available for the community to validate our findings.


\end{document}